\documentclass[g5paper,10pt,noblurb,electronic]{kthesis}
\usepackage{graphicx}
\usepackage{amsmath}
\usepackage{amssymb}
\usepackage[colorlinks=true,urlcolor=blue,linkcolor=black]{hyperref}
\usepackage{caption}
\usepackage{subfig}
\usepackage{color}
\usepackage[squaren]{SIunits}
\usepackage[numbers,sort&compress]{natbib}
\usepackage{textcomp}
\usepackage{bbm}
\usepackage{amsfonts} 
\usepackage{pifont} 
\usepackage{amssymb}

\usepackage{numprint}
\npthousandsep{,}\npthousandthpartsep{}\npdecimalsign{.}

\usepackage{array}
\usepackage{longtable}
\usepackage{threeparttable}
\usepackage{multirow}

\usepackage{algorithmic}

\definecolor{dkgreen}{rgb}{0,0.6,0}
\definecolor{gray}{rgb}{0.5,0.5,0.5}

\newcommand{\cref}[1]{Chapter~\ref{#1}}
\newcommand{\sref}[1]{Section~\ref{#1}}
\newcommand{\eref}[1]{Equation~\ref{#1}}
\newcommand{\aref}[1]{Appendix~\ref{#1}}

\usepackage{tikz}
\usetikzlibrary{shapes,arrows}

\begin{document}

\title{Partitioning Graph Databases}
\subtitle{A Quantitative Evaluation}
\author{Alex Averbuch \& Martin Neumann}
\date{\today}
\thesistype{Master of Science Thesis}
\imprint{Stockholm, Sweden 2010}
\trita{TRITA-ICT-EX-2010:275}
\publisher{Universitetsservice US AB}
\address{KTH School of Computer Science and Communication\\ SE-100 44 Stockholm\\ SWEDEN}
\kthlogo{images/logos/kth_svv_comp_science_comm}
\maxsecnumdepth{subsection}
\setsecnumdepth{subsection}
\maxtocdepth{subsection}
\settocdepth{subsection}
\renewcommand{\partnamefont} {\usefont{T1}{lmss}{sbc}{n}\boldmath\huge}
\renewcommand{\partnumfont} {\usefont{T1}{lmss}{sbc}{n}\boldmath\huge}
\renewcommand{\parttitlefont}{\usefont{T1}{lmss}{sbc}{n}\boldmath\Huge}
\renewcommand{\chapnamefont} {\usefont{T1}{lmss}{sbc}{n}\boldmath\huge}
\renewcommand{\chapnumfont} {\usefont{T1}{lmss}{sbc}{n}\boldmath\huge}
\renewcommand{\chaptitlefont}{\usefont{T1}{lmss}{sbc}{n}\boldmath\Huge}
\setsecheadstyle {\usefont{T1}{lmss}{bx}{n}\boldmath\Large\raggedright}
\setsubsecheadstyle{\usefont{T1}{lmss}{bx}{n}\boldmath\large\raggedright}
\setparaheadstyle {\normalsize\bfseries\boldmath}
\setsubparaheadstyle{\normalsize\bfseries\boldmath}
\makeevenhead{headings}
{\normalfont\small\thepage}{}{\normalfont\small\leftmark}
\makeoddhead{headings}
{\normalfont\small\rightmark}{}{\normalfont\small\thepage}
\setlength{\parskip}{0pt}

\maketitle

\pagenumbering{roman}

\addcontentsline{toc}{chapter}{Acknowledgements}
\chapter*{Acknowledgements}

We would like to thank everyone in the Neo4j team. 
They gave us the opportunity to work on this project in the first place,
and financially supported us for the duration of it. Without their help this would not have been possible.
The Neo4j team made time to meet with us and answer questions on a regular basis, 
in spite of their many other commitments.
Most importantly, they introduced us to the wonderful world of graph shaped data!

Thanks to our supervisor, \v{S}ar\={u}nas Girdzijauskas.
\v{S}ar\={u}nas made time to discuss our problems, guided us in setting goals, and kept us on schedule.
We also appreciate his help in proof reading our work and always providing constructive criticism --- 
despite the fact that it usually resulted in more work for us.

Information regarding the practical uses of graphs and graph algorithms is not yet widely available,
making it difficult to gain an appreciation for the benefits of graph processing.
We would like to thank Marko Rodriguez for sharing his extensive experience in graph processing systems,
and for always providing a fresh perspective on problems.

Thanks too to Wojtek Galuba for providing the sample Twitter dataset,
and to Henning Meyerhenke, 
for answering our questions regarding implementation of the DiDiC algorithm.

Lastly, thanks to our friends and families for keeping us sane during the course of this project.
\cleardoublepage

\addcontentsline{toc}{chapter}{Abstract}
\chapter*{Abstract}

The amount of globally stored, electronic data is growing at an increasing rate. 
This growth is both in size and connectivity,
where connectivity refers to the increasing presence of, and interest in, 
relationships between data \cite{ref:graph_general72}.
An example of such data is the social network graph created and stored by Twitter \cite{ref:general49}.

Due to this growth, demand is increasing for technologies that can process such data.
Currently relational databases are the predominant data storage technology,
but they are poorly suited to processing connected data as they are optimized for index-intensive operations.
Conversely, the storage engines of graph databases are optimized for graph computation as they store records adjacent to one another, linked by direct references.
This enables retrieval of adjacent elements in constant time, regardless of graph size,
and allows for relationships to be followed without performing costly index lookups.
However, as data volume increases these databases outgrow the resources available on a single computer,
and partitioning the data becomes necessary. 
At present, few graph databases are capable of doing this \cite{ref:graph_db71}.

In this work we evaluate the viability of using graph partitioning algorithms as a means of partitioning graph databases,
with focus on the Neo4j graph database \cite{ref:graph_db37}.
For this purpose, a prototype partitioned database was developed. 
Then, three partitioning algorithms were explored and one implemented.
During evaluation, three graph datasets were used: two from production applications, and one synthetically generated.
These were partitioned in various ways and the impact on database performance was measured.
To gauge this impact, 
we defined one synthetic access pattern per dataset and executed each one on the partitioned datasets.
Evaluation took place in a simulation environment, 
which ensured repeatability and made it possible to measure certain metrics, such as network traffic and load balance.

Simulation results show that, compared to random partitioning, 
use of a graph partitioning algorithm reduced inter-partition traffic by 40--90\,\%, depending on dataset.
Executing the algorithm intermittently during database usage was shown to maintain partition quality,
while requiring only 1\,\% the computation time of initially partitioning the datasets.
Finally, a strong correlation was found between theoretic graph partitioning quality metrics and the generated inter-partition traffic under non-uniform access patterns.
Our results suggest that use of such algorithms to partition graph databases can result in significant performance benefits, and warrants further investigation.

\cleardoublepage


\tableofcontents*{}
\cleardoublepage

\listoffigures
\cleardoublepage

\listoftables
\cleardoublepage

\pagenumbering{arabic}

\chapter{Introduction}
\label{cha:intro}

\section{Motivation}

Modern social networking services, such as Facebook and Twitter, 
store and process petabytes of data each day \cite{ref:graph_general73}.
This data can naturally be modeled as a graph, 
where vertices represent users and edges represent the personal relationships between them.
Moreover, the operations performed on this data are often in the form of graph traversals.
That is, they walk along elements (vertices and edges) of the graph in order to find the content of interest. 
An example of such an operation is searching for all friends-of-friends for a given user.
Similar to social networks, search providers like Google and Yahoo! also manage petabytes of data,
however they do it to capture and analyze the structure of the Internet.
Again, this data (and the Internet itself) is a massive graph,
but in this case websites are the vertices and the hyperlinks between websites are edges.
Interestingly, the type and scale of data exemplified here is not unique to social networks and search providers.
These are simply examples of a growing trend:
the exponential growth of publicly available, highly connected (graph-like) data \cite{ref:graph_general72}.
Consequently, there is an increasing demand for storage systems that are capable of processing large graph-like datasets.

Traditional storage technologies, like relational databases, are capable of storing this type of data.
However, to perform traversals they must frequently move between tables, 
which translates to join and index operations. 
%
%
%
As efficiency of these operations is dependent on the size of tables, this solution scales poorly. 
In contrast, graph databases are optimized for the storage and exploration of graph data structures.
By storing connected records adjacent to one another, linked by direct references,
graph databases model graphs explicit in their storage format. 
This allows them to retrieve adjacent entities in constant time, regardless of dataset size, 
making them the more scalable solution when modeling and computing on highly connected data.

Nevertheless, although the ability to model highly connected data is an advantage in many respects, 
it introduces challenges for graph databases when a dataset grows beyond the resources of a single computer and must be partitioned. 
Due to its high connectivity, such data naturally contains many dependencies between data items. 
When partitioning, some of these dependencies will connect data items on different partitions. 
These inter-partition dependencies then lead to communication overhead between partitions when they are traversed.
To minimize the performance impact caused by this communication the number of inter-partition dependencies should be kept minimal.
In graph theory, this problem is formalized as the Graph Partitioning Problem and is known to be NP-complete. 
Due to the inherent difficulty, very few \cite{ref:graph_db71} graph databases are capable of performing intelligent partitioning at present.

Fortunately, graph partitioning is a well studied problem.
The benefit of this is that much research into graph partitioning already exists, 
and a multitude of heuristic-based partitioning algorithms have been developed.
However, although many partitioning algorithms exist, they are often developed with different assumptions in mind.
For example, some algorithms are developed under the assumption that the entire graph is accessible at all times. 
In reality, when working with large partitioned datasets the performance penalty incurred due to network latency makes it impractical to use algorithms based on this assumption.
Additionally, a limitation that applies to most graph partitioning research is that algorithms are evaluated using theoretic quality metrics only.
Those theoretic evaluations rarely measure how the performance of a graph database is affected when the dataset it stores is partitioned.
In particular, they assume that every dependency (edge) is of equal importance,
which is unlikely to be the case for data stored in graph databases. 
Some edges stored by a graph database will inevitably be traversed with a higher frequency than others.
Therefore, it is of greater importance that those edges do not cross between two partitions, 
as they will create a relatively high amount of traffic.
Due to these points, it is difficult to determine which graph partitioning algorithms --- if any --- 
are well suited to partitioning graph databases, or if they are competitive with naive methods like random partitioning. 

In the interest of answering these questions, 
this work evaluates the viability of various graph partitioning algorithms as a means of partitioning graph databases.
To achieve this, a set of properties were defined, which we deem desirable for such an algorithm to have. 
For example, the algorithm should not depend on a view of the entire graph (as previously mentioned), 
and it shall be able to maintain the quality of partitions in a dynamic environment 
(when the graph undergoes modifications at runtime). 
Using the defined properties, three graph partitioning algorithms \cite{ref:partition27, ref:partition53, ref:partition26}
were identified and assessed in detail.
During this assessment, the algorithms were compared to one another, 
with special emphasis placed on their suitability to partitioning graph databases. 
Finally, one algorithm \cite{ref:partition26} (DiDiC) was selected and its impact on graph database performance was compared against that of more naive partitioning methods, including random partitioning.

To ensure a thorough comparison was performed,
each partitioning method was used to partition three different graph datasets:
two from production applications, and one synthetically generated.
The synthetic dataset was modeled against the tree structure of a file system,
while the others originated from a Geographic Information System (GIS), and by crawling Twitter \cite{ref:graph_general65}.
These datasets represent vastly different application domains and differ greatly in structure,
allowing us to analyze the partitioning methods under a broad range of conditions.
To gauge how well the datasets were partitioned, three synthetic access patterns were defined, and later executed on the partitioned datasets.
Note that, to capture the uniqueness of our datasets, each access pattern was tailored to the domain from which the dataset originated.
For example, access patterns on the Twitter dataset were defined as friend-of-a-friend operations as these are common in social networks.
Finally, to execute these access patterns a prototype partitioned database and simulation environment were developed.
These were built as an extension of Neo4j \cite{ref:graph_db37},
an open-source graph database that currently lacks the ability to perform intelligent partitioning.
The simulation environment was used to execute access patterns in a repeatable manner, 
while measuring certain metrics, including the generated inter-partition traffic (number of operations that require two different partitions to communicate).

Results from our evaluation showed that, compared to random partitioning, 
use of the DiDiC algorithm reduced inter-partition traffic by 40--90\,\%, depending on dataset.
This indicates that the structural properties of a dataset, such as graph topology, have an influence on how easily it can be partitioned.
Furthermore, a strong correlation was found between theoretic graph partitioning quality metrics, such as edge cut (see \cref{cha:graph_partitioning}), 
and the generated inter-partition traffic under arbitrary access patterns.
The value of this correlation is that it makes it possible to map the theoretic results obtained from other bodies of research, to real world performance metrics.
Ultimately, our results suggest that use of graph partitioning algorithms to partition graph databases can result in significant performance benefits, and warrants further investigation.

We are not aware of other work related to the application of graph partitioning algorithms to graph databases
that contains evaluation of the same nature or level of detail.

\section{Related Work}

Published work related to partitioning graph databases is scarce.
Despite this, research applied to distributed graph processing frameworks, graph storage in the cloud, 
and partitioning of online social networks (OSNs) is more common.


YARS2 \cite{ref:graph_db16} 
presents an architecture for a semantic search and query engine over distributed graph structured data,
placing particular focus on the design of distributed indexes for such an application.
In contrast to YARS2, we assume an efficient index structure exists and place most importance on optimizing index-free, graph traversal operations.

Plasma \cite{ref:traversal13}, a graph based distributed computing model,
describes a platform designed to support federated graph repositories.
In Plasma queries are performed locally and appear local, 
but are transparently distributed by the framework query planner.
Unfortunately, the Plasma paper lacks specifics and implementation details.
Furthermore, it addresses the federation of graph repositories rather than the problem of reducing network traffic between those repositories.
As we focus on partitioning and distributing a single graph database instance, 
the contributions made by Plasma were not applicable to this thesis.

Pegasus \cite{ref:graph_db36} is an open source peta-scale graph mining library implemented on top of Hadoop \cite{ref:scalable40}, the open source version of MapReduce \cite{ref:scalable41}.
The Pegasus project attempts to express graph algorithms using the MapReduce paradigm, 
thereby benefiting from the transparent data distribution performed by Hadoop ---
albeit with limited control over data placement or on-disk file format.
Our work explicitly considers the affects of data placement,
and attempts to maintain backwards compatibility with the Neo4j API.
For these reasons Hadoop was deemed unsuitable.

Pregel \cite{ref:graph_db38} is a graph computational model developed by Google,
and designed for scalable and fault-tolerant operation on thousands of commodity computers.
It has been shown to scale to billions of vertices and trillions of edges. 
Pregel was inspired by the Bulk Synchronous Parallel model \cite{ref:scalable31};
a computational model for the execution of parallel algorithms on top of multiple sequential von Neumann machines.
Pregel programs are expressed as a sequence of iterations,
in each of which vertices receive messages sent in the previous iteration, send messages to other vertices, 
and modify their own state. 
The Pregel paper presents impressive results with regards to performance and scalability,
however these results were achieved using a random partitioning scheme. 
Results using other partitioning methods were not given.

Surfer \cite{ref:partition29} is a distributed computing framework for graphs on heterogeneous clouds.
Observing that cloud infrastructures often have network topologies resembling a tree,
Surfer models the computer network as a graph, models the dataset as a graph,
then partitions both using recursive bisection.
The graphs are then mapped to each other, by assigning dataset partitions to computer clusters.
This is referred to as bandwidth-aware partitioning as it attempts to minimize inter-cluster 
communication and maximize intra-cluster communication.
Surfer supports MapReduce and a custom approach, Propagation, for interacting with the graph, 
but notes MapReduce is poorly suited to graph computations as it results in excessive network traffic.
Similar to Surfer, 
our work makes use of graph partitioning algorithms to aid in the distribution of graph datasets.
However, as Surfer makes no attempt to adapt the partitioning when the dataset changes,
the algorithm used in Surfer was not well suited to our problem.

The authors of \cite{ref:partition48} explore properties of OSNs;
specifically, how to use them in guiding the choice of partitioning algorithms.
Several partitioning algorithms are evaluated.
They observe that, given selection of an appropriate algorithm,
traditional graph partitioning techniques can be effectively used to reduce network traffic in OSN applications.
A later paper \cite{ref:partition47} by the same authors builds on this work.
It evaluates the combination of an algorithm from \cite{ref:partition48} with a replication scheme that replicates vertices on the boundary of partitions.
Evaluation focuses on measuring the overhead of performing the actual replication, 
in terms of additional storage used and network bandwidth consumed.
Many simplifying assumptions are made, including: only OSN graphs are considered, 
graph traversal operations are assumed to extend only one hop,
the partitioning framework requires access to the entire graph when performing replication,
and the graph is assumed to be static.

\section{Thesis Outline}

The remainder of this document is structured as follows:
\cref{cha:graphs_and_graphdbs} covers graph basics, how graphs are queried and persisted,
and their applications;
\cref{cha:graph_partitioning} describes graph partitioning algorithms,
and their application to graph databases;
\cref{cha:prototype_partitioning_alg} summarizes the graph partitioning algorithms that were considered for use in evaluation;
\cref{cha:prototype_partitioned_db} details the design of our prototype partitioned graph database;
\cref{cha:eval_method} outlines the evaluation method used;
\cref{cha:eval_results} presents results and supporting reasoning;
finally, \cref{cha:conclusion} provides the conclusion.
\chapter{Graphs \& Graph Databases}
\label{cha:graphs_and_graphdbs}

Graphs are fundamental data structures in computer science. 
The purpose of this chapter is to cover graph basics, the various types of graphs and differences between them, techniques for querying graph data, and to provide an introduction to graph databases.
This chapter is essentially a summary of \cite{ref:graph_db35,ref:graph_db43}. 
For greater detail please refer to those papers.

\section{Graph Types}
\label{sec:graph_types}

A graph is a data structure composed of vertices and edges, where edges express relationships between vertices.
The most basic example is a simple graph; composed of a set of vertices, a set of edges, edges are undirected, 
each edge must connect two unique vertices, and no two vertices may have more than one edge between them.

\begin{figure}[htbp]
\begin{center}
\includegraphics[scale=1.15]{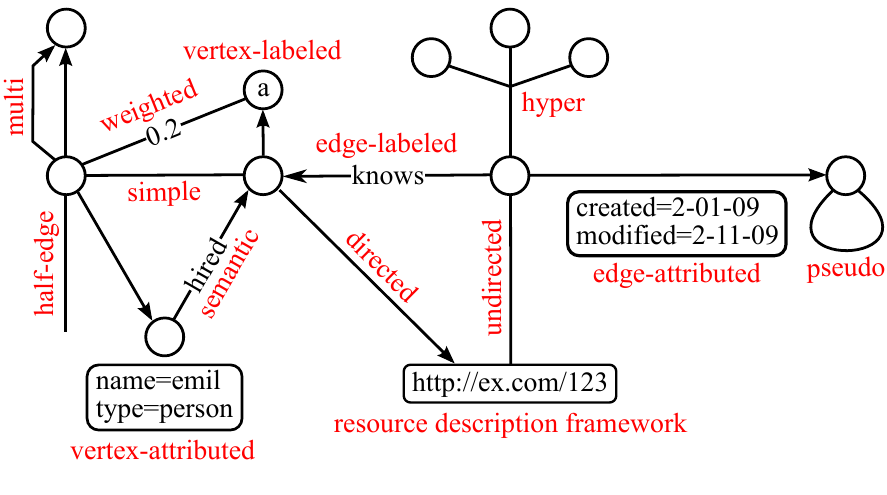} 
\caption{Graph types}
\label{img:graph_types}
\end{center}
\end{figure}

Although already a powerful tool for expressing objects and their relationships to one another,
this model can be enriched in numerous ways. For example, vertices may be given names, 
and edges extended to have weights and directions. The result is more expressive graphs.
\tref{tab:graph_types} contains a list of graph types, obtained from \cite{ref:graph_db43}.
The list is not exhaustive but intended to provide a general overview of ways in which graphs may be extended.
Note that in some cases combinations of these graph types may be constructed; 
they are not necessarily mutually exclusive.
\begin{table}[htbp]
\centering
\begin{threeparttable}

\begin{tabular}
{|>{\raggedright}p{3.0cm}|>{\raggedright\arraybackslash}p{9.0cm}|}

  \hline \textbf{Graph Type} & \textbf{Description} \tabularnewline  \hline \hline   
  Simple & Prototypical graph. An edge connects two vertices and no loops are allowed \tabularnewline \hline  
  Undirected & Typical graph. Used when relationships are symmetric \tabularnewline \hline  
  Multi & Allows multiple edges between the same two vertices \tabularnewline \hline 
  Weighted & Represents strength of ties or transition probabilities \tabularnewline \hline   
  Semantic & Models cognitive structures such as the relationship between concepts and the instances of those concepts \tabularnewline \hline 
  Vertex-attributed & Allows non-relational meta-data to be appended to vertices \tabularnewline \hline   
  Vertex-labeled &  Allows vertices to have labels (e.g. identifiers) \tabularnewline \hline 
  Edge-attributed & Allows non-relational meta-data to be appended to edges \tabularnewline \hline   
  Edge-labeled & Denotes the way in which two vertices are related (e.g. friendships) \tabularnewline \hline   
  Directed & Orders the vertices of an edge to denote edge orientation \tabularnewline \hline   
  RDF\rlap{\tnote{a}}\ \ \cite{ref:graph_db42} & Graph standard, developed by W3C\rlap{\tnote{b}}\ . 
  Vertices and edges are denoted using URI\rlap{\tnote{c}}\ \tabularnewline \hline   
  Half-edge & A unary edge (i.e. an edge connects to one vertex only) \tabularnewline \hline   
  Pseudo & Used to denote reflexive relationships \tabularnewline \hline   
  Hypergraph & An edge may connect an arbitrary number of vertices \tabularnewline \hline   
\end{tabular}
     \begin{tablenotes}
       \item[a] Resource Description Framework       
       \item[b] World Wide Web consortium
       \item[c] Uniform Resource Identifiers
     \end{tablenotes}     
\end{threeparttable}
\caption{Graph types}
\label{tab:graph_types}
\end{table}	
\fref{img:graph_types}, also obtained from \cite{ref:graph_db43}, 
graphically illustrates the difference between these various graph types.

\begin{figure}[htbp]
\centering
\includegraphics[scale=1.15]{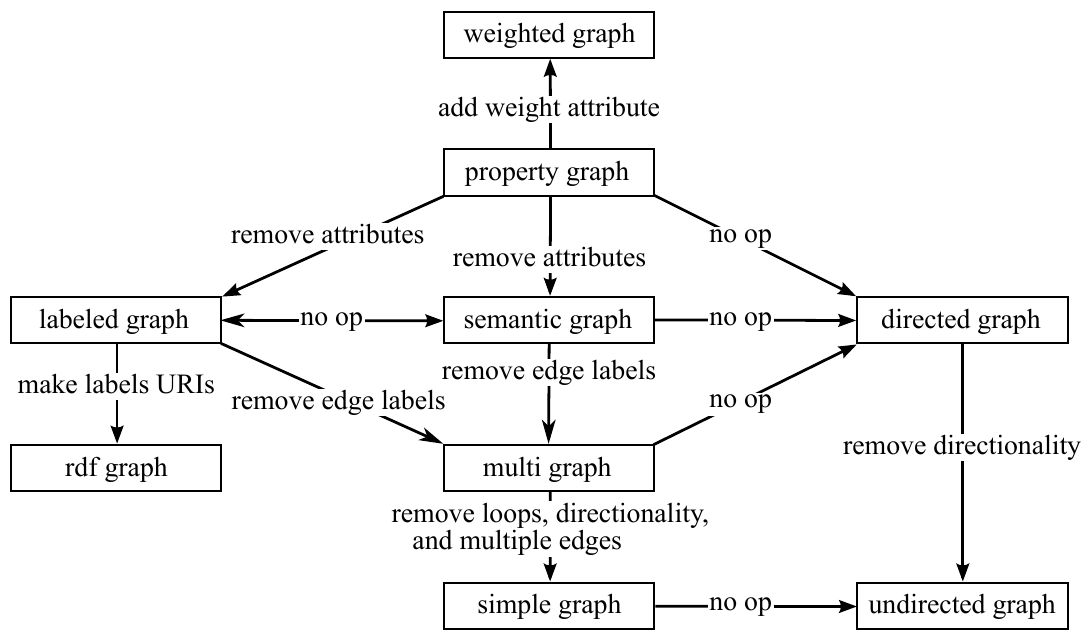} 
\caption{Graph type morphisms}
\label{img:graph_types_morphisms}
\end{figure}

Many graph systems support directed, labeled, attributed, multi-graphs.
This amalgamation is commonly referred to as a \textit{property graph}.
The popularity of property graphs is largely due to their flexibility in being able to express other graph types.
For example, by not making use of vertex/edge attributes a semantic graph is generated, 
or by adding weight attributes to edges a weighted graph is generated.
This process of creating new graph types by extending/restricting other graph types is referred to as 
\textit{graph type morphism} in \cite{ref:graph_db43}, and illustrated in \fref{img:graph_types_morphisms}.

\section{Graph Traversal Pattern}
\label{sec:graph_access_patterns}

The primary way to compute on graphs is using graph traversals.
In essence, traversals walk along the elements of a graph and are the fundamental building blocks of graph algorithms \cite{ref:graph_db15}; algorithms that either quantify aspects of a graph, alter the graph state,
or solve problems that are a function of the graph structure.
This unique problem-solving style is coined the \textit{graph traversal pattern} by \cite{ref:graph_db35}.

As mentioned previously, property graphs are a predominant graph type.
However, property graphs have labeled edges, whereas most graph algorithms were developed for unlabeled graphs. 
Consequently, when vertices are related to each other in many different ways, 
the meaning of results returned by standard graph algorithms become ambiguous. 
A novel solution to this problem is to interpret a path through a graph as a virtual edge,
thereby creating a virtual graph over the original graph.
The virtual graph then contains only the edges that represent certain paths in the original graph.
It becomes equivalent to an unlabeled graph where all edges have the same meaning,
that of the path they represent.
Using the technique just described, 
it becomes possible to express standard graph algorithms on rich property graphs. 
The benefit of using edge-labeled (property) graphs is they contain many types of rankings;
as many as the types of paths that exist in the graph.

Complex applications, such as a social networks, might use a large range of different traversals.
In these cases it is unlikely that all vertices and edges are accessed with the same frequency.
Depending on the application, graph topology, and traversal types there will be a pattern of how frequently entities are visited. We define this as the \textit{access pattern}. 
Note, even if such patterns exist it may be difficult to measure or describe them.



Perhaps most importantly, the graph traversal pattern does not require a global analysis of data. 
For many problems only local subsets of the graph need to be traversed. 
By structuring the graph in such a way as to minimize traversal steps, 
limit the use of external indices, and reduce the number of set-based operations, 
users of the graph traversal pattern gain efficiency that would be difficult to obtain with other methods.

\section{Graph Databases}
\label{sec:graph_databases}

The interlinked table structure of relational databases \cite{ref:general_db44}
has been the predominant information storage and retrieval model during the last 50 years. 
With the recent growth of highly connected data and desire to process it, 
new data management systems have been developed. 
As defined in \cite{ref:graph_db43}, 
a graph database \cite{ref:graph_db46} is one optimized for graph traversals. 
The Neo4j \cite{ref:graph_db37} graph database is one such database.

In contrast to the index-intensive, set-theoretic operations of relational databases, 
graph databases make use of index-free, local traversals. 
The property graph type, supported by most graph databases, 
may still make use of indices to allow for the retrieval of elements from property values.
However, the index is only used to retrieve start elements,
from which point an index-free traversal is executed through the graph.
In graph databases graph elements are adjacent to one another by direct references.
Vertices are adjacent to their incoming/outgoing edges, 
and edges are adjacent to their incoming/outgoing vertices. 
The advantage afforded by this is constant time retrieval of adjacent elements;
complexity of local read operations remains constant regardless of graph size.

In principle, any database can be used to represent and process a graph. 
However, when traversals are the ultimate use case for the graph, graph databases are the optimal solution.
Relational databases can efficiently join tables, 
in order to move between tables that are linked by certain columns.
One of their limitations though, is the graph they model is not explicit in the relational structure, 
it must be inferred through a series of index-intensive operations. 
Another drawback, while a subset of the data may be desired, 
join operations require all data in all queried tables be examined in order to extract the desired subset.

By definition graph databases provide index-free adjacency, 
therefore they excel when traversals span multiple steps and unite disparate vertices by a path. 
This is because no index-lookup operations are performed during traversals.
Additionally, graph databases perform no explicit join operations, traversing an edge is analogous to joining. 
As a result, graph databases allow for operations that would not be feasible using other storage solutions.
Traversals comprised of complex paths become possible,
and the type of path taken can be used to define inferred relationships between the two vertices it connects.

\begin{figure}[htbp]
\begin{center}
\includegraphics[scale=1.15]{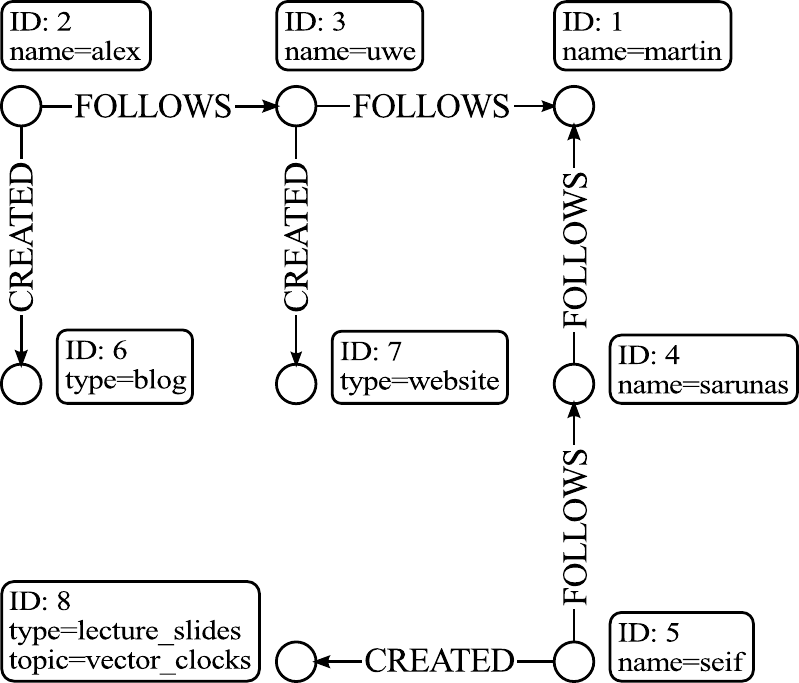} 
\caption{Example social network graph}
\label{img:example_graph}
\end{center}
\end{figure}

To illustrate how a graph database can be queried consider a property graph that models a social network (\fref{img:example_graph}). Vertices in this graph represent one of two things, a person or content (e.g. web page) created by a person. The graph contains edges of two types, \texttt{follows} and \texttt{created}. The \texttt{follows} edge type connects two people in the social network, and represents a follows relationship as used in the Twitter \cite{ref:general49} micro-blogging application. The \texttt{created} edge type connects a person with the content they created. To learn more about the interests of members in their social group, users of this graph may want to know what content was created by the followers of their followers. This knowledge can be obtained by traversing the graph. 

For this purpose the Neo4j graph database provides a native Java API and the Gremlin \cite{ref:graph_db34} graph programming language. 
Gremlin is a relatively simple, database agnostic, language with a syntax similar to XPath \cite{ref:general64}.
Using Gremlin, graph queries can be expressed in a succinct way.
A limited subset of the Gremlin syntax is presented in \tref{tab:gremlin_syntax}.

\begin{table}[htbp]
\extrarowheight = 0.5mm
\begin{center}
\begin{tabular}
{|>{\raggedright}p{0.3\textwidth}|>{\raggedright\arraybackslash}p{0.63\textwidth}|}

	\hline
	\textbf{Function} &
	\textbf{Description} \\
	\hline
	\hline

	\texttt{g:id-v(id)} &
	Retrieves a vertex by ID from graph \texttt{g}\\
	\hline
	
	\texttt{inE} &
	Returns all incoming edges attached to a vertex \\
	\hline
	
	\texttt{outE} &
	Returns all outgoing edges attached to a vertex \\
	\hline
	
	\texttt{bothE} &
	Returns all edges attached to a vertex \\
	\hline
	
	\texttt{inE[$@$\mbox{label}=\mbox{`type'}]} &
	Returns all incoming edges of a certain type, that are attached to a vertex.
	Also applies to the \texttt{outE} and \texttt{bothE} functions.\\
	\hline
	
	\texttt{inV} &
	Returns the source vertex of an edge \\
	\hline
	
	\texttt{outV} &
	Returns the destination vertex of an edge \\
	\hline

	\texttt{bothV} &
	Returns source and destination vertices of an edge \\
	\hline
	
\end{tabular}
\end{center}
\caption{Subset of the Gremlin syntax --- Version 0.2.2}
\label{tab:gremlin_syntax}
\end{table}

\newpage

For the example graph presented in \fref{img:example_graph}, the question ``what content did Martin's followers' followers create?'' can be expressed in Gremlin as:
\begin{center}
\begin{tabular}
{>{\raggedright\arraybackslash}p{12.0cm}}
\texttt{g:id-v(1)/inE[$@$\mbox{label}=\mbox{`follows'}]/outV/inE[$@$\mbox{label}=\mbox{`follows'}]/outV/} \\
\texttt{outE[$@$\mbox{label}=\mbox{`created'}]/inV} \\
\end{tabular} 
\end{center}
This expression defines the type of traversal/path that will be taken through the graph, and would return the vertices at the end of that path. In this case the result would be the \textit{blog} vertex with ID 6 and the \textit{lecture\_notes} vertex with ID 8.

\section{Applications}

As mentioned in \sref{sec:graph_access_patterns}, 
using graph databases is most beneficial when queries can be expressed as traversals over local regions of a graph. 
Problems that are well suited to this approach are scoring, recommending, searching, and ranking.

Scoring refers to applications that, given some vertices and a path description, 
will return a score for those vertices.
For example, ``Score user $X$ given the other users in their immediate social group''.
Searching refers to applications that, 
given some starting vertices and a path description, will return the vertices at the end of that path.
An example of this is the query presented in \sref{sec:graph_databases}.
Recommendation refers to applications that provide users with suggestions, 
such as ``If you like product $X$, you may also like product $Y$''.
Finally, ranking refers to applications that,
given some vertices and a path description, return a map of scored vertices.
An example of this is the PageRank algorithm \cite{ref:scalable39} used by Google. 

\newpage{}
These are all examples of traversals that are defined as abstract paths through a graph,
where the paths taken determine the meaning of a rank, score, recommendation, or search result.
Because graph databases maintain direct references between related data,
they are most efficient when performing this type of local data analysis.
They allow many of these traversals to be computed in real-time.
Other storage technologies, such as relational databases, can also be used to execute such operations.
However, they are impractical because when the length of traversals grows the cost of performing index lookups 
--- often of $O(log(n))$ complexity --- will dominate computation time.

In summary, a graph database may be the optimal technology to solve a problem if the solution to that problem:
can be expressed using local traversals within a larger data structure,
can be represented as being with respect to a set of root elements,
or does not require a global analysis of the data.


%

\chapter{Graph Partitioning}
\label{cha:graph_partitioning}

As previously mentioned, currently few graph databases are capable of performing intelligent partitioning.
However, as graph partitioning is an established field, many graph partitioning algorithms already exist.
In our work we try to reuse these algorithms to partition graph databases.

As the field is broad, 
we grouped partitioning algorithms into categories for the faster identification of algorithms that may be 
suitable for a given type of problem.
In our work this was achieved by first identifying properties that are common to a large number of partitioning algorithms. 
We consider the properties in \tref{tab:partitioning_properties} to be desirable, 
as each provides an advantage in performance, scalability or expressiveness.

\begin{table}[htbp]
\extrarowheight = 1.0mm
\begin{center}
\begin{threeparttable}
\begin{tabular}
{|>{\centering}p{3.0cm}|>{\raggedright\arraybackslash}p{8.7cm}|}
  \hline 

  \textbf{Property} & 
  \multicolumn{1}{c|}{\textbf{Description}} \tabularnewline  
  
  \hline   
  \hline   
  
  \multirow{4}{*}{Local View} &  
  At no point does the algorithm require access to all of the graph state. Additionally, 
  complexity of all algorithm operations must be bounded by the size of subgraph explored in those operations, 
  with at most weak dependency on the size of the graph.
  \tabularnewline \hline   
  
  \multirow{2}{*}{Distributed} &
  Algorithm is capable of executing on multiple networked computers, concurrently.
  \tabularnewline \hline   
  
  \multirow{2}{*}{Parallel} &
  Algorithm is capable of executing on multiple processor cores of the same computer, concurrently.
  \tabularnewline \hline   
  
  \multirow{2}{*}{Multi-Constraint} &  
  Algorithm optimizes more than one constraint (see \sref{sec:definitions}) when computing a partitioning.
  \tabularnewline \hline   

  \multirow{3}{*}{Dynamic} &
  Algorithm is capable of maintaining the quality of a partitioning in the presence of dynamism;
  when vertices and/or edges are added and/or removed from the graph.
  \tabularnewline \hline   
  
  \multirow{4}{*}{Smooth} &   
  Generally considered only for dynamic algorithms. 
  Given atomic changes to the graph, algorithm keeps consecutive partitionings similar, 
  preserving as much of the previous partitioning as possible.
  \tabularnewline \hline   
  
  \multirow{4}{*}{Iterative} &  
  Rather than attempting to compute an optimal partitioning and then terminating,
  the algorithm continues in iterations, indefinitely.
  Each subsequent iteration improves the partitioning quality beyond that of the previous iteration.
  \tabularnewline \hline    
   
  \multirow{3}{*}{Weighted} &
  Edge weights are considered when performing a partitioning.
  Algorithm attempts to cut higher weighted edges with lower probability.
  \tabularnewline \hline   
  
  \multirow{3}{*}{Complexity} &
  Algorithms with a lower computational complexity are more desirable.
  They allow for the partitioning of larger graphs.
  \tabularnewline \hline   

\end{tabular}
\end{threeparttable}
\caption{Partitioning algorithm properties}
\label{tab:partitioning_properties}
\end{center}
\end{table}	

The remainder of this chapter condenses numerous graph partitioning works into a summarized form,
placing specific focus on the aspects considered most important to our work.

\section{Partitioning Graph Databases}	
\label{sec:partitioning_graphdbs}

One potential application of graph partitioning algorithms is to partition graph databases.
However, graph databases represent specific graph implementations,
operating in unique environments, with particular requirements.
For example, in many cases graph databases are used in long living applications,
subjected to continuous changes throughout their lifetime.
These characteristics must be considered.

As explained in \cref{cha:graphs_and_graphdbs}, 
a primary reason for using graph databases is their support for efficient execution of graph traversals.
As a consequence, it is reasonable to assume that access patterns executed on the database will also be in the form of graph traversals.
One goal when partitioning graph databases is then,
to reduce the number of traversal operations that are required to cross partition boundaries.

Another goal is to balance the load across database partitions, 
not just with respect to partition size, but traffic per partition also;
size and traffic are only equivalent in the event that access patterns are uniformly random.

\begin{figure}[htbp]
\begin{center}
\includegraphics[scale=1.0]{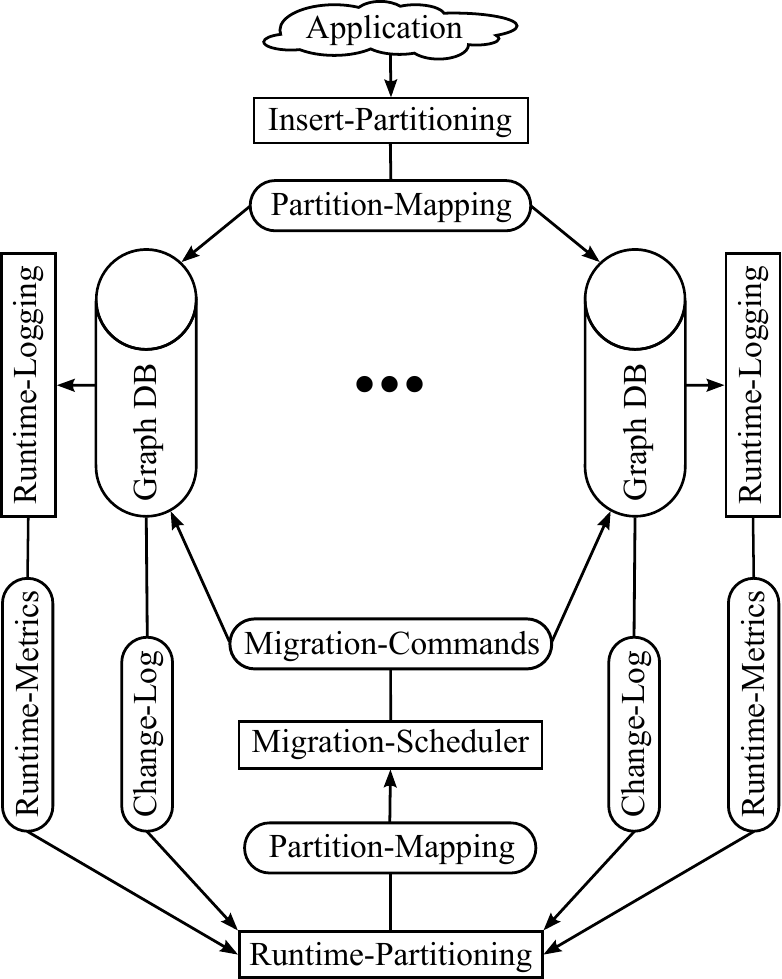} 
\caption{Graph database partitioning architecture}
\label{img:partitioned_graphdb}
\end{center}
\end{figure}

To solve these problems using a graph partitioning algorithm we submit that a supporting framework is necessary.
To outline the requirements of such a framework, we define the following abstractions,
which we consider necessary to distribute graph databases
(a visual illustration of the architecture for this proposed partitioning framework is presented in
\fref{img:partitioned_graphdb}):

\begin{description}
\item[Insert-Partitioning:] 
If the graph topology and/or access patterns are known, 
data can be allocated to partitions at the time it is written to the database.
This is beneficial as it requires only low computational complexity.
To abstract this partition allocation logic into a decoupled component,
we define the Insert-Partitioning component.

\begin{tabular}
{l>{\raggedright\arraybackslash}p{8.0cm}}
\textbf{Inputs:} & Insert-Partitioning-Function, Data-Entities \\ 
\textbf{Outputs:} & Partition-Mapping \\ 
\end{tabular} 

\item[Runtime-Logging:]
Many applications are designed to model dynamic, continuously evolving domains. 
For example, graphs maintained by social networks change every time a user starts a relationship, 
moves to another city, or communicates using an existing relationship.

Insert-Partitioning is computationally inexpensive and adequate in some situations, 
but it assumes the general graph topology and/or access patterns will remain unchanged.
Partition allocation may be optimal when performed, 
then invalidated later due to changes to the graph or the way it is accessed.

Metrics such as access patterns, partition sizes, 
and traffic per partition can be used to recognize when partitioning quality has degraded. 
We define a Runtime-Logging component that encapsulates the task of collecting such metrics.

\begin{tabular}
{l>{\raggedright\arraybackslash}p{8.0cm}}
\textbf{Inputs:} & Runtime-Logging-Function \\ 
\textbf{Outputs:} & Runtime-Metrics \\ 
\end{tabular} 

\item[Runtime-Partitioning:] 
As the topology of a graph or its access patterns evolve, partitioning quality may degrade, 
and previously inserted data may need to be reallocated to new partitions.
The data allocation performed by Insert-Partitioning must be updated at runtime and, 
as with Insert-Partitioning, the way in which this is implemented is domain specific.

Note, although the use of graph partitioning algorithms may be suited to this task,
they are not mandatory.

For the responsibility of allocating entities to partitions, 
we define the Runtime-Partitioning component.

\begin{tabular}
{l>{\raggedright\arraybackslash}p{8.0cm}}
\textbf{Inputs:} & Runtime-Partitioning-Function, Runtime-Metrics, Change-Log \\ 
\textbf{Outputs:} & Partition-Mapping \\ \\ \\
\end{tabular} 

\item[Migration-Scheduler:]
Partition reallocation may be beneficial, 
but if the data migration occurs during peak traffic, database performance will be affected;
the process becomes counter productive. 

Partition-Mapping, produced by the Runtime-Partitioning component, 
is a set of instructions specifying where to migrate data, 
but says nothing about \textit{when} to migrate.

For that purpose we need a module that is responsible for deciding when data migration should occur, 
and then commands the partition servers to perform migration.
To perform this task, we define the Migration-Scheduler component.
  
\begin{tabular}
{l>{\raggedright\arraybackslash}p{8.0cm}}
\textbf{Inputs:} & Migration-Scheduler-Function, Partition-Mapping \\ 
\textbf{Outputs:} & Migration-Commands \\ 
\end{tabular} 
\end{description}	

	
\section{Definitions}			
\label{sec:definitions}

Graph partitioning algorithms partition a graph, $G$, into subgraphs, known as partitions.
For the purposes of this work a partitioning, $\Pi$, is a function that divides $V$, 
the vertices in $G$, into $k$ disjoint subsets (edges reside on the partition of their start vertex).
\eref{eq:partitioning} and \eref{eq:partitioning_result} formally define a partitioning.
\begin{equation}\label{eq:partitioning}
	\Pi = V \rightarrow \lbrace \pi_1,\ldots,\pi_k \rbrace 
\end{equation}
Such that,
\begin{equation}\label{eq:partitioning_result}
	V = \pi_1 \cup \pi_2 \cup \ldots \cup \pi_k
\end{equation}

When working with partitioning algorithms, and graphs in general, 
it becomes evident certain concepts reoccur often.
As they will be repeatedly referred to in later chapters, 
\tref{tab:partitioning_general} provides a central point of reference to these concepts 
and their definitions.

\begin{table}[htbp]
\extrarowheight = 1.0mm
\begin{center}
\begin{threeparttable}
\begin{tabular}
{|>{\centering}p{3.0cm}|>{\raggedright\arraybackslash}p{8.6cm}|}
  \hline 

  \textbf{Concept} & \multicolumn{1}{c|}{\textbf{Description}} \tabularnewline  
  
  \hline   
  \hline   
  
  \multirow{4}{*}{Simple} \multirow{4}{*}{Graph} \linebreak
  \multirow{4}{*}{$G$} & 
  Set of vertices, $V$, and undirected edges, $E$. 
  \begin{eqnarray}
  	G = (V,E)
  \end{eqnarray}
  \tabularnewline \hline   
  
  Edge Weight\linebreak
  $wt(e)$ & 
  Weight value associated with edge $e$. 
  \tabularnewline \hline   

  \multirow{4}{*}{Degree} \linebreak
  \multirow{4}{*}{$d(v)$} & 
  Sum of the weights of edges connected to vertex $v$. 
  \begin{eqnarray}
  	d(v)= \sum_{e = \lbrace \cdot , v \rbrace \in E} wt(e)
  \end{eqnarray}
  \tabularnewline \hline   
  
  Partition \linebreak 
  $\pi$ & 
  Subset of vertices in the graph, $\pi \subseteq V$. \linebreak
  \tabularnewline \hline   
    
  \multirow{3}{*}{Partition} \\
  \multirow{3}{*}{Compliment} \\
  \multirow{3}{*}{$\pi^c$} & 
  All vertices in $V$ that are not in $\pi$.
  \begin{eqnarray}
  	\pi^c = V \setminus \pi
  \end{eqnarray}
  \tabularnewline \hline   

  \multirow{4}{*}{Volume} \linebreak 
  \multirow{4}{*}{$\mu(S)$} & 
  Sum of the degrees of vertices in a set, S.
  \begin{eqnarray}
  	\mu(S) = \sum_{v \in S} d(v)
  \end{eqnarray}
  \tabularnewline \hline   

  \multirow{4}{*}{Intra-Weight} \linebreak 
  \multirow{4}{*}{$iw(S)$} & 
  Sum of the weights of edges connecting a set of vertices.
  \begin{eqnarray}
  	iw(S)= \sum_{e = \lbrace u,v \rbrace \in E} wt(e) \mbox{ ; } u,v \in S
  \end{eqnarray}
  \tabularnewline \hline   

  \multirow{4}{*}{Partition} \multirow{4}{*}{Degree} \linebreak
  \multirow{4}{*}{$\partial(\pi)$} & 
  Sum of weights of edges leaving $\pi$.
  \begin{eqnarray}
  	\partial(\pi)= \sum_{e = \lbrace u,v\rbrace \in E} wt(e) \mbox{ ; } u \in \pi \wedge v \in \pi^c  
  \end{eqnarray}
  \tabularnewline \hline   
  
\end{tabular}
\end{threeparttable}
\caption{Partitioning concepts}
\label{tab:partitioning_general}
\end{center}
\end{table}	

While partitioning a graph, a partitioning algorithm attempts to optimize one or more parameters.
These parameters, often referred to as constraints, 
can also be used to measure the success, or quality of a partitioning.
\tref{tab:partitioning_constraints} contains definitions for a number of common constraints.
It presents the name and description of each constraint, 
and specifies whether it is desirable to minimize or maximize its value.

\begin{table}[htbp]
\extrarowheight = 1.0mm
\begin{center}
\begin{threeparttable}
\begin{tabular}
{|>{\centering}p{2.0cm}|>{\centering}p{0.9cm}|>{\raggedright\arraybackslash}p{8.5cm}|}
  \hline 

  \textbf{Constraint} & 
  \textbf{Goal} & 
  \multicolumn{1}{c|}{\textbf{Description}} \tabularnewline  
  
  \hline   
  \hline   
  
  \multirow{6}{*}{Edge Cut} 
  \multirow{6}{*}{$ec(G)$} & 
  \multirow{6}{*}{Min} & 
  Sum of weights of edges that cross partition boundaries. 
  \begin{eqnarray}\label{eq:constraint_edgecut}
  ec(G) = \sum^{\vert \Pi \vert}_{i=1} \left( \sum_{e = \lbrace u,v\rbrace \in E} wt(e) \mbox{ ; } 	
  u \in \pi_i \wedge v \in \pi^c_i \right) 
  \end{eqnarray}
  \tabularnewline \hline   

  \multirow{7}{*}{Conductance} 
  \multirow{7}{*}{$\phi(G)$} & 
  \multirow{7}{*}{Min} & 
  Ratio of Partition Degree, $\partial(\pi)$, to Partition Volume, $\mu(\pi)$. 	
  \begin{eqnarray}\label{eq:constraint_conductance}
	\phi(\pi) & = & \dfrac{\partial(\pi)}{\mu(\pi)} \nonumber \\
	\phi(G) & = & \min\limits_{\pi \in \Pi} \left( \phi(\pi) \right)
  \end{eqnarray}
  \tabularnewline \hline   
    
  \multirow{6}{*}{Modularity} 
  \multirow{6}{*}{$Mod(\Pi)$} & 
  \multirow{6}{*}{Max} & 
  Ratio of actual Edge Cut to expected Edge Cut given a random partitioning. 
  \begin{eqnarray}\label{eq:constraint_modularity}
  Mod(\Pi) = 
  \sum^{\vert \Pi \vert}_{i=1} \left( \dfrac{iw(\pi_i)}{iw(G)}- \left( \dfrac{\sum_{v\in \pi_i d(v)}}{2iw(G)}\right)^2 \right) 
  \end{eqnarray}
  \tabularnewline \hline   
    
  \multirow{4}{*}{Partition} \\
  \multirow{4}{*}{Count} \\
  \multirow{4}{*}{$\zeta(\Pi)$} & 
  \multirow{5}{*}{Min} & 
  Difference between number of created partitions, $\Pi_C$, and number of desired partitions, $\Pi_D$. 
  \begin{eqnarray}\label{eq:constraint_partitioncount}
	\zeta(\Pi) = abs( | \Pi_C | - | \Pi_D | )  
  \end{eqnarray}	
  \tabularnewline \hline   
  
  \multirow{4}{*}{Partition} \\
  \multirow{4}{*}{Size} \\
  \multirow{4}{*}{$\beta(\Pi)$} & 
  \multirow{5}{*}{Min} &
  Standard deviation of sizes of all created partition. 
  \begin{eqnarray}\label{eq:constraint_partitionsize}
  	\Pi & = & V \rightarrow \lbrace \pi_1,\ldots,\pi_k \rbrace \nonumber \\	
  	\beta(\Pi) & = & stdev( |\pi_1| \cdots |\pi_k| )
  \end{eqnarray}
  \tabularnewline \hline   
  
\end{tabular}
\end{threeparttable}
\caption{Partitioning constraints}
\label{tab:partitioning_constraints}
\end{center}
\end{table}	

\section{Algorithms Overview}
\label{sec:algorithm_overview}

As mentioned, graph partitioning is an established field and well studied problem.
Published research in this area is numerous, stemming back half a century.
Among the most well known early works are the Kernighnan-Lin \cite{ref:partition20} 
heuristic procedure for partitioning graphs and the Fiduccia-Mattheyses \cite{ref:partition21} 
linear time heuristic for improving network partitions.

A field closely related to graph partitioning is graph clustering.
A summary of the last 50 years in graph clustering is given by \cite{ref:partition22},
and a thorough overview and state of the art by \cite{ref:partition25}.

In contrast to early algorithms, which were largely sequential and assumed the graph to be static,
more recent research has focused on the development of parallel partitioning algorithms, 
capable of maintaining a partitioning in the presence of changes made to the underlying graph.
A number of such algorithms are discussed in \cite{ref:partition5,ref:partition23}.
These, based on disturbed diffusion, are all dynamic, parallel and multi-constraint.
They still exhibit key drawbacks though.
They are not local view algorithms and, as a consequence, no distributed implementations of them exist.

In \cite{ref:partition1} the authors build on the work of \cite{ref:partition23} to design a new algorithm.
The algorithm is one of many \cite{ref:graph_partitioning66,ref:graph_partitioning67,ref:graph_partitioning68, ref:graph_partitioning69,ref:graph_partitioning70} that attempt to optimize the modularity of a partitioning.
It finds partitions of higher quality, while reducing dependence on a global view of the graph.
However, as global view is still needed in certain cases, 
it can not be regarded as a true local view algorithm and is not distributed.

\paragraph*{}
An increasing amount of research has focused on algorithms that partition very large graphs, 
such as those created by social network applications.
By necessity these are all true local view algorithms,
because for graphs of such a size it is impractical to maintain a global view.

To partition these large graphs, a novel algorithm design is introduced in \cite{ref:graph_partitioning60}, 
and later extended in \cite{ref:partition28}.
It is comprised of two routines, Nibble and Partition,
which may be implemented in any way.
The Nibble subroutine takes a seed vertex as input,
explores the surrounding subgraph to identify its host partition,
then removes that partition from the graph.
The Partition routine repeatedly calls Nibble, 
until all partitions have been found and the entire graph has been partitioned.
As implemented in \cite{ref:graph_partitioning60,ref:partition28} this results in very fast algorithms, 
with nearly-linear time complexity (with respect to vertex count).
This same algorithm design has also been adopted by other researchers.
Namely in \cite{ref:graph_partitioning61},
where Nibble was implemented using personalized PageRank vectors.
These algorithms represent some of the fastest known graph partitioning algorithms.
They are not dynamic, parallel, or distributed though.
Moreover, lack of sufficiently detailed descriptions makes them difficult to implement.

\section{Summary}

Before attempting to design new graph partitioning algorithms it is beneficial to explore the many that already exist.
By gaining a basic understanding of the differences between partitioning algorithms ---
the problems each addresses, the assumptions each makes, and their respective limitations ---
the process of searching for a partitioning algorithm that suits a particular problem is simplified.

This chapter presents only a brief overview of the graph partitioning field;
more detail is beyond the scope of this work.
However, the various definitions and concepts introduced here provide a valuable base on which to conduct further research; and will be used in later chapters to identify and evaluate those algorithms that appear most relevant.

Lastly, the architecture of an abstract graph database partitioning framework was presented in this chapter. This too will be referred to and further defined in later chapters.
\chapter{Prototype Graph Partitioning Algorithms	}
\label{cha:prototype_partitioning_alg}

To perform all subsequent evaluation, a graph partitioning algorithm had to be selected first.
Although various algorithms were introduced during the initial literature review (\sref{sec:algorithm_overview}),
they all displayed limitations that rendered them unsuitable for our purposes.
This chapter covers the three most promising algorithms in more detail,
including the benefits and limitations that were identified for each algorithm.

To aid in the exploration of these algorithms two libraries were developed,
graph\_gen\_utils and graph\_cluster\_utils. 
The graph\_gen\_utils library (\aref{apx:graph_gen_utils}) provides functionality to persist graph data structures using various formats,
load graphs into main memory for faster computation,
and log various graph and partitioning metrics.
Whereas graph\_cluster\_utils (\aref{apx:graph_cluster_utils}) contains implementations of some of the explored algorithms, as well as part of the functionality described in \fref{img:partitioned_graphdb}.


\section{Considered Algorithms}

Three partitioning algorithms were selected as possible candidates for use during evaluation.
A summary of each of these algorithms, along with their respective benefits and limitations, 
is presented in this section.

	\subsection{EvoPartition}
	
\paragraph*{} Algorithm EvoPartition \cite{ref:partition27} uses a subroutine, EvoCut.
EvoCut is a local view algorithm, based on the Evolving Set Process\footnote{Briefly, 
ESP is a Markov chain in which states are subsets of the vertex set, $V$. 
These sets grow and shrink between state transitions.} (ESP) \cite{ref:partition30}, 
and is used to find vertex sets with low conductance.
EvoPartition uses EvoCut to create a fast algorithm for finding partitions of balanced size and low conductance.

Basically, EvoCut starts from random starting 
vertex\footnote{Vertices with higher degree are selected as starting vertex with greater probability,
increasing the likelihood of starting near the center of a vertex cluster with low conductance.}, $v$, 
and initializes its vertex set to $S_0 = \lbrace v \rbrace$. 
It then performs the Evolving Set Process and,
on termination, if a set with sufficiently low conductance and high volume is found,
it is returned. If such a set is not found the empty set is returned.
EvoPartition and EvoCut are implementation of the Partition and Nibble routines, respectively,
that were introduced in \sref{sec:algorithm_overview}.

EvoPartition calls EvoCut and allocates the returned set to a partition.
This is performed repeatedly, until the majority of vertices in $V$ have been assigned to partitions.
More specifically, EvoCut computes the ESP vertex sets as defined by \eref{eq:next_set},
\eref{eq:prob_next_set} and \eref{eq:prob_walk}.
\begin{eqnarray}
\label{eq:next_set}
S_{i+1} & = & \lbrace v:p(v,S_i)\geq U \rbrace \\
\label{eq:prob_next_set}
p(v,S) & = & \sum_{u \in S} p(v,u) = p_{stay} \cdot \left( \frac{e(v,S)}{d(v)} + 1 \cdot (v \in S) \right) \\
\label{eq:prob_walk}
p(v,u) & = & 
\left\{  
\begin{array}{l l}
  p_{stay} & \quad \mbox{if } v = u \\
  (1-p_{stay}) \cdot \frac{1}{d(v)} & \quad \mbox{if } \lbrace v,u \rbrace \in E \\
  0 & \quad \mbox{otherwise} \\ 
\end{array} \right. 
\end{eqnarray}
Value $U$ is a threshold, selected uniformly at random from the interval $[0,1]$.
Vertex $v$ is selected uniformly at random from the boundary\footnote{Boundary vertices include all vertices
on the border of a vertex set, including those inside \& outside of the set 
i.e $\delta(S) =
\lbrace u \mbox{ ; } u \in S \mbox{ } \wedge \mbox{ } ec(\lbrace u \rbrace,S^c) > 0 \rbrace 
\mbox{ } \cup \mbox{ } 
\lbrace v \mbox{ ; } v \in S^c \mbox{ } \wedge \mbox{ } ec(\lbrace v \rbrace,S) > 0 \rbrace$}
vertices, $\delta(S_i)$, of the current set, $S_i$.
Function $p(v,S_i)$ defines the probability of vertex $v$ being in set $S_{i + 1}$.
Finally, function $p(v,u)$ gives the probability that a random walk traverses from one vertex, $v$, 
to another, $u$.
The probability of staying at the starting vertex, $v$, is known as the \textit{staying probability},
and denoted by $p_{stay}$. For the purposes of EvoCut, $p_{stay} = \frac{1}{2} $.

\paragraph*{}EvoCut continues until a stopping condition, $\tau$, is reached.
At that time a sample path, $(S_0, \ldots, S_\tau)$, has been generated and $S_\tau$ is returned.
Time $\tau$ is defined to be the first time $S_i$ has sufficiently low conductance and high volume,
or that the work performed exceeds a specified limit.

\paragraph*{}Because EvoCut only performs computation on the boundary vertices of a set, 
EvoPartition is very fast, having nearly-linear time complexity (with respect to vertex count).
This is formally shown in \eref{eq:complexity_esp},
\begin{equation}
\label{eq:complexity_esp}
O(|E| + |V|\phi^{-1/2}) \cdot O(\mbox{polylog}(|V|))
\end{equation}
where $\phi$ is an input parameter defining the upper bound on the conductance of the returned partitioning.
		
		\subsubsection*{Benefits \& Limitations}

EvoPartition is a fast algorithm, 
its computational complexity is among the lowest of modern graph partitioning algorithms.
This is largely due to the fact EvoCut is a true local view algorithm;
the cost of finding one partition is only dependent on the size of that partition.

EvoPartition is a multi-constraint algorithm. It returns partitions of approximately equal size, 
explicitly optimizes conductance, and implicitly optimizes edge cut. 
Unfortunately, the number of returned partitions can not be controlled,
which renders it less suitable for partitioning graph databases.

Among other limitations of EvoPartition: it does not consider edge weights, can not cope with dynamism,
and was not designed for parallel or distributed execution.
Finally, EvoPartition is a complex, recent algorithm that lacks a sufficiently detailed description,
so producing a reliable implementation is difficult.
		
	\subsection{Dynamic Cut-Cluster Algorithm}
	
The Dynamic Cut-Cluster Algorithm (DCCA) \cite{ref:partition53} is based on the Cut-Cluster Algorithm \cite{ref:partition56}, with extensions that allow it to be used in dynamic environments. 
Both algorithms are based on the maximum flow theorem \cite{ref:partition54}, 
which uses minimum cut trees \cite{ref:partition55} to identify partitions.
The DCCA algorithm attempts to minimize conductance of partitions.
Note, as minimum cut trees are only defined for undirected, weighted graphs, 
DCCA may be applied only to these graph types. 

\begin{figure}[htbp]
 \centering
\subfloat[Input graph --- $G$]
{\includegraphics[scale=1.0]{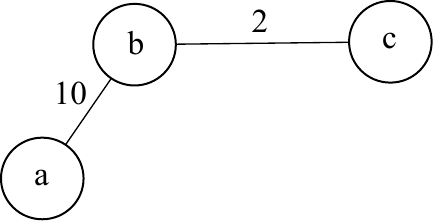}
\label{fig:cut_cluster_1}}
\subfloat[Sink added --- $G'$]
{\includegraphics[scale=1.0]{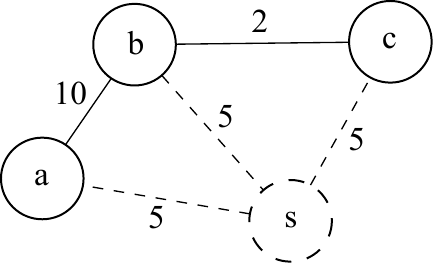}
\label{fig:cut_cluster_2}}
\qquad
\subfloat[Minimum cut tree of $G$]
{\includegraphics[scale=1.0]{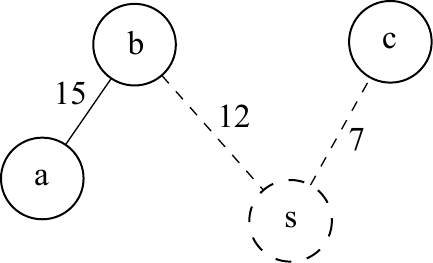}
\label{fig:cut_cluster_3}}
\subfloat[Sink removed to create partitions]
{\includegraphics[scale=1.0]{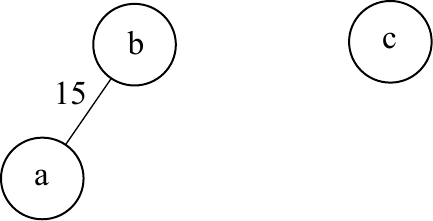}
\label{fig:cut_cluster_4}}
\caption{Example use of minimum cut tree algorithm to partition a graph}
\label{fig:cut_cluster} 
\end{figure}

\paragraph*{}
Given an input graph, $G$, DCCA first adds an additional \textit{sink} vertex, $s$, to $G$
and connects it to all other vertices. 
The new graph --- with $s$ added --- is referred to as $G'$.
All edges connected to $s$ have a weight of $ \alpha $, where $ 0 \leqslant \alpha < \infty $. 
\fref{fig:cut_cluster_1} shows an example input graph,
and \fref{fig:cut_cluster_2} shows the same graph with $s$ added and $ \alpha = 5 $. 

After constructing the minimum cut tree of $G'$,  
removal of $s$ from the minimum cut tree results in division of the tree into disjoint subgraphs.
The vertices of each subgraph form a partition, $\pi_{i}$, in the original graph.
\fref{fig:cut_cluster_3} shows the minimum cut tree of $G'$,
and \fref{fig:cut_cluster_4} show the partitioning created by removing $s$.

Together with $\alpha$, the structure of the graph defines the number of partitions returned.
The complexity of this process is defined by the minimum cut tree generation,
for which recent algorithms \cite{ref:partition53} have a complexity of $O(|V| \cdot |E|)$. 

A unique characteristic of DCCA is that all partition are guaranteed to be of a certain quality, 
given by $\partial(\pi) \leqslant \alpha \leqslant iw(\pi)$.

\paragraph*{}
To maintain the partitioning quality, DCCA is executed each time the graph is modified.
This behavior defines it as a \textit{Reactive} algorithm.
When the graph is modified DCCA updates the minimum cut tree, and therefore the partitioning, accordingly. 
The complexity of these updates depends on the type of modifications performed.
\tref{tab:DCCA_complexity} contains the possible modification operations,
their complexity, and whether or not they guarantee to be smooth.

Each process only modifies a subset of the vertices in the graph, 
so a view of the entire tree is never needed. 
Unfortunately, not all update operations guarantee smoothness; 
some may result in costly changes to the partitioning but not improve it.

\begin{table}[htbp]
\extrarowheight = 0.5mm
\begin{center}
\begin{tabular}
{|l|l|c|}
\hline
\textbf{Update Operation} 	& \textbf{Complexity} & \textbf{Smooth}	\\ 
\hline
\hline
Add vertex   				&      $0$ 										& yes	\\
\hline
Delete vertex				&      $0$ 										& yes	\\
\hline
Add intra-partition edge	&      $0$ 										& yes	\\
\hline
Delete intra-partition edge	&      $ O((|\Pi|+|\pi_i|) \cdot |V|^{3/2})$ 	& no	\\
\hline
Add inter-partition edge	&      $ O((|\pi_i|+|\pi_j|) \cdot |V|^{3/2})$ 	& no	\\
\hline
Delete inter-partition edge	&      $ O(|\Pi| \cdot |V|^{3/2})$				& yes	\\
\hline
\end{tabular}
\end{center}
\caption{Complexity of DCCA graph modification operations}
\label{tab:DCCA_complexity}
\end{table}

		\subsubsection*{Benefits \& Limitations}	

Benefits of DCCA include:
it is one of few algorithms we are aware of that guarantee the quality of identified partitions;
the reactive nature of DCCA limits its computational overhead;
and, after the initial partitioning, DCCA only needs a local view. 
Furthermore, because DCCA is a local view algorithm,
distributed or parallel versions of the algorithm may be easier to implement.
However, at present we are not aware of any literature regarding such implementations.

The DCCA algorithm has a number of limitations.		
It is a single constraint algorithm, providing no guarantees on the number or size of identified partitions.
As the number of partitions depends on graph structure,
when the graph is modified the number of partitions produced by DCCA may also change.
The value of $\alpha$ cannot be changed at runtime to compensate for graph modifications.
Finally, some update operations do not guarantee smoothness, so unnecessary changes in the partitioning may occur.
		
	\subsection{Distributed Diffusive Clustering}	
	\label{sec:partitioning_alg_didic}

The Distributed Diffusive Clustering algorithm (DiDiC) \cite{ref:partition26} is a local view heuristic, 
which attempts to optimize the modularity of a partitioning.
It is based on the method of disturbed diffusion \cite{ref:partition1,ref:partition23}.
DiDiC was designed to perform load balancing on a P2P 
supercomputer\footnote{This supercomputer performs parallel computations in the 
Bulk Synchronous Parallel (BSP) \cite{ref:scalable31} style.}.
It identifies subsets of machines among which network bandwidth is high,
in doing so dividing the network into partitions.
Machines are modeled as vertices and network connections as weighted edges,
where weights are proportional to available bandwidth.

\paragraph*{}Basically, disturbed diffusion is the process of disseminating load across the vertices of a graph; 
it shares similarities with gossip algorithms and random walks \cite{ref:partition32,ref:partition33}.
As with random walks, a diffusion process tends to stay within dense graph regions.

Disturbed diffusion is made up of two diffusive systems;
a primary system, $\mbox{\textit{diff}}_{P}$, and a secondary system, $\mbox{\textit{diff}}_{S}$.
Diffusion system $\mbox{\textit{diff}}_{P}$ exploits the properties of diffusion to identify dense graph regions.
Diffusion system $\mbox{\textit{diff}}_{S}$ is used to disturb $\mbox{\textit{diff}}_{P}$ by introducing a bias that prevents it from 
converging to a uniform distribution, where all vertices contain the same load.
The intended result is a distribution in which high concentrations of load are located near 
the center of dense graph regions, thereby identifying partitions.

\paragraph*{}
DiDiC starts from a random configuration and then converges towards a higher quality partitioning.
It creates a disturbed diffusion system for each of the desired number of partitions, $k$.
Every vertex participates in every diffusion system and, accordingly, 
every vertex stores two load vectors of size $k$.
One load vector for $\mbox{\textit{diff}}_{P}$, $w \in \mathbb{R}$, and one for $\mbox{\textit{diff}}_{S}$, $l \in \mathbb{R}$.
Each load vector element represents the diffusion system of a partition.
E.g. $w_v(c)$ returns the primary load value of partition $\pi_c$ for vertex $v$.
Load vectors are initialized as in \eref{eq:didic_init}.

\begin{equation}\label{eq:didic_init}
w_u(c)=l_u(c)= \left\lbrace
\begin{array}{l l}
  100 & \quad u \in \pi_c \\
  0 & \quad u \not\in \pi_c\\ 
\end{array} \right.
\end{equation}

\paragraph*{} Diffusion in DiDiC is iterative and consists of $T$ iterations,
where $t$ denotes the current iteration.
For each iteration of DiDiC, $\psi$ iterations of $\mbox{\textit{diff}}_{P}$ are performed, where $s$ denotes the current iteration.
Finally, 
for each iteration of $\mbox{\textit{diff}}_{P}$, $\rho$ iterations of $\mbox{\textit{diff}}_{S}$ are performed, where $r$ denotes the current iteration.
E.g. $w^s_v(c)$ gives the primary load value of partition $\pi_c$ for vertex $v$ at iteration $s$ of $\mbox{\textit{diff}}_{P}$.
\eref{eq:didic_diff_p} and \eref{eq:didic_diff_s} define the diffusion process for one iteration.
\begin{eqnarray}\label{eq:didic_diff_p}
x^{s-1}_{e= \lbrace u,v \rbrace}(c) & = & wt(e)\cdot \alpha(e)(w^{s-1}_u(c)-w^{s-1}_v(c)) \nonumber \\
w^s_u(c) & = & w^{s-1}_u(c)+l^s_u(c)-\sum_{e=\lbrace u,v \rbrace \in E}x^{s-1}_e(c)
\end{eqnarray}
\begin{eqnarray}\label{eq:didic_diff_s}
y^{r-1}_{e= \lbrace u,v \rbrace}(c) & = & 
wt(e)\cdot \alpha(e) \left( \frac{l^{r-1}_u(c)}{b_u(c)}- \frac{l^{r-1}_v(c)}{b_v(c)}\right) \nonumber \\
l^r_u(c) & = & l^{r-1}_u(c)-\sum_{e=\lbrace u,v \rbrace \in E}y^{r-1}_e(c)  \nonumber \\
b_u(c) & = & \left\lbrace
\begin{array}{l l}
  1 & \quad u\not\in \pi_c \\
  10 & \quad \mbox{otherwise}\\ 
\end{array} \right.
\end{eqnarray}
Function $\alpha$ represents the \textit{flow scale} of an edge, 
which regulates how much load is transmitted across that edge.
Function $wt(e)$ returns the weight of edge $e$ and is in the interval [0, 1].
Lastly, $b$ denotes \textit{benefit}.
Function $b$ ensures vertices not in $\pi_i$ send the majority of their load in $l$ to any neighboring vertices that do belong to $\pi_i$.
Function $b$ is the mechanism that produces disturbance.

After every time step each vertex, $v$, selects the partition it belongs to, $\pi_i$.
It does this by selecting the partition that corresponds to the highest primary load value,
because if the load value of a partition is high it is likely that neighboring vertices also belong to that partition.
The selection process is defined formally in \eref{eq:didic_partition_affiliation}.
\begin{equation}
\label{eq:didic_partition_affiliation}
\pi_i = 
\max\limits_{i=1}^{\vert \Pi \vert} \left( w_v(i) \right)
\end{equation}

To deal with dynamism\footnote{
A dynamic environment is one in which vertices may be added or removed from the graph at runtime.}, 
when a vertex is deleted the neighboring vertices receive an equal share of its load and
when a vertex is added it is assigned to a random partition.
Dynamism only affects the partitions it is applied to.

For pseudocode of the DiDiC algorithm refer to \fref{alg:pseudo_didic}.
Per iteration, the time complexity of DiDiC is $O(k \cdot \psi \cdot \rho \cdot 2 \cdot |E|)$.

\begin{figure}[htbp]
\begin{center}
\parbox{10.0cm}{
\begin{algorithmic}
\STATE $\pi = \mbox{RandomValue}(1,k)$
\STATE initializeLoadVectors()
\FOR[algorithm iterations]{$t = 1$ to $T$} 
	\FOR[partitions]{$c = 1$ to $k$} 
		\FOR[primary diffusion system, $\mbox{\textit{diff}}_P$]{$s = 1$ to $\psi$} 
			\STATE $\overline{w}_v(c) = w_v(c)$
			\FOR[secondary diffusion system, $\mbox{\textit{diff}}_S$]{$r = 1$ to $\rho$} 
				\STATE $\overline{l}_v(c) = l_v(c)$
				\FORALL[neighbors of \ensuremath{u}]{$v$ neighbor of $u$} 
					\STATE $\overline{l}_v(c) = \overline{l}_v(c) - \alpha(e) \cdot wt(e) \cdot 
							\left( \frac{l_v(c)}{b_v(c)} - \frac{l_u(c)}{b_u(c)} \right) $
				\ENDFOR
			\ENDFOR
			\FORALL[neighbors of \ensuremath{u}]{$v$ neighbor of $u$} 
				\STATE $\overline{w}_v(c) = \overline{w}_v(c) - \alpha(e) \cdot wt(e) \cdot 
						\left( w_v(c) - w_u(c) \right) $
			\ENDFOR
			\STATE $w_v(c) = \overline{w}_v(c) + \overline{l}_v(c)$
			\STATE $l_v(c) = \overline{l}_v(c)$
		\ENDFOR
	\ENDFOR
	\STATE $\pi = \mbox{argmax}_{c=1 \ldots k} w_v(c)$
	\STATE adaptToGraphChanges()
\ENDFOR
\end{algorithmic}
}
\caption{DiDiC pseudocode, from the point of view of one vertex}
\label{alg:pseudo_didic}
\end{center}
\end{figure}

		\subsubsection*{Benefits \& Limitations}

Even when initialized with a random partitioning, 
DiDiC is capable of converging towards a high quality partitioning.
This means a database can be partitioned randomly when data is first inserted,
then improved later when computational resources are available, or when necessity dictates.
		
According to the constraint definitions in \tref{tab:partitioning_constraints},
DiDiC is a multi-constraint algorithm; 
it enforces an upper bound on the number of created partitions,
explicitly optimizes modularity, 
and implicitly optimizes edge cut.
The balance of partition sizes is not explicitly optimized however.

DiDiC was originally designed to run in a dynamic, distributed environment.
It therefore elegantly copes with dynamism,
can be implemented to execute on multiple processors or computers concurrently,
and each instance on each processor/computer is a local view algorithm.
	
\section{Conclusion}

Three different graph partitioning algorithms were explored,
each of which displayed a different approach to the graph partitioning problem.
The properties of these algorithms were identified and their benefits and limitations explored.
A concise comparison of the properties offered by each algorithm is listed in \tref{tab:algorithms_comparison}.

\begin{table}[htbp]
\begin{center}
\begin{threeparttable}

\begin{tabular}
{c c c c }
 & \textbf{DiDiC} & \textbf{DCCA} & \textbf{EvoPartition} \\ 

\textbf{Multi-Constraint} & 
\checkmark & 
\ding{55} & 
\checkmark \\ 

\textbf{Local View} & 
\checkmark & 
\checkmark & 
\checkmark \\ 

\textbf{Distributed} & 
\checkmark & 
\ding{55} & 
\ding{55} \\ 

\textbf{Parallel} & 
\checkmark\tnote{a} & 
\ding{55} & 
\ding{55} \\ 

\textbf{Iterative} & 
\checkmark & 
\ding{55} &
\ding{55} \\ 

\textbf{Dynamic} & 
\checkmark & 
\checkmark & 
\ding{55} \\ 

\textbf{Smooth} & 
\ding{55} & 
\checkmark\tnote{b} & 
\ding{55} \\ 

\textbf{Weighted} & 
\checkmark & 
\checkmark & 
\ding{55} \\ 
\end{tabular}
     \begin{tablenotes}
       \item[a] We assume a parallel implementation is trivial
       \item[b] Most supported operations are smooth
     \end{tablenotes}     
\end{threeparttable}
\end{center}
\caption{Property comparison of candidate graph partitioning algorithms}
\label{tab:algorithms_comparison}
\end{table}

EvoPartition is very fast and a true local view algorithm.
It ensures partitions are of similar sizes, explicitly minimizes conductance, and implicitly minimizes edge cut.
However, it makes no guarantee about how many partitions will be created, 
does not take edge weights into consideration, and is difficult to implement.
Its greatest limitation, though, is that it is not a dynamic algorithm.
Datasets contained in graph databases are likely to experience frequent modifications,
for that reason we consider EvoPartition to be unsuitable for our requirements.

The DCCA algorithm has a number of appealing characteristics.
It is dynamic, requires only a local view, makes use of edge weights, 
and guarantees the quality of created partitions.
Another key benefit is its reactive nature, 
as it means the algorithm uses no computational resources when they are not required.
However, because DCCA only optimizes for conductance, the number and size of partitions can not be controlled.
Most importantly the number of partitions may change when the graph structure is modified, 
making a mapping between server and partition more complicated.
We regard that to be a major drawback, consequently we deem DCCA unsuitable for our purposes.

A very desirable trait of DiDiC is its ability to start from a random partitioning,
then iteratively improve at a later time.
It explicitly optimizes for modularity and implicitly optimizes for edge cut.
Additionally, because it was specifically designed to operate in a dynamic, distributed environment,
it requires only a local view and can cope with dataset modifications.
A limitation of DiDiC is that it does not guarantee to create equal sized partitions.
Another limitation is, although it can enforce an upper bound on the number of created partitions,
it cannot control the exact number.
In spite of these limitations, the benefits provided DiDiC to make it the most suitable algorithm for our needs.
For that reason, DiDiC was the partitioning algorithm used in our evaluation.
\chapter{Prototype Partitioned Graph Database}
\label{cha:prototype_partitioned_db}

\section{Introduction}
To test partitioning algorithms and explore their impact on actual systems we chose the Neo4j graph database. 
Neo4j is an embedded graph database, implemented in Java and available under the GNU AGPL license. It stores data in the form of a property graph (\sref{sec:graph_types})
in which all vertices and edges, called nodes and relationships respectively, 
have unique IDs and store properties as key-value pairs.
Additionally, vertices and edges can be indexed, by ID or properties.

Many utility packages and third-party libraries are available for Neo4j,
but for this thesis we consider only its core functionality. 
This core functionality is encapsulated by several interfaces.
\texttt{Relationship} and \texttt{Node}, represent edges and vertices respectively (see \tref{tab:neo4j_relationship_api} and \tref{tab:neo4j_node_api}). 
\texttt{GraphDatabaseService} represents the graph database as a whole.
It provides the functionality to create vertices and to retrieve vertices and edges by their ID
(see \tref{tab:neo4j_graph_database_api}). 
Note that vertices are created using \texttt{GraphDatabaseService}, 
while edges are created using the \texttt{Node} interface. 


\begin{table}[htbp]
\extrarowheight = 0.5mm
\begin{center}
\begin{tabular}
{|>{\raggedright}p{0.35\textwidth}|>{\raggedright\arraybackslash}p{0.55\textwidth}|}

	\hline
	\textbf{Function} &
	\textbf{Description} \\
	\hline
	\hline
	
	\texttt{delete()} &	 
	Deletes this relationship \\
	\hline
	
	\texttt{getEndNode()} &
	Returns the end Node of this Relationship \\
	\hline
	
	\texttt{getId()} & 
	Returns the unique ID of this Relationship \\
	\hline
	
	\texttt{getStartNode()} & 
	Returns the start Node of this Relationship \\
	\hline
	
	\texttt{getType()} & 
	Returns the type of this Relationship \\
	\hline
	
	\texttt{getProperty(key)} &
	Returns the property value associated with the given key \\
	\hline
	
	\texttt{getPropertyKeys()} &
	Returns all existing property keys, or an empty iterable if this property container has no properties \\
	\hline
	
	\texttt{removeProperty(key)} &
	Removes the property associated with the given key and returns the old value \\
	\hline
	
	\texttt{setProperty(key, value)} &
	Sets the property value for the given key to value \\
	\hline
	
\end{tabular}
\end{center}
\caption{Subset of the Neo4j Relationship API --- Version 1.0}
\label{tab:neo4j_relationship_api}
\end{table}

\begin{table}[htbp]
\extrarowheight = 0.5mm
\begin{center}
\begin{tabular}
{|>{\raggedright}p{0.45\textwidth}|>{\raggedright\arraybackslash}p{0.45\textwidth}|}

	\hline
	\textbf{Function} &
	\textbf{Description} \\
	\hline
	\hline
	
	\texttt{createRelationshipTo(otherNode, relationshipType)} &
	Creates a Relationship between this Node and another \texttt{otherNode} \\
	\hline
	
	\texttt{delete()} &
	Deletes this Node if it has no Relationships attached to it \\
	\hline
	
	\texttt{getId()} &
	Returns the unique ID of this Node \\
	\hline
	
	\texttt{getRelationships()} &
	Returns all the Relationships attached to this Node \\
	\hline
	
	\texttt{getRelationships(direction)} &
	Returns outgoing or incoming Relationships attached to this Node \\
	\hline
	
	\texttt{getRelationships(\\relationshipType, direction)} &
	Returns all Relationships with the given type and direction that are attached to this Node \\
	\hline
	
	\texttt{getProperty(key)} &
	Returns the property value associated with the given key \\
	\hline
	
	\texttt{getPropertyKeys()} &
	Returns all existing property keys, or an empty iterable if Node has no properties\\
	\hline
	
	\texttt{removeProperty(key)} &
	Removes the property associated with the given key and returns the old value \\
	\hline
	
	\texttt{setProperty(key, value)} &
	Sets the property value for the given key to \texttt{value} \\
	\hline
	
\end{tabular}
\end{center}
\caption{Subset of the Neo4j Node API --- Version 1.0}
\label{tab:neo4j_node_api}
\end{table}

\begin{table}[htbp]
\extrarowheight = 0.5mm
\begin{center}
\begin{tabular}
{|>{\raggedright}p{0.4\textwidth}|>{\raggedright\arraybackslash}p{0.5\textwidth}|}

	\hline
	\textbf{Function} &
	\textbf{Description} \\
	\hline
	\hline
	
	\texttt{beginTx()} &
	Starts a new transaction and associates it with the current thread \\
	\hline
	
	\texttt{createNode()} &
	Creates a new Node \\
	\hline
	
	\texttt{getAllNodes()} &
	Returns all Nodes in the node space \\
	\hline
	
	\texttt{getNodeById(ID)} &
	Looks up a Node by ID \\
	\hline
	
	\texttt{getRelationshipById(ID)} &
	Looks up a Relationship by ID \\
	\hline
	
	\texttt{getRelationshipTypes()} &
	Returns all Relationship types currently in the underlying store \\
	\hline
	
	\texttt{shutdown()} &
	Shuts down Neo4j \\
	\hline
	
\end{tabular}
\end{center}
\caption{Subset of the Neo4j GraphDatabaseService API --- Version 1.0}
\label{tab:neo4j_graph_database_api}
\end{table}

\section{Extending GraphDatabaseService}

As an open source project Neo4j is easy to extend and use in other projects. 
Also, Neo Technology and a growing community actively provide support.
These benefits partially outweigh the limitation that Neo4j v1.0 has no support for dataset partitioning 
and does not provide all metrics desired for runtime logging (\sref{sec:partitioning_graphdbs}).
To support the functionality defined by our partitioning framework (\fref{img:partitioned_graphdb}) 
it was necessary to extend the \texttt{GraphDatabaseService} interface. 
The \texttt{PGraphDatabaseService} --- denoting partitioned \texttt{GraphDatabaseService} --- 
interface and associated abstract framework were created. 
Besides the new functions in \texttt{PGraphDatabaseService} (\tref{tab:neo4j_partitioned_graph_database_api}), 
all changes are transparent to the user, 
providing backwards compatibility between \texttt{GraphDatabaseService} and \linebreak \texttt{PGraphDatabaseService}.

\begin{table}[htbp]
\extrarowheight = 0.5mm
\begin{center}
\begin{tabular}
{|>{\raggedright}p{0.36\textwidth}|>{\raggedright\arraybackslash}p{0.54\textwidth}|}

	\hline
	\textbf{Function} &
	\textbf{Description} \\
	\hline
	\hline

	\texttt{getNumInstances()} &
	Returns the number of partitions \\
	\hline
	
	\texttt{getInstancesIDs()} &
	Returns a array of partition identifyer (PID) \\	
	\hline
	
	\texttt{getInstanceInfoFor(PID)} &
	Returns an InstanceInfo object for the partition with given PID or null if none exists \\
	\hline

	\texttt{getPlacementPolicy()} &
	Returns the current PlacementPolicy used \\
	\hline
	
	\texttt{setPlacementPolicy(\\PlacementPolicy)} &
	Replaces the current palcement policy \\
	\hline
	
	\texttt{addInstance()} &
	Creates a new partition and returns it identifier \\ 
	\hline

	\texttt{removeInstance(PID)} &
	Remove the Partition with the given PID if empty and returns if the operation was successfull \\
	\hline
	
	\texttt{createNodeOn(PID)} &
	Creates and returns a Node on the Partition with the given PID \\
	\hline
	
	\texttt{moveNodes(Iterable<Node>, PID)} &
	Moves all nodes of the set to the partition with the given PID \\
	\hline
	
\end{tabular}
\end{center}
\caption{Subset of the PGraphDatabaseService API}
\label{tab:neo4j_partitioned_graph_database_api}
\end{table}

To transparently migrate data between partitions the index structure is extended internally. 
Partitions are assigned a globally unique ID, called PID. 
Vertices and edges continue to use the standard Neo4j IDs, but these are renamed to LID, denoting local ID. 
This is because their uniqueness is only guaranteed within the context of a partition. 
Combined, PID and LID define the globally unique position (POS) of a vertex/edge. 
Finally, each entity is assigned a globally unique ID (GID), which is mapped to the POS of that entity internally. 
To perform this mapping the \texttt{GIDService} interface was created (see \tref{tab:GIDService}). 
From the perspective of a user, GID replaces the standard Neo4j ID; 
users of \texttt{PGraphDatabaseService} need only know the GID of an entity, 
which is automatically resolved to a vertex or edge.

\begin{table}[htbp]
\extrarowheight = 0.5mm
\begin{center}
\begin{tabular}
{|>{\raggedright}p{0.3\textwidth}|>{\raggedright\arraybackslash}p{0.6\textwidth}|}
    \hline
	\textbf{Function} &
	\textbf{Description} \\
	\hline
	\hline
	
	\texttt{addNode(GID, POS)} & Updates the position of a vertex with the given GID to the given POS\\
	\hline
	
	\texttt{addRela(GID, POS)} & Updates the position of an edge with the given GID to the given POS\\
	\hline
	
	\texttt{remNode(GID)} & Unregisters a vertex from the GID service and frees its GID\\
	\hline
	
	\texttt{remRela(GID)} & Unregisters an edge from the GID service and frees its GID\\
	\hline
	
	\texttt{createNodeGID()} & Returns a globally unique GID for a vertex\\
	\hline
	
	\texttt{createRelaGID()} & 	Returns a globally unique GID for an edge\\
	\hline
	
\end{tabular}
\end{center}
\caption{GIDService interface}
\label{tab:GIDService}
\end{table}

To provide the functionality of runtime logging, partitions store and maintain an \texttt{InstanceInfo} object. 
This object contains the following information: 
number of vertices and edges in a partition; 
amount of local traffic, equivalent to the number of served requests; 
and amount of global traffic, equivalent to the number of requests that resulted in communication with another partition.

In a partitioned graph database an edge may connect vertices belonging to different partitions. 
We refer to an edge, $e=\lbrace u,v \rbrace$, as an inter-edge, $e^{\mathcal{E}}$, 
if $u$ and $v$ are on different partitions, and as an intra-edge, $e^{\mathcal{A}}$, otherwise.
\eref{eq:intra_inter} defines this formally. 

\begin{equation}
\label{eq:intra_inter}
  e = \lbrace u,v \rbrace  = 
  \begin{cases}
  e^{\mathcal{A}} & u,v \in \pi_i \\
  e^{\mathcal{E}} & u \in \pi_i \wedge v \in \pi^c_i
  \end{cases}
\end{equation}

\newpage
Inter-edges must be handled differently to intra-edges, as they require partitions to communicate with each other.
The model used to achieve this is left to the implementation of the abstract framework.
At present it is assumed that \texttt{Node} and \texttt{Relationship} instances are retrieved from the database as direct references. 
In a truly distributed implementation this approach would likely need to be revised. 
This problem, as well as that of developing distributed replication, consistency, and transactional models lies beyond the scope of our thesis. 

\section{PGraphDatabaseService Implementations}
As part of this work two implementations of the \texttt{PGraphDatabaseService} interface were created. 
A complete prototype implementation of the abstract framework,
and simpler implementation that emulates the functionality of the prototype.

\subsection{Prototype}
\begin{figure}[htbp]
 \centering
\subfloat[Original Graph]
{\includegraphics[scale=1.0]{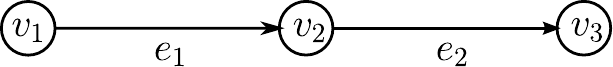}
\label{fig:simple_ghost_construct_1}}
\qquad
\subfloat[Partitioned Graph --- Shadow Construct]
{\includegraphics[scale=1.0]{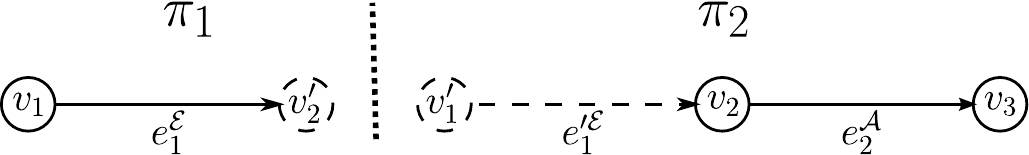}
\label{fig:simple_ghost_construct_2}}
\caption{Simple example of the Shadow Construct}
\label{fig:ghost_construct} 
\end{figure}

The prototype implementation (called \texttt{PGraphDatabaseServicePrototype}) 
has a \texttt{GraphDatabaseService} instance for each partition. 
The \texttt{GIDService} is realized as a central service using BerkleyDB \cite{ref:general59}. 
All partitions reside in the same JVM and are managed by the \texttt{PGraphDatabaseService} implementation. 
Communication between partitions is modeled using direct function calls. 
As previously mentioned, to support inter-edges a suitable model needed to be designed.
To implement such a model in \texttt{PGraphDatabaseServicePrototype} we created the Shadow Construct. 
\fref{fig:ghost_construct} shows how an inter-edge is modeled using such a construct.
Inter-edge $e^{\mathcal{E}}_1$ connects two vertices, $v_1 \in \pi_1$ and $v_2 \in \pi_2$,
where $v_1$ is the start vertex of $e^{\mathcal{E}}_1$.
As our implementation enforces that edges are stored on the same partition as their start vertex,
$e^{\mathcal{E}}_1$ resides on $\pi_1$.
Making up the Shadow Construct for this example:
a shadow of $v_1$, $v'_1$, is stored on $\pi_2$;
a shadow of $v_2$, $v'_2$, is stored on $\pi_1$;
and a shadow of $e^{\mathcal{E}}_1$, $e'^{\mathcal{E}}_1$, is stored on $\pi_2$.
%

Shadow entities can be seen as references to the original entity, they store the GID and POS of their original. 
In contrast, original entities are not aware of their shadows, 
but shadows can be obtained when necessary by traversing the graph. 
An entity can have at most one corresponding shadow entity on any given partition. 
Moreover, shadow entities only exist when they are part of at least one Shadow Construct; they are deleted when no longer needed. 

When traversing the graph, encountered shadow entities are resolved using their POS. 
Instead of a GID lookup of order $\log(|V|)$ only a limited lookup of order $\log(|\pi_i|) $ is needed, 
where $\pi_i$ is the partition the traversal needs to be moved to. 
This partly meets the requirement of index-free adjacency. 
Shadow entities are only used internally, they have no GID and are not visible to the user.

\subsection{Emulator}
After completing \texttt{PGraphDatabaseServicePrototype} a simple test was run to determine performance.
A graph was created with two partitions, two vertices, and one edge. 
The vertices were on different partitions and the edge connected them. 
The test comprised of traversing this edge \numprint{1000000} times,
and time required to complete the test was measured.

The prototype was found to be approximately 15 times slower than an unpartitioned Neo4j instance. 
Because it was built on top of Neo4j, encapsulating multiple instances,
the performance loss can be partly explained by the increased software stack.
However, to find the true source of this performance bottleneck the design has to be explored in greater detail, 
and changes to the core of Neo4j may be necessary. 
This is beyond the scope of this thesis and left for future research.

Due to its poor performance \texttt{PGraphDatabaseServicePrototype} was deemed inadequate for use during evaluation.
To decrease complexity and, therefore, allow for the execution of larger experiments,
a second \texttt{PGraphDatabaseService} implementation (called \texttt{PGraphDatabaseServiceEmulator}) was created.

In \texttt{PGraphDatabaseServiceEmulator} all graph entities are stored in a single \texttt{GraphDatabaseService} instance, 
avoiding the need for Shadow Construct and \texttt{GIDService} implementations.
\texttt{PGraphDatabaseServiceEmulator} emulates the behavior of \texttt{PGraphDatabaseServicePrototype} by assigning partition identifiers to vertices and edges (partitions are a logical concept only), but is still capable of measuring the same metrics. 
The reduced complexity of this implementation resulted in improved performance.
\tref{tab:performance_traversal} shows how much time each database implementation took to complete our simple performance test. 
\begin{table}[htbp]
\extrarowheight = 0.5mm
\begin{center}
\begin{tabular}{|c|c|c|}
\hline 
\textbf{Database} & \textbf{Inter-edge} & \textbf{Intra-edge}\\ 
\hline \hline
\texttt{Neo4j} & n/a & \unit{\numprint{1869}}{\milli\second} \\
	\hline
\texttt{PGraphDatabaseServiceEmulator} & \unit{\numprint{3679}}{\milli\second} & \unit{\numprint{3428}}{\milli\second} \\
	\hline	
\texttt{PGraphDatabaseServicePrototype} & \unit{\numprint{29493}}{\milli\second} & \unit{\numprint{28033}}{\milli\second} \\
	\hline
\end{tabular} 
\caption{Time to perform \numprint{1000000} traversals over one edge}
\label{tab:performance_traversal}
\end{center}
\end{table}		


\newpage
\section{Summary}
Neo4j was chosen as our platform for evaluating the partitioning methods discussed in \sref{cha:prototype_partitioning_alg}. 
As Neo4j v1.0 has no native support for partitioned data, 
the \texttt{PGraphDatabaseService} interface and associated abstract framework were created on top of the existing Neo4j API. 
Besides implementing functionality required by runtime logging, 
these provide the general structure of a partitioned Neo4j graph database and could be used as a base for further exploration in this area. 
The prototype implementation of this framework explores some of the concrete challenges for a partitioned graph database, such as indexing and inter-partition communication. 
However, for the purpose of evaluating partitioning methods this prototype has proven to be too inefficient. 
To allow for experiments of a larger scale an emulator implementation of the framework was implemented. 
This emulator was used in all subsequent evaluations. 


\chapter{Evaluation Methodology}
\label{cha:eval_method}

The purpose of this evaluation was to compare the DiDiC algorithm against two baseline partitioning methods.
To do this, each method was used to partition a number of graph databases.
Then, the results were compared using various performance metrics.
This chapter covers specifics of the evaluation environment, details of the experiments performed, 
and an explanation of the performance metrics that were measured.


\section{Overview}

In an attempt to test the partitioning methods across a broad range of scenarios,
experiments were performed on datasets with varying characteristics.
Each dataset was unique with respect to the domain it modeled, the data it contained, 
and the topology of the graph it represented.


Due to the fact that each dataset describes a different application domain and no production logs were available,
a unique access pattern was defined for each dataset. 
The correlation between edge cut and network traffic may be affected if certain edges are accessed more frequently than others.
To investigate if this is the case, access patterns were designed to mimic this, non-uniform, behavior.
The effect of each partitioning method was tested by a series of read operations, further noted as evaluation logs,
on the partitioned datasets. 
Each evaluation log was created using the associated access pattern definition. 

While performing the evaluation logs, performance metrics were logged.
These metrics included: 
the number of operations that involved communication between partitions;
the load balance across partitions, with respect to edge storage;
the load balance across partitions, with respect to vertex storage;
and the load balance across partitions, with respect to traffic performed.

As it is common for datasets to undergo regular changes,
experiments were designed to test the partitioning methods while exposed to differing levels of dynamism.
These were intended to measure what effect --- if any --- dynamism had on the measured performance metrics,
and if those effects could be negated by using a graph partitioning algorithm.

To perform the access patterns in a controlled environment a basic simulator
(\aref{apx:neo4j_access_simulator}) was developed.
The simulator allowed for detailed access patterns to be defined, 
evaluation logs to be created from those definitions,
and experiments to be performed by executing the evaluation logs.
All experiments were performed using the simulator, ensuring that they were deterministic and repeatable.

\section{Datasets \& Access Patterns}	
\label{sec:datasets_and_access_patterns}

Three datasets were used during evaluation:
an synthetically generated dataset, intended to model the general folder structure of a file system;
a dataset acquired from a Geographic Information System (GIS) application,
which models the major transport systems in Romania;
and a dataset obtained by crawling Twitter \cite{ref:general49},
which models users of Twitter as well as the \textit{Follows} relationships between them.

The three datasets --- vastly different from each other in many respects --- 
and their respective access patterns are described further in the remainder of this section. 
To provide insight on the structure of each graph we include analysis of average degree, 
in-degree (number of incoming edges), out-degree (number of outgoing edges), 
and clustering coefficient (connectivity between neighboring vertices).

Access patterns are created based on this structural knowledge and influenced by usage of existing applications, they were artificially generated and not based on production logs.



	\subsection{File System}
	\label{sec:dataset_and_access_fstree}

As mentioned previously, this dataset was artificially generated.
It was created with the aim of resembling the general structure of a regular file system.

\paragraph*{Structure:}
Vertices are used to model files, folders, users, organizations, and certain events.
For example, the action of creating a file may be modeled as an event that contains meta data about the event.
Edges are used to model folder hierarchy --- relationships between files and their parent folders ---
as well as associating files or folders with their event vertices.

\begin{figure}[htbp]
\centering
\includegraphics[scale=1.0]{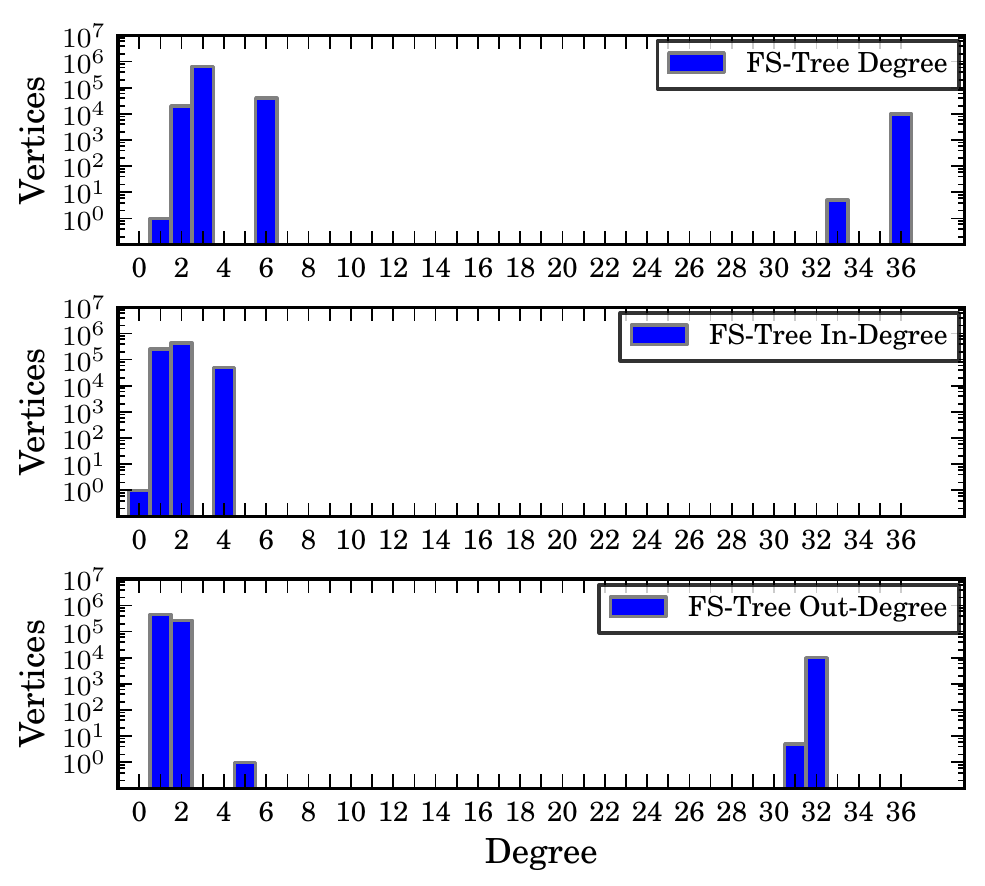} 
\caption{Degree distribution -- File System}
\label{fig:fstree_degree_dist_all}
\end{figure}

The graph contains \numprint{730027} vertices, \numprint{1310041} edges,
and has a clustering coefficient of \numprint{0.116905}.
With respect to out-degree 
there are basically two types of vertices in this graph,
those with an out-degree of 1--2 and those with an out-degree of 30--32.
The vertices with a out-degree of 1--2 are those that model files, users, organizations and events. 
Vertices with an out-degree of 30--32 model folders.
\fref{fig:fstree_degree_dist_all} illustrates the in-degree, out-degree, and degree distributions
of all vertices in the graph.

\begin{figure}[htbp]
\begin{center}
\includegraphics[scale=1.0]{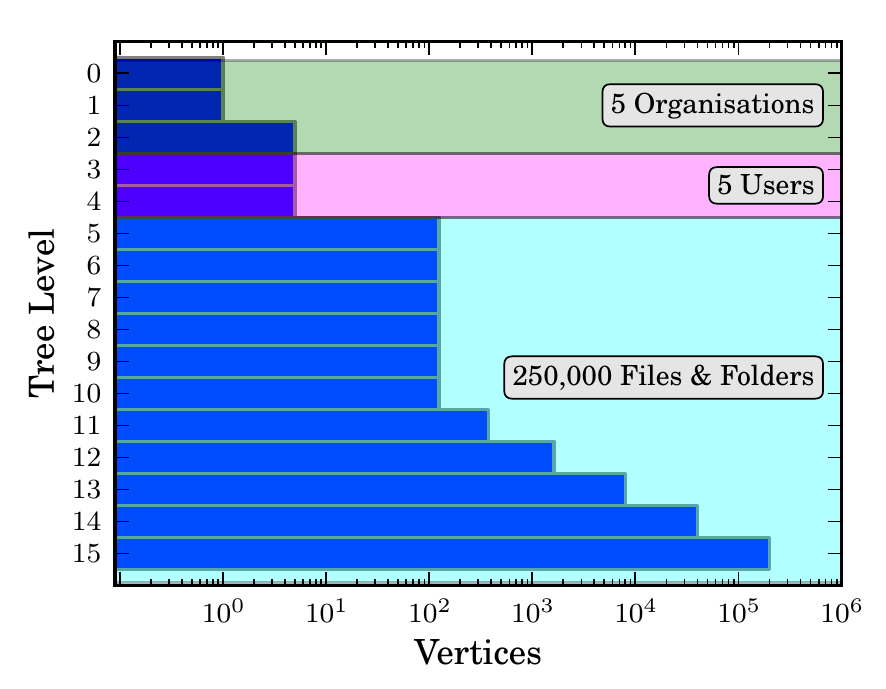} 
\caption{Nodes at each level of the tree -- File System}
\label{fig:fstree_nodes_at_levels_annotated}
\end{center}
\end{figure}

With the exception of edges that connect event vertices, the file system graph is acyclic --- a tree.
\fref{fig:fstree_nodes_at_levels_annotated} shows the vertices at each level of the tree, 
along with annotations that specify how many of those vertices are files, folders, users, or organizations.

To give a clearer perspective of topology, 
a smaller --- \numprint{742} vertices and \numprint{1327} edges --- 
sample graph was generated and visualized (\fref{fig:sampeTree_1000_v742_10,blue__e1327_2,black_})
using igraph \cite{ref:general50}.

\begin{figure}[htbp]
\begin{center}
\includegraphics[scale=0.2]
{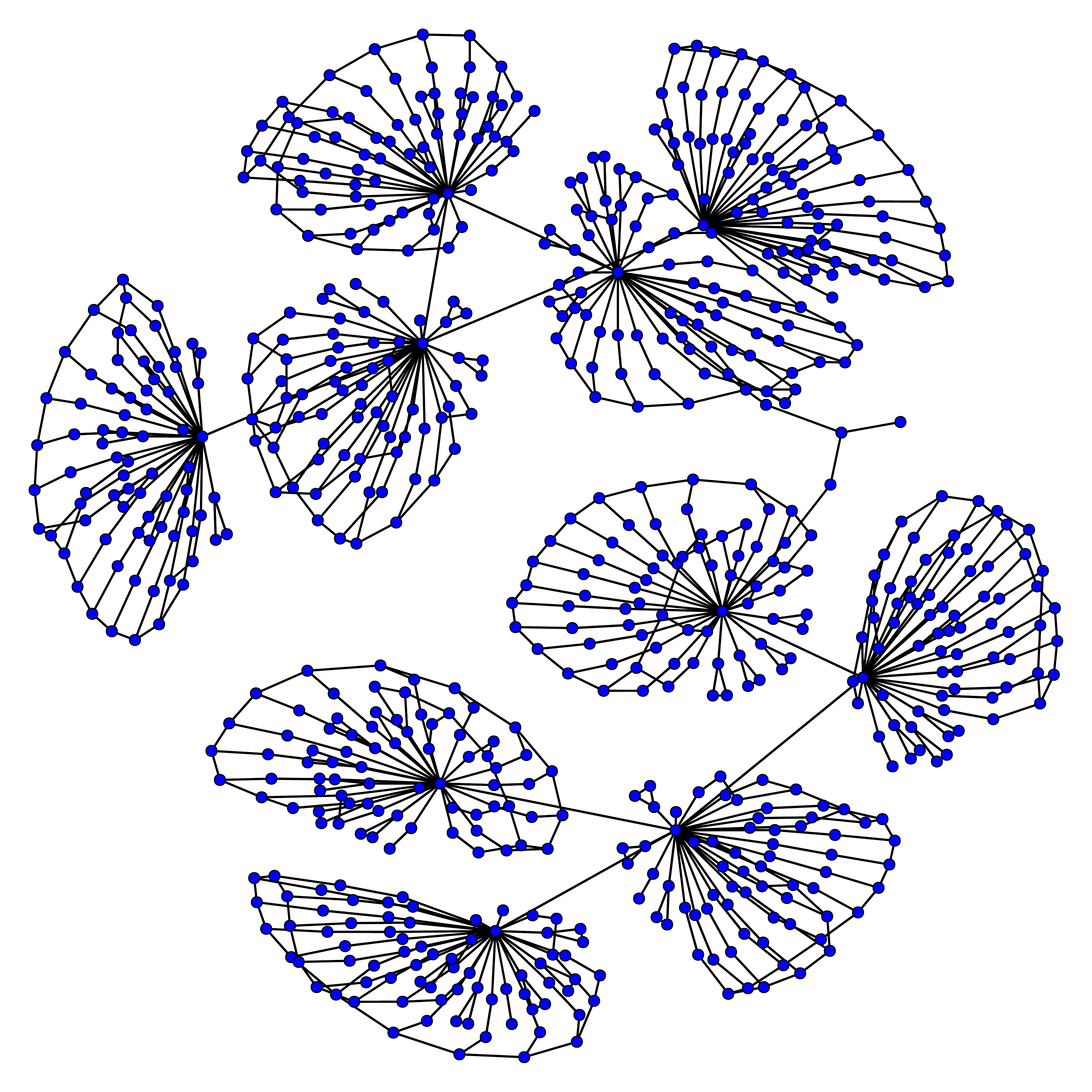} 
\caption{Sample graph -- File System}
\label{fig:sampeTree_1000_v742_10,blue__e1327_2,black_}
\end{center}
\end{figure}		

\paragraph*{Access Patterns:}		
For the file system dataset the defined access pattern was a search operation, using Breadth First Search,
that starts from some folder and ends at some lower level file or folder.
Only vertices representing files or folders were considered in these search operations.

The end point was selected first, randomly, where the probability of a vertex being selected as end point is
proportional to the degree of that vertex. 
Consequently, the probability of choosing a folder as the end point is greater than that of choosing a file. The start point was selected by performing a walk of random length up the tree, starting from the end point, then choosing the vertex on which the walk ends. The walk length was selected uniformly at random, 
such that it ends at some level between the end point and the root folder of the current user.

\begin{figure}[htbp]
\begin{center}
\includegraphics[scale=1.0]{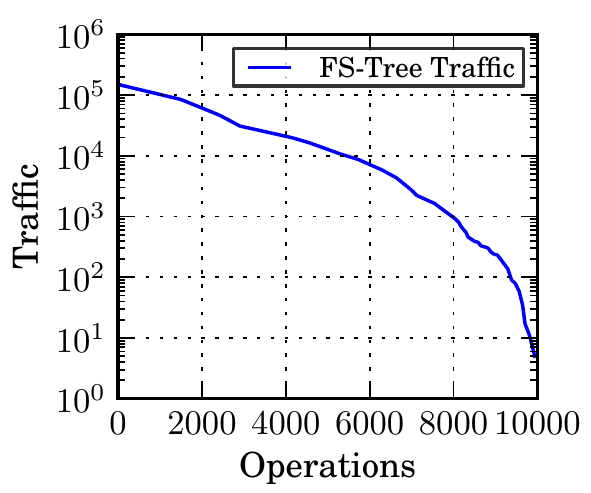} 
\caption{Traffic distribution of access pattern -- File System}
\label{fig:fstree_traf}
\end{center}
\end{figure}

We generated \numprint{10000} of these operations and wrote them to an operation log.
The log was then replayed for all experiments on the file system dataset.
\fref{fig:fstree_traf} displays the traffic generated by each of these operations.
Note, operations are sorted by the amount of traffic they generate, from most to least.
A unit of traffic is equivalent to one of the following actions:
performing an index lookup, for a vertex or edge;
retrieving a property value, of a vertex or edge;
retrieving an edge connected to the current vertex;
or retrieving start/end vertex of the current edge.

\tref{tab:fstree_graph_actions} lists the sequence of graph actions performed at each step of the access pattern,
and identifies which of these actions can potentially generate network traffic.
This will be referenced in later discussions.

\begin{table}[htbp]
\extrarowheight = 0.5mm
\begin{center}
\begin{tabular}{|l|l|}
\hline 
\textbf{Graph Action} & \textbf{Potential Network Traffic} \\ 
\hline \hline

\texttt{Vertex.getId()} &
\multicolumn{1}{c|}{No} \\ \hline

\texttt{Vertex.getEdge()} &
\multicolumn{1}{c|}{Yes} \\ \hline

\texttt{Edge.getEndVertex()} &
\multicolumn{1}{c|}{No} \\ \hline

\end{tabular} 
\caption{Access pattern graph actions --- File System}
\label{tab:fstree_graph_actions}
\end{center}
\end{table}		

Given that the y-axis is log scaled, the traffic distribution plot shows near-exponential decay;
few operations generate a large amount of traffic and most generate a considerable amount less.
Traffic is greatest when end point is a leaf vertex and start point a root folder,
but this is unlikely to occur. 
Leaves are all file vertices with low degree, so are not likely to be selected as end point.
In the rare case that a leaf vertex is an end point,
the probability of selecting the root folder as start point is at minimum, 
as the walk length from end point to start point is at maximum.
On the other hand, the probability of selecting the end point from a higher level is considerably greater, 
as this is where the folders reside.
				
\newpage				
	\subsection{Geographic Information System}

Geographic Information Systems encode the locations of transportation routes,
businesses, public venues and private residences.
They can be regarded as digital maps, providing the means to search for the coordinates of a given venue
or calculate the route between two points.
The GIS dataset used in this evaluation encodes geographic information for the country of Romania.
It was provided by Neo Technology \cite{ref:general51}.
				
\paragraph*{Structure:}
Vertices are used to model discrete geographic locations.
Each vertex contains a number of properties, including: 
longitude of location, latitude of location, name of location, and date that the database entry was created.
Edges are used to model transportation routes between two vertices (locations).
Each edge contains a number of properties, including: name of route, type of route, 
and weight of route --- the time required to travel across that route. 

\begin{figure}[htbp]
\centering
\includegraphics[scale=1.0]{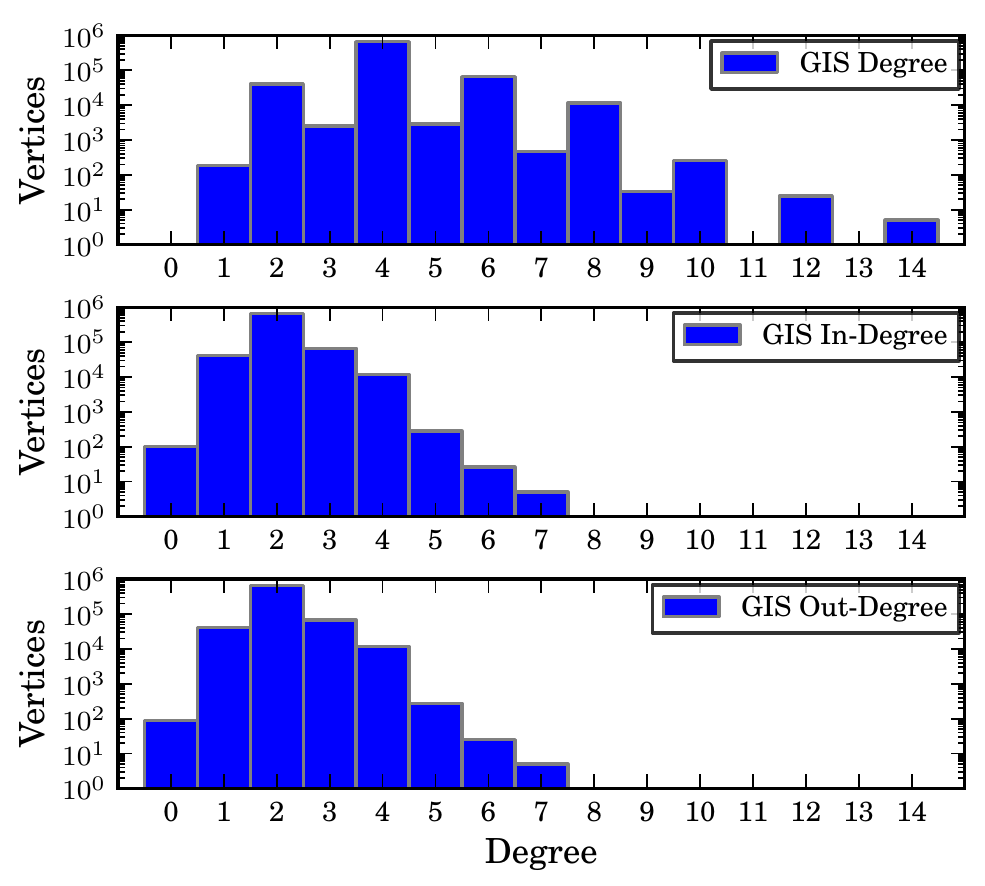} 
\caption{Degree distribution -- GIS}
\label{fig:gis_degree_dist_all}
\end{figure}

The graph contains \numprint{785891} vertices, \numprint{1621138} edges,
and has a clustering coefficient of \numprint{0.000884}.
Note, that clustering coefficient appears low because it was calculated for the entire graph.
As large regions of the graph model sparsely populated areas --- containing mostly rural roads --- of Romania, 
clustering coefficient within cities is expected to be higher.
The degree distribution supports this statement.
\fref{fig:gis_degree_dist_all} shows a large number of vertices with degree of 1--3,
which are likely points along rural highways. 
Conversely, there are at least as many vertices with degree of 4--14,
which represent points along inner-city streets.

\begin{figure}[htbp]
\begin{center}
\includegraphics[scale=1.0]{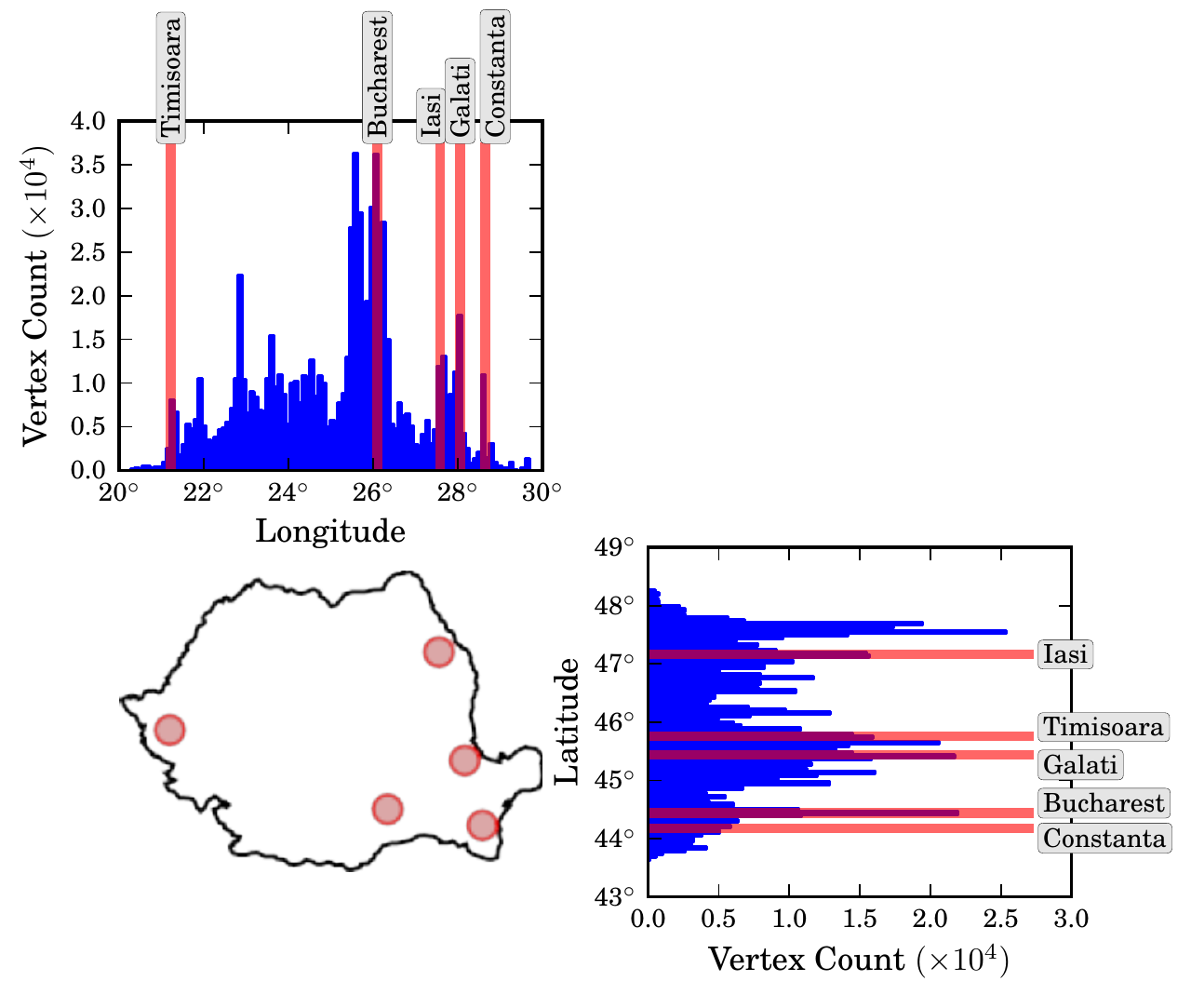} 
\caption{Vertex concentrations by coordinates -- GIS}
\label{fig:gis_density_map_lon_lat}
\end{center}
\end{figure}

\fref{fig:gis_density_map_lon_lat} displays the number of vertices at specific latitudes and longitudes,
with annotations to signify where certain cities are located.
This shows vertices are concentrated near the coordinates of large cities,
and provides additional evidence of the heterogeneous nature of vertex distribution, 
with respect to coordinates.

To give a clearer perspective of topology, 
a subset --- \numprint{2519} vertices and \numprint{5035} edges --- 
of the GIS dataset was visualized (\fref{fig:sampleGIS_1000_v2519_6,blue__e5035_2,black_}) using igraph.
This data represents a \unit{185}{\squaren{\kilo\metre}} region around the center of Bucharest.

\begin{figure}[htbp]
\begin{center}
\includegraphics[scale=0.2]
{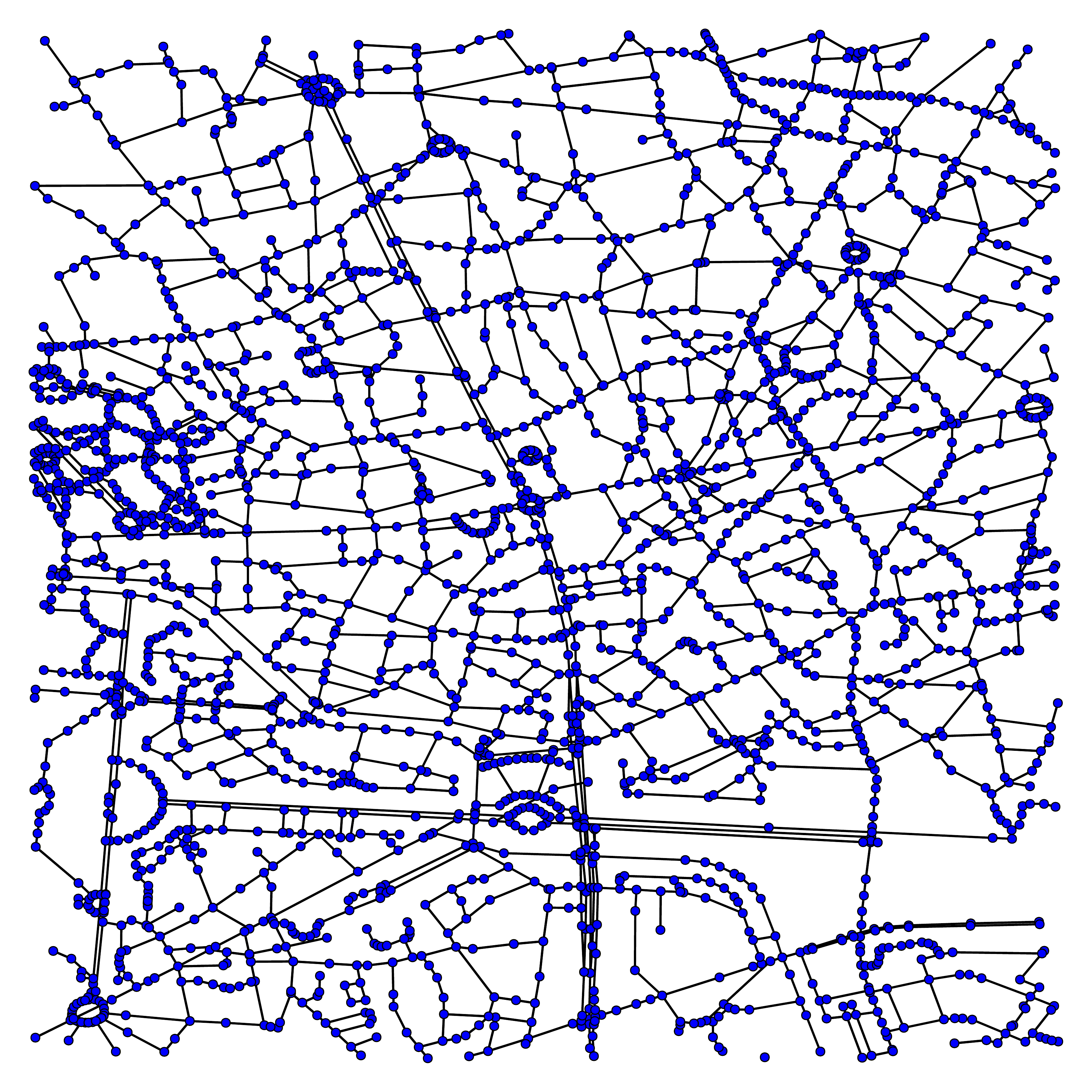} 
\caption{Sample graph -- GIS}
\label{fig:sampleGIS_1000_v2519_6,blue__e5035_2,black_}
\end{center}
\end{figure}		

\paragraph*{Access Patterns:}
GIS dataset access patterns were defined as a shortest path search, 
using the A* \cite{ref:general57} algorithm, between two geographic points.
Two variations of this access pattern were defined, short and long.
Short operations were designed to --- with higher probability --- start and end within one city,
whereas long operations were intended to be between different cities.

Start point was selected randomly, 
where probability of selecting a vertex as start point was proportional to its distance from the nearest city. 
The following cities were considered:
Bucharest, Iasi, Galati, Timisoara, and Constanta.

For long operations the end point was selected using the same method used for selecting the start point.
For short operations the end point was selected by performing a random walk from the start point, 
then choosing the vertex on which the walk ends.
Walk length was selected randomly using an exponential distribution with mean 11.
The reason for using this mean follows.
Graph diameter was calculated for the subgraph representing the inner \unit{100}{\squaren{\kilo\metre}} of Bucharest, which was equal to 22.
This value was then halved, 
because operations were assumed to start in the center of a city and remain in the same city.

We generated \numprint{10000} of these operations and wrote them to an operation log.
The traffic each of them generated is displayed in \fref{fig:gis_traf}, in sorted order.

\begin{table}[htbp]
 \centering
\subfloat[Short Operations]
{\includegraphics[scale=1.0]{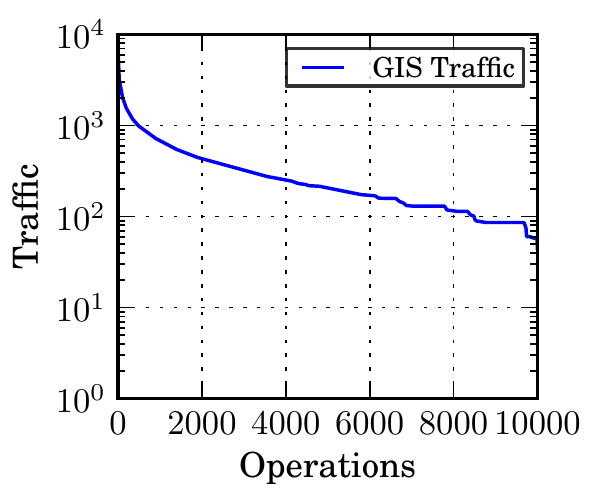}
\label{fig:gis_traf_short}}
\subfloat[Long Operations]
{\includegraphics[scale=1.0]{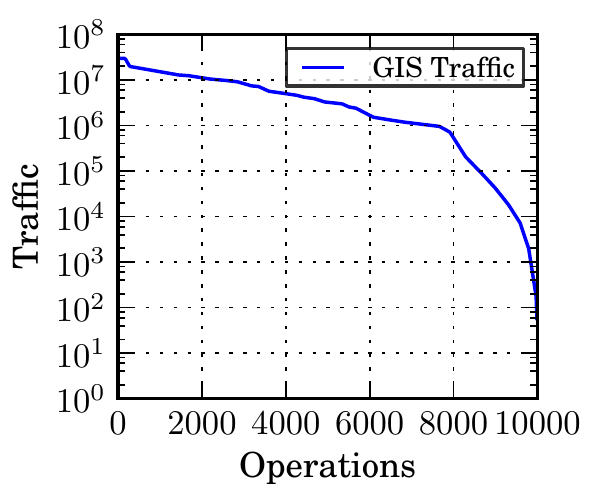}
\label{fig:gis_traf_long}}
\caption{Traffic distribution of access pattern -- GIS}
\label{fig:gis_traf} 
\end{table}
						
\tref{tab:fstree_graph_actions} lists the sequence of graph actions performed at each step of the access pattern,
and identifies which of these actions can potentially generate network traffic.
Note that \texttt{Vertex.getId()} appears twice in the table.
This is because the A* algorithm repeatedly compares the destination vertex against other vertices.

\begin{table}[htbp]
\extrarowheight = 0.5mm
\begin{center}
\begin{tabular}{|l|l|}
\hline 
\textbf{Graph Action} & \textbf{Potential Network Traffic} \\ 
\hline \hline

\texttt{Vertex.getId()} &
\multicolumn{1}{c|}{No} \\ \hline

\texttt{Vertex.getId()} &
\multicolumn{1}{c|}{No} \\ \hline

\texttt{Vertex.getProperty(LATITUDE)} &
\multicolumn{1}{c|}{No} \\ \hline

\texttt{Vertex.getProperty(LONGITUDE)} &
\multicolumn{1}{c|}{No} \\ \hline

\texttt{Vertex.getEdge()} &
\multicolumn{1}{c|}{Yes} \\ \hline

\texttt{Edge.getId()} &
\multicolumn{1}{c|}{No} \\ \hline

\texttt{Edge.getProperty(WEIGHT)} &
\multicolumn{1}{c|}{No} \\ \hline

\texttt{Edge.getStartVertex()} &
\multicolumn{1}{c|}{No} \\ \hline

\texttt{Edge.getEndVertex()} &
\multicolumn{1}{c|}{No} \\ \hline

\end{tabular} 
\caption{Access pattern graph actions --- GIS}
\label{tab:gis_graph_actions}
\end{center}
\end{table}		

\fref{fig:gis_traf_short} shows most operations generate similarly little traffic, 
with few generating considerably more.
With high probability short operations start in highly clustered areas (cities).
Because random walks tend to stay in highly clustered areas \cite{ref:partition32},
start and end points will remain close.
The relative few larger operations can be explained by the exponential distribution used to select walk length.

\fref{fig:gis_traf_long} shows 80\,\% of operation generating a very large amount of traffic,
and the remainder generating considerably less.
This is because only five cities are considered when selecting start and end points.
As a result, the start and end points for 20\,\% of all operations will reside in or near the same city.
				
	\subsection{Social Network}		

Twitter is a social networking service that enables users to publish messages and read those of other users. 
Users can subscribe to receive messages that were published by specific users;
this is known as \textit{following} users. 
The Twitter dataset used in this evaluation modeled the ``follows'' relationships between users.
It was acquired by crawling the Twitter service, and provided by the authors of \cite{ref:graph_general65}.
			
\paragraph*{Structure:}
Vertices model users and edges model the ``follows'' relationships between them.
An out-going edge indicates a ``following'' relationship, 
and an in-coming edge indicates a ``being followed'' relationship.
In this case edges are uni-directional,
If user $a$ is following user $b$ it does not imply the opposite is also true. \\

\begin{figure}[htbp]
\begin{center}
\includegraphics[scale=1.0]{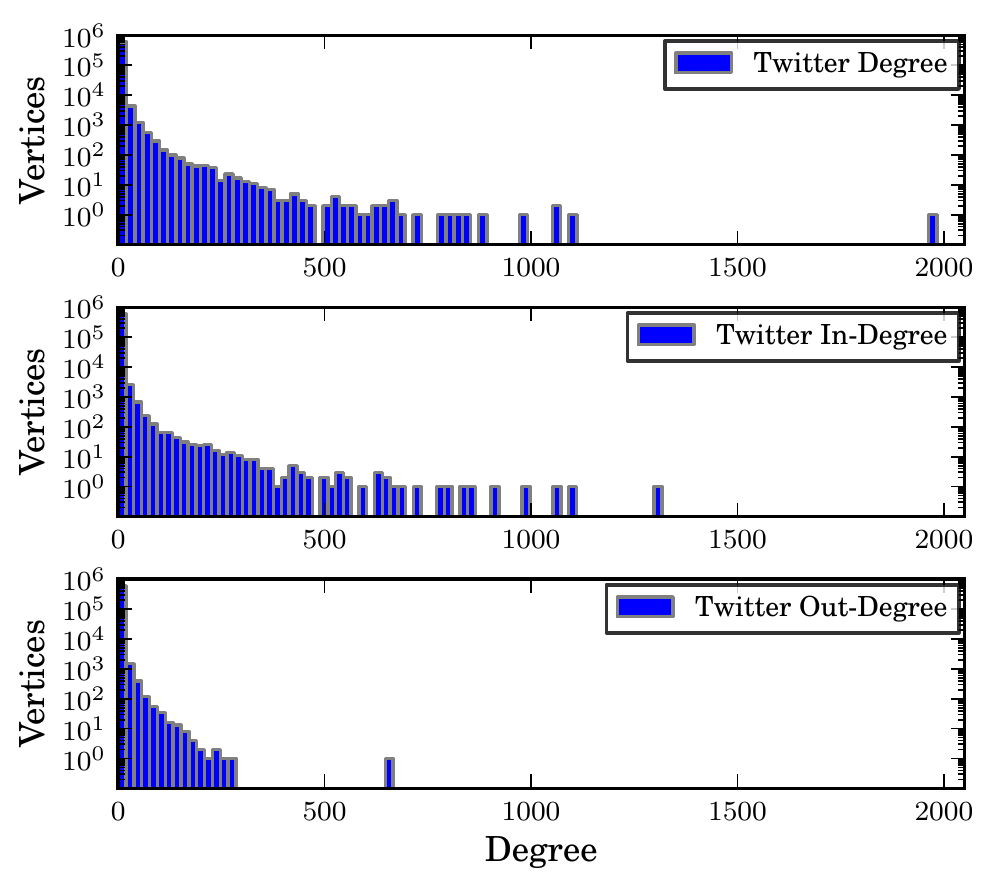} 
\caption{Degree distribution of Twitter}
\label{fig:twitter_degree_dist_all}
\end{center}
\end{figure}

The graph contains \numprint{611643} vertices, \numprint{851799} edges, 
and has a clustering coefficient of \numprint{0.000311}.
The relatively low clustering coefficient indicates that clusters are harder to identify, 
as they do not differ much from the rest of the graph.
The vertex degree follows an exponential distribution (\fref{fig:twitter_degree_dist_all}). 
Graphs with such a distribution are referred to as scale-free graphs and are common for social networks.

To give a clearer perspective of topology, 
a subset --- \numprint{1283} vertices and \numprint{1340} edges --- 
of the Twitter dataset was visualized (\fref{fig:sampleTwitterStart_1000_v1283_10,blue__e1340_2,black_}) 
using igraph.
This subgraph was created by selecting a random vertex, 
then performing Breadth First Search to retrieve nearby vertices.

\begin{figure}[htbp]
\begin{center}
\includegraphics[scale=0.2]
{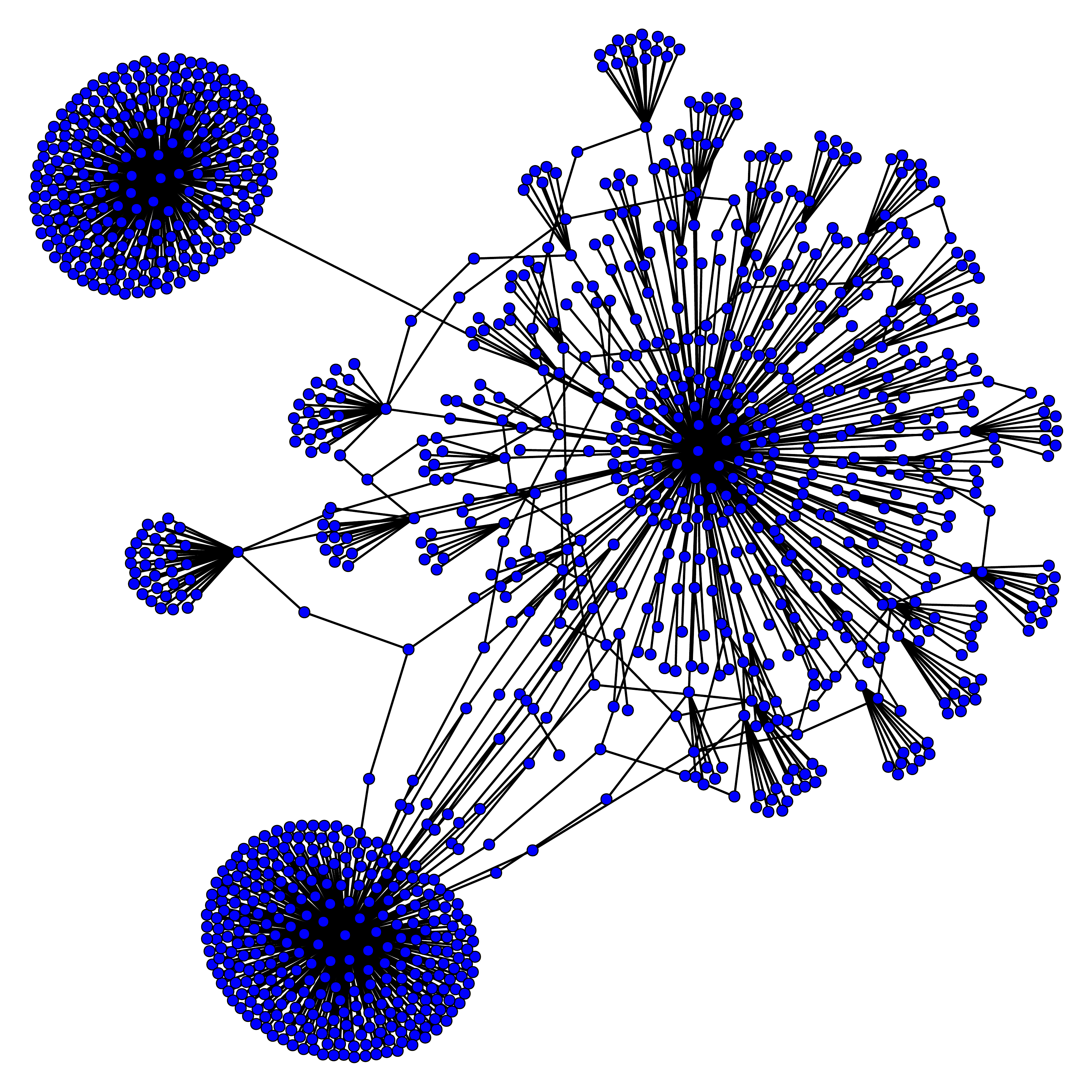} 
\caption{Sample graph -- Twitter}
\label{fig:sampleTwitterStart_1000_v1283_10,blue__e1340_2,black_}
\end{center}
\end{figure}		

\paragraph*{Access Patterns:}
Access patterns for the Twitter dataset were defined as friend-of-a-friend operations,
`retrieve all users being followed by the users I follow'.
Once a start vertex is selected, a Breadth First Search operation is performed using only out-going edges,
and extending up to two traversal steps.
The start point for these operations was selected randomly, 
where probability of selecting a vertex as start point is proportional to the out-degree of that vertex.

\begin{figure}[htbp]
\begin{center}
\includegraphics[scale=1.0]{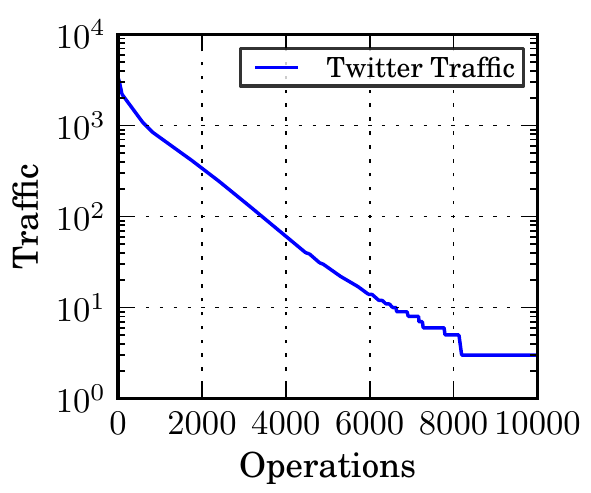} 
\caption{Traffic distribution of access pattern -- Twitter}
\label{fig:twitter_traf}
\end{center}
\end{figure}

We generated \numprint{10000} of these operations and wrote them to an operation log.
The traffic each generated is displayed in \fref{fig:twitter_traf}, in sorted order.

\tref{tab:fstree_graph_actions} lists the sequence of graph actions performed at each step of the access pattern,
and identifies which of these actions can potentially generate network traffic.

\begin{table}[htbp]
\extrarowheight = 0.5mm
\begin{center}
\begin{tabular}{|l|l|}
\hline 
\textbf{Graph Action} & \textbf{Potential Network Traffic} \\ 
\hline \hline

\texttt{Vertex.getId()} &
\multicolumn{1}{c|}{No} \\ \hline

\texttt{Vertex.getEdge()} &
\multicolumn{1}{c|}{Yes} \\ \hline

\texttt{Edge.getEndVertex()} &
\multicolumn{1}{c|}{No} \\ \hline

\end{tabular} 
\caption{Access pattern graph actions --- Twitter}
\label{tab:twitter_graph_actions}
\end{center}
\end{table}		

It is evident that the traffic distribution of Twitter operations follows an exponential decay,
where most operations generate a relatively small amount of traffic and only very few generate a large amount.
Despite the fact that start vertices are selected based on their out-degree,
the results show that start vertices are unlikely to have a high out-degree. 
This is because a very small percentage of the vertices have a high out-degree. Additionally, 
the degree distribution indicates that neighbors of the start vertex are also unlikely to have high out-degree.
When combined, 
these points suggest that likelihood of encountering vertices with high out-degree during an operation is low.
Given that traffic is at its maximum only when vertices with high out-degree are encountered during the Breadth First Search, this explains the traffic distribution plot.

	\subsection{Datasets --- Summary and Comparison}

The datasets presented in this section were all of a similar size, but each possessed a different graph topology.
Topologies varied between the tree-like structure of the file system dataset,
a scale-free topology for the Twitter dataset,
and a sparse graph with highly connected dense regions in the GIS dataset.

For each dataset an non-uniform access pattern was defined mimicing possible usage. Breadth First Search on the file system, shortest path algorithm on the GIS dataset, and friend-of-a-friend search for the Twitter dataset a. 
None of these patterns were based on actual production logs. 


\section{Partitioning Methods}
\label{sec:partitioning_methods}
The primary focus of this thesis was partitioning of graph databases.
To present a more thorough coverage of this topic, 
three general partitioning methods were compared in our experiments:
random partitioning, DiDiC partitioning, and hardcoded partitioning.

\begin{description}
\item[Random partitioning:] \ \\
Was the baseline partitioning method. 
Using random partitioning, the probability of any given vertex being assigned to any given partition is equal.

\item[DiDiC partitioning:] \ \\
Creates partitions by running the DiDiC graph partitioning algorithm for 100 iterations.

\item[Hardcoded partitioning:] \ \\
Refers to application specific partitioning methods.
Many applications have sufficient domain knowledge to make partitioning decisions within application logic. 
If topology and access patterns are known they can be considered when assigning data to partitions.

Given that our datasets model vastly different domains, unique partitioning methods were defined for each of 
them.

To limit the number of variables, 
a requirement of all hardcoded partitioning methods was to keep partition sizes equal.
\end{description}
A description of hardcoded partitioning methods follows.

\newpage
		\subsubsection*{File System Hardcoded Partitioning}
For the file system dataset, graph topology and access patterns were both known.
The dataset had a tree topology, and search operations were known to be confined within subtrees.
For these reasons, the hardcoded partitioning attempted to partition the dataset into subtrees.

All leaf folders were located and added to a list,
ensuring that folders close to each other --- part of same subtree --- 
were placed adjacent to each other in the list.
The list was then divided into as many segments as there were partitions,
ensuring segments were of equal size and, with high probability, 
neighboring folders were placed in the same segment.
Note, leaf vertices were partitioned first as they accounted for the majority of graph entities 
(refer to \fref{fig:fstree_nodes_at_levels_annotated});
when leaves are distributed evenly, the entire partitioning becomes balanced.
Then, moving up the tree, higher level folders were assigned to the same partition as their child folders,
and non-folder vertices to the same partition as their parent folder.

		\subsubsection*{GIS Hardcoded Partitioning}
For the GIS dataset, graph topology was not well known but domain knowledge and access patterns were.
Every vertex had longitude and latitude properties, so geographic location of every vertex was known.
In our case access patterns consisted only of shortest path search operations between two points.
Given these facts, the hardcoded partitioning attempted to keep geographically close vertices on the same partition.

\begin{figure}[htbp]
\begin{center}
\includegraphics[scale=1.0]{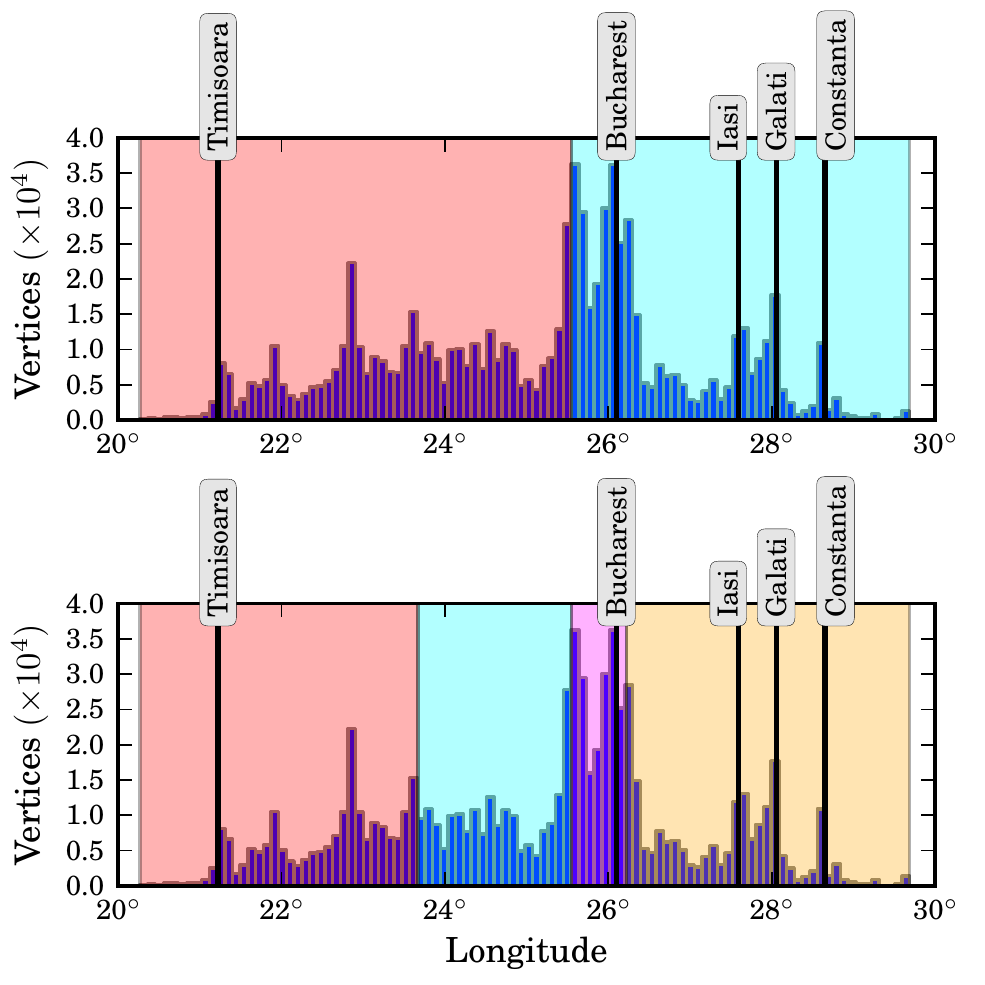} 
\caption{Hardcoded partitioning method for GIS dataset.
Upper image shows two partitions, lower image shows four partitions.}
\label{fig:gis_partition_lon_balanced}
\end{center}
\end{figure}

To simplify implementation, it was decided to only use longitude coordinates when partitioning.
The total number of vertices, $\vert V \vert$, range of longitude values, 
$20\degree-30\degree$, and number of desired partitions, $\vert \Pi \vert$, were known.
The partitioning method scanned over all vertices from east to west, assigning them to partitions in the following way.
The first $\vert V \vert / \vert \Pi \vert$ vertices encountered were written to the first partition,
the second $\vert V \vert / \vert \Pi \vert$ vertices to the second partition, etc.,
until all vertices had been assigned to partitions.
Refer to \fref{fig:gis_partition_lon_balanced} for a graphical illustration.

		\subsubsection*{Twitter Hardcoded Partitioning}
For the Twitter dataset little was known about the graph topology,
and no domain knowledge was encoded in the properties of vertices or edges.
Due to insufficient available data, no hardcoded partitioning was performed.

\newpage
\section{Dynamism}
\label{sec:dynamism}

As mentioned, some experiments were designed to test how dynamism affected the partitioning methods.
Usually dynamism refers to the act of changing a (graph) data structure by means of adding or removing (graph) entities.
For the purposes of this evaluation, the goal of dynamism was to degrade partitioning quality.

For a fair comparison between the original datasets and those that had been exposed to dynamism,
evaluation was always performed using identical evaluation logs.
However, evaluation logs are guaranteed to return the same results only if executed on the same graph.
Therefore, a requirement of the dynamism creation process was that it not change the graph structure of a dataset,
regardless of the amount of dynamism applied.
To meet that requirement, vertices and edges were never actually added or removed.
Instead, they were assigned to different partitions,
simulating the act of removing and then reinserting the same graph entities.

For the purpose of measuring the quantity of dynamism,
we defined one unit of dynamism as the act of moving one vertex,
from its original partition to any given partition (including its original partition).
As an extension, when expressed as a percentage, dynamism refers to:
\begin{equation}
\mbox{dynamism} = \frac{\mbox{units of dynamism performed}}{\vert V \vert}
\label{eq:dynamism}
\end{equation}
This method of creating dynamism ensures graph structures remain unchanged, 
and meets the goal of degrading partition quality.
The definition given in \eref{eq:dynamism} makes dynamism quantifiable and easier to reason about.

\paragraph*{}
To create dynamism, three Insert-Partitioning methods (\sref{sec:partitioning_graphdbs})
were compared in our experiments.
For every method, 
the vertices to move were selected uniformly at random from among all vertices in a dataset.
For selecting partitions to move vertices to, the three evaluated methods were:
random, fewest vertices, and least traffic.

\begin{description}
\item[Random:] \ \\
The probability of moving a vertex to any given partition is equal.
Baseline Insert-Partitioning method for creating dynamism.

\item[Fewest Vertices:] \ \\
Vertices are moved to the partition with fewest vertices.
Intended to balance partition sizes.

\item[Least Traffic:] \ \\
Vertices are moved to the partition that has had the least amount of total traffic.
Intended as a naive approach to balance traffic on partitions.
\end{description}

\section{Experiments}
\label{sec:experiments}

This section explains all experiments that were performed during evaluation.
For each experiment it briefly describes the reason for, and method of, performing it.

\begin{description}
\item[Static experiment:] \ \\
Designed to compare different partitioning methods (see \sref{sec:partitioning_methods}).

The partitioned datasets were used as input, no dynamism was applied,
evaluation logs were applied on each dataset, then performance was measured and logged.

\item[Insert experiment:] \ \\
Designed to measure how partitioning quality degrades when dynamism is applied to partitioned datasets.

Five levels of dynamism --- 1\,\%, 2\,\%, 5\,\%, 10\,\%, and 25\,\% --- were explored, using the three insert methods described in \sref{sec:dynamism}.
Since Least Traffic insert partitioning method relies on traffic distribution, dynamism was interleaved with read operations, as defined by the dataset's access pattern. 


The DiDiC partitionings were used as input. Dynamism was applied and snapshots taken.
Evaluation logs were executed on the snapshots. Finally, performance was measured and logged.

\item[Stress experiment:] \ \\
Designed to measure the ability of DiDiC in repairing a partitioning after dynamism had degraded its quality.

The snapshots created during insert experiments were used as input to this experiment.
One iteration of DiDiC was performed on each snapshot to improve its partitioning.
Evaluation logs were performed on the repartitioned snapshots.
Finally, performance was measured and logged.

\item[Dynamic experiment:] \ \\
Designed to measure the ability of DiDiC in coping with ongoing dynamism.

The 25\,\% dynamism operations from insert experiments were re-applied for this experiment.
However, they were divided into five equal-sized logs, each equivalent 5\,\% of dynamism.
The DiDiC partitionings were used as input and 5\,\% dynamism was applied five times.
After each application of dynamism: 
one iteration of DiDiC was performed, evaluation logs executed, then performance measured and logged.
\end{description}	

\section{Summary}	
The intention of this chapter was to cover the evaluation process in depth,
and present the environment in which all experiments were performed.

The evaluated datasets were covered, including their source and topology.
For each dataset the associated access patterns were described, 
including how evaluation logs where created, the traffic they generated, 
and a brief reasoning about what caused the observed traffic.
An introduction of partitioning method was given, 
stating motivation for, and implementation of, each one where necessary.
The concept of dynamism was outlined. 
This included the motivation for creating dynamism,
the requirements placed on dynamism-generating methods, 
and implementation of the methods considered during this evaluation.
Finally, the experiments performed as part of this evaluation, their purpose, 
and their implementation were briefly detailed.
\chapter{Evaluation Results}
\label{cha:eval_results}

As covered in \cref{cha:eval_method}, numerous experiments were performed as part of this evaluation.
The results from those experiments, along with associated discussions, are presented in this chapter.

\section{Measurements}	

The data obtained from our experiments was designed to measure load balance across partitions 
and network traffic between partitions.
Load balance refers to how vertices, edges, and traffic, are distributed among the partitions.
Network traffic refers to how often read operations cause two partitions to communicate.

\paragraph*{}
Load balance values are presented using the coefficient of variation, $c_v$.
Load at each partition is measured, standard deviation of these load values is calculated,
standard deviation is divided by the mean of all load values,
then this value is presented as a percentage.
Coefficient of variation is standard deviation as a percentage of the mean, 
as given by \eref{eq:coefficient_of_variation}.

\begin{equation}
\label{eq:coefficient_of_variation}
c_v = \frac{\sigma}{\mu}
\end{equation}

Network traffic is presented using two measurements:
edge cut, as a theoretic estimate of expected network traffic;
and a metric we refer to as Percentage Global, $\mathcal{T}^{\%}_G$.
As explained in \sref{sec:datasets_and_access_patterns}, 
for each dataset and evaluation log the total amount of generated traffic was already known,
we refer to this as Total Traffic, $\mathcal{T}_T$.
Then, during experimentation we logged the number of operations that caused two partitions to communicate,
we refer to this as Global Traffic, $\mathcal{T}_G$.
Percentage Global is then a ratio of these values, as given by \eref{eq:perc_global}.

\begin{equation}
\label{eq:perc_global}
\mathcal{T}^{\%}_G = \frac{\mathcal{T}_G}{\mathcal{T}_T}
\end{equation}

\section{Edge Cut}
\label{sec:experiment_edge_cut}

The edge cut was calculated for all partitioned datasets.
This was done to investigate if a correlation exists between theoretic partitioning quality metrics 
(refer to \tref{tab:partitioning_constraints})
and the network traffic generated when accessing a partitioned graph database.

The edge cut values presented in \tref{tab:edgecut} show the effectiveness of each partitioning method at reducing edge cut. A few interesting results are visible here.
The hardcoded partitioning methods perform extremely well,
showing that --- when it exists --- it is beneficial to make use of domain knowledge during partitioning.
DiDiC partitioning compares well to hardcoded partitioning, 
in spite of the fact that no domain knowledge was used when assigning vertices to partitions.

Edge cut results of DiDiC partitioning for the Twitter dataset are considerably higher, 
highlighting an important point. Not all graph topologies can be partitioned equally well. 
For random partitioning edge cut can be estimated as $1-(1/\vert \Pi \vert)\times 100$, 
following a logarithmic growth rate with respect to partition count. 
Results showed the other partitioning methods following a similar growth rate, 
though the sample size is too small to generalize this observation.

%
%

\begin{table}[htbp]
\extrarowheight = 0.5mm
\begin{center}
\begin{tabular}{|c|c|c|c|}
\hline 
\textbf{Dataset} & \textbf{Partitions} & \textbf{Partitioning} & \textbf{Edge Cut} \\ 
\hline \hline

\multirow{6}{*}
{File System} & 

\multirow{3}{*}
{2} & 

Random &
50\,\% \\ \cline{3-4}

&
&
DiDiC &
2.4\,\% \\ \cline{3-4}

&
&
Hardcoded &
0.05\,\% \\ \cline{2-4}

&
\multirow{3}{*}
{4} & 

Random &
75\,\% \\ \cline{3-4}

&
&
DiDiC &
3.6\,\% \\ \cline{3-4}

&
&
Hardcoded &
0.07\,\% \\ \hline \hline

\multirow{6}{*}
{GIS} & 

\multirow{3}{*}
{2} & 

Random &
50\,\% \\ \cline{3-4}

&
&
DiDiC &
1.9\,\% \\ \cline{3-4}

&
&
Hardcoded &
0.01\,\% \\ \cline{2-4}

&
\multirow{3}{*}
{4} & 

Random &
75\,\% \\ \cline{3-4}

&
&
DiDiC &
3.2\,\% \\ \cline{3-4}

&
&
Hardcoded &
0.04\,\% \\ \hline \hline

\multirow{4}{*}
{Twitter} & 

\multirow{2}{*}
{2} & 

Random &
50\,\% \\ \cline{3-4}

&
&
DiDiC &
25\,\% \\ \cline{2-4}

&
\multirow{2}{*}
{4} & 

Random &
75\,\% \\ \cline{3-4}

&
&
DiDiC &
37\,\% \\ \hline

\end{tabular} 
\caption{Edge cut for all datasets and partitioning methods}
\label{tab:edgecut}
\end{center}
\end{table}		

\section{Static Experiment}	

Results from the static experiments (as described in \sref{sec:experiments}) are presented in this section.
For clarity, each dataset is covered in a different subsection.

	\subsection{File System}
	\label{sec:experiment_static_fstree}
	
Load balance results of the file system dataset are presented in \tref{tab:fstree_bal_std}.
As expected, random partitioning effectively balanced vertex and edge distributions across partitions.
A promising result is that DiDiC too, produced very good results.
Hardcoded partitioning performed well, but resulted in slightly unbalanced load.
This is likely because the dataset contained five organizations, 
making it difficult for the partitioning logic to allocate vertices evenly across two and four partitions.

For traffic distribution, random and DiDiC partitionings again did very well,
with hardcoded partitioning performing markedly worse.
To understand why, the operations that generate this traffic must be considered.
For the file system access patterns, 
start vertices were selected with a probability proportionate to their out-degree.
Out-degree is approximately 35 for folders and one for files.
Assuming that all partitions contain the same ratio of folders to files, 
it follows that traffic distribution will be governed by vertex and edge distributions.
Vertex and edge distributions are most unbalanced in the hardcoded partitioning,
resulting in an unbalanced traffic distribution.

\begin{table}[htbp]
\extrarowheight = 0.5mm
\begin{center}
\begin{tabular}{|c|c|c|c|c|}
\hline 
\textbf{Partitions} & \textbf{Partitioning} & \textbf{Traffic} & \textbf{Vertices} & \textbf{Edges} \\ 
\hline \hline

\multirow{3}{*}
{2} & 

Random &
1.87\,\% & 
0.39\,\% & 
0.58\,\% \\ \cline{2-5}

&
DiDiC & 
3.24\,\% & 
0.24\,\% & 
0.23\,\% \\ \cline{2-5}

&
Hardcoded & 
13.96\,\% & 
1.80\,\% & 
1.82\,\% \\ \hline

\multirow{3}{*}
{4} & 

Random &
1.87\,\% & 
0.15\,\% & 
0.62\,\% \\ \cline{2-5}

&
DiDiC & 
0.78\,\% & 
0.51\,\% & 
0.48\,\% \\ \cline{2-5}

&
Hardcoded & 
15.52\,\% & 
2.54\,\% & 
2.58\,\% \\ \hline

\end{tabular} 
\caption{Load balance dislayed as Coefficient of Variation --- File System}
\label{tab:fstree_bal_std}
\end{center}
\end{table}		

\begin{figure}[htbp]
 \centering
\subfloat[2 partitions]
{\includegraphics[scale=1.0]{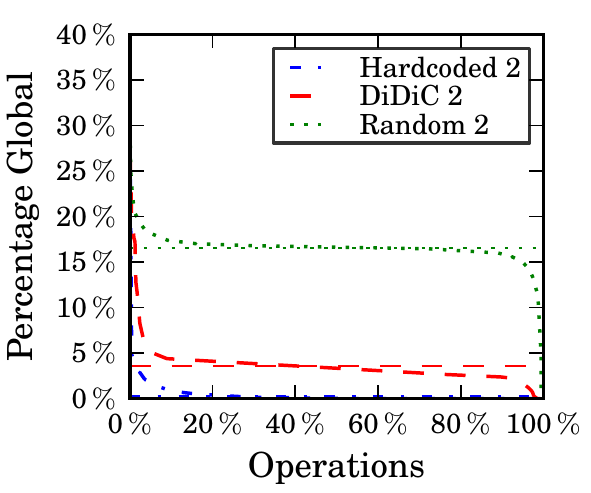}
\label{fig:fstree2_g_l_traf}}
\subfloat[4 partitions]
{\includegraphics[scale=1.0]{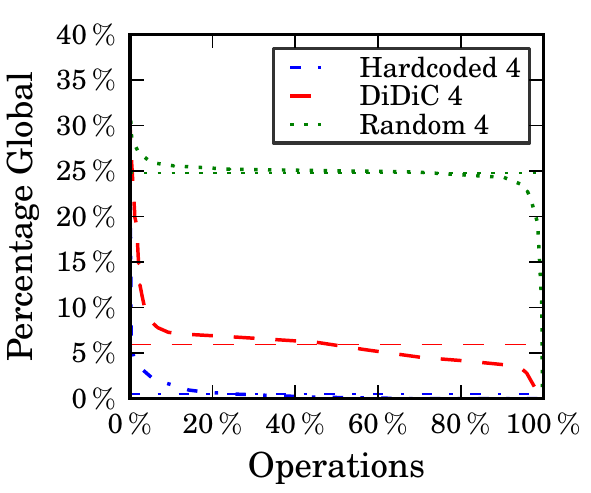}
\label{fig:fstree4_g_l_traf}}
\caption{Global traffic as percentage of total traffic --- File System}
\label{fig:fstree_g_l_traf} 
\end{figure}

The plots presented in \fref{fig:fstree_g_l_traf} display percentage global values for all operations,
ordered from highest to lowest. They show that, in comparison with random partitioning,
DiDiC partitioning reduced network traffic by approximately 80\,\%.
This same level of improvement is seen with two partitions and four partitions.
Hardcoded partitioning was even more effective, nearly eliminating network traffic.
The reason for this is that, although the DiDiC partitioning reduced edge cut tremendously,
it is unlikely that the vertices of any given partition, $\pi_i$,
formed a single connected component.
It is improbable that a path existed between two vertices of partition $\pi_i$,
that consisted only of vertices belonging to $\pi_i$.
For the hardcoded partitioning this is much more likely to occur,
so operations are significantly less likely to cross partition boundaries.

To explain the observed percentage global values, 
we claim a correlation exists between access patterns, percentage global, and edge cut.
\tref{tab:fstree_graph_actions} presents the graph actions performed at each step of a file system read operation.
It shows the number of actions that may incur network traffic, 
denoted by $\mathcal{T}_{PG}$ for Potential Global Traffic.
It also shows the number of actions guaranteed to never incur network traffic, 
denoted by $\mathcal{T}_L$ for Local Traffic.
Note that $\mathcal{T}_{PG}$ actions only incur network traffic when they span multiple partitions,
and the probability of this occurring is defined by edge cut.
Given an edge cut of $ec(\Pi)=x\,\%$
and the simplifying assumption that all edges are equally likely to be traversed,
results in $x\,\%$ of all $\mathcal{T}_{PG}$ actions incurring network traffic.
\eref{eq:perc_global_ext} formalizes this explanation as an extension of \eref{eq:perc_global}.
\begin{equation}\label{eq:perc_global_ext}
\mathcal{T}^{\%}_G = \frac{\mathcal{T}_{PG} \times ec(\Pi)}{\mathcal{T}_L + \mathcal{T}_G}
\end{equation}
Using \eref{eq:perc_global_ext}, the graph actions defined by \tref{tab:fstree_graph_actions},
edge cut data from \tref{tab:edgecut}, 
and mean percentage global values (illustrated by dashed horizontal lines in \fref{fig:fstree_g_l_traf}),
the correlation between percentage global and edge cut can be tested.
For two partitions and random partitioning, 
\fref{fig:fstree2_g_l_traf} shows that measured $\mathcal{T}^{\%}_G$ is 0.1652, 
and \eref{eq:perc_global_ext_fstree2} confirms the calculated value matches closely to measurements.
\begin{eqnarray}\label{eq:perc_global_ext_fstree2}
\mathcal{T}_{PG} & = & 1 \nonumber \\
\mathcal{T}_{L} & = & 2 \nonumber \\
ec(\Pi) & = & 0.50 \nonumber \\
\mathcal{T}^{\%}_G & = & \frac{1 \times 0.50}{2 + 1} \nonumber \\
& = & 0.1666 \nonumber \\
& \approx & 0.1652
\end{eqnarray}
For four partitions and random partitioning, 
\fref{fig:fstree4_g_l_traf} shows that measured $\mathcal{T}^{\%}_G$ is 0.2481, 
and \eref{eq:perc_global_ext_fstree4} confirms the calculated value matches closely to measurements.
\begin{eqnarray}\label{eq:perc_global_ext_fstree4}
\mathcal{T}_{PG} & = & 1 \nonumber \\ 
\mathcal{T}_{L} & = & 2 \nonumber \\ 
ec(\Pi) & = & 0.75 \nonumber \\ 
\mathcal{T}^{\%}_G & = & \frac{1 \times 0.75}{2 + 1} \nonumber \\ 
& = & 0.2500 \nonumber \\ 
& \approx & 0.2481
\end{eqnarray}
The flat nature of these percentage global plots can be explained in a similar way.
For all file system operations the values of $\mathcal{T}_{PG}$ and $\mathcal{T}_{L}$ were constant.
Fluctuations at left and right ends of the plot are due to fluctuations in the encountered edge cut;
the number of inter-partition edges encountered by these operations varied slightly from the global average. 
Note also that the calculated values assume a uniform access pattern, 
while non-uniform access patterns were used for all experiments. 

	\subsection{Romania GIS}
	
Load balance results of the GIS dataset are presented in \tref{tab:gis_bal_std}.
As with the file system dataset, 
random and DiDiC partitionings effectively balanced vertex and edge distributions.
In this case the hardcoded partitioning performed equally well.
This is because coordinates of all vertices were known, 
which made it trivial to partition them equally by longitude.
Again, random partitioning performed well with regard to traffic distribution.
DiDiC performed slightly worse but still achieved near-equal traffic distribution.
The hardcoded partitioning performed poorly here, and \fref{fig:gis_partition_lon_balanced} helps explain why.
Despite partition sizes being equal, the number of cities in each partition is not.
Because GIS access patterns selected start, and sometimes end, 
vertices with a probability proportionate to their distance from a city, this led to imbalanced traffic.

The difference in traffic imbalance between short and long operations is due to the nature of the operations.
End vertices in short operations were considerably closer to their start vertices than they were in long operations.
Consequently, if the distribution of start vertices across partitions was imbalanced, traffic imbalance followed.
Because long operations traversed more of the graph and generated more traffic, 
start vertex selection had a lower impact on traffic distribution.
\begin{table}[htbp]
\setlength{\belowcaptionskip}{0pt}   
\extrarowheight = 0.50mm
\begin{center}
\begin{tabular}{|c|c|c|c|c|c|}
\hline 
\textbf{Operation} & 
\textbf{Partitions} & 
\textbf{Partitioning} & \textbf{Traffic} & \textbf{Vertices} & \textbf{Edges} \\ 
\hline \hline

\multirow{6}{*}{Short} & 

\multirow{3}{*}{2} & 

Random &
0.17\,\% & 
0.06\,\% & 
0.06\,\% \\ \cline{3-6}

&
&
DiDiC & 
0.24\,\% & 
0.77\,\% & 
0.06\,\% \\ \cline{3-6}

&
&
Hardcoded & 
98.77\,\% & 
0.12\,\% & 
1.91\,\% \\ \cline{2-6}

&
\multirow{3}{*}{4} & 

Random &
0.98\,\% & 
0.18\,\% & 
0.21\,\% \\ \cline{3-6}

&
&
DiDiC & 
6.62\,\% & 
0.91\,\% & 
0.21\,\% \\ \cline{3-6}

&
&
Hardcoded & 
92.57\,\% & 
0.67\,\% & 
2.02\,\% \\ \hline \hline

\multirow{6}{*}{Long} & 

\multirow{3}{*}{2} & 

Random &
0.37\,\% & 
0.06\,\% & 
0.06\,\% \\ \cline{3-6}

&
&
DiDiC & 
0.67\,\% & 
0.77\,\% & 
0.06\,\% \\ \cline{3-6}

&
&
Hardcoded & 
30.47\,\% & 
0.12\,\% & 
1.91\,\% \\ \cline{2-6}

&
\multirow{3}{*}{4} & 

Random &
0.15\,\% & 
0.18\,\% & 
0.21\,\% \\ \cline{3-6}

&
&
DiDiC & 
1.83\,\% & 
0.91\,\% & 
0.21\,\% \\ \cline{3-6}

&
&
Hardcoded & 
24.89\,\% & 
0.67\,\% & 
2.02\,\% \\ \hline
\end{tabular} 
\caption{Load balance dislayed as Coefficient of Variation --- GIS}
\label{tab:gis_bal_std}
\end{center}
\end{table}		

\begin{figure}[htbp]
 \centering
\subfloat[2 partitions - short operations]
{\includegraphics[scale=1.0]{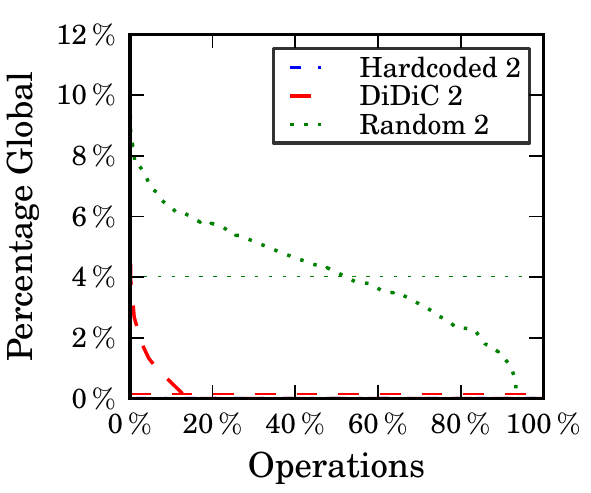}
\label{fig:gis2_g_l_traf_short}}
\subfloat[4 partitions - short operations]
{\includegraphics[scale=1.0]{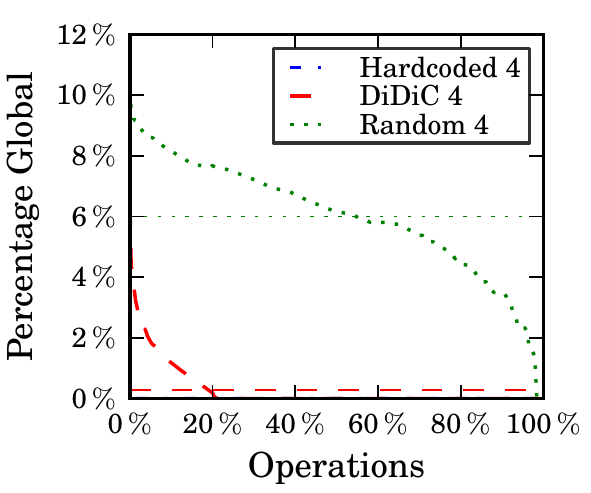}
\label{fig:gis4_g_l_traf_short}}
\qquad
\subfloat[2 partitions - long operations]
{\includegraphics[scale=1.0]{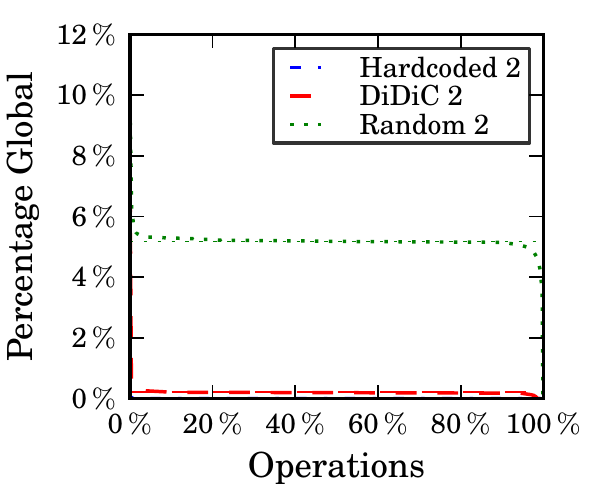}
\label{fig:gis2_g_l_traf_long}}
\subfloat[4 partitions - long operations]
{\includegraphics[scale=1.0]{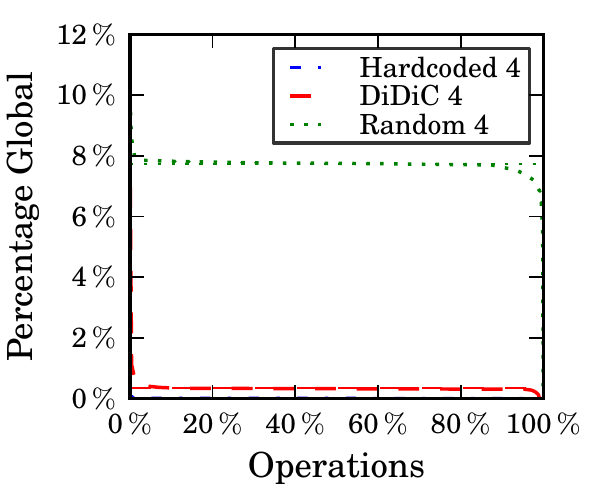}
\label{fig:gis4_g_l_traf_long}}
\caption{Global traffic as percentage of total traffic --- GIS}
\label{fig:gis_g_l_traf} 
\end{figure}

\fref{fig:gis_g_l_traf} displays the percentage global plots for all GIS operations,
ordered from highest to lowest.
They show that DiDiC reduced network traffic by over 90\,\% when compared to random partitioning,
and hardcoded partitioning performed even better, nearly eliminating network traffic.
The reason that hardcoded partitioning produced such impressive results is the same as was explained for the file system dataset.
DiDiC reduced edge cut well, but it is unlikely that the vertices of a partition, $\pi_i$, 
formed a single connected component. With the hardcoded partitioning this was more likely to occur.
The percentage global values, too, can be explained as they were for the file system dataset;
using the identified correlation between access patterns, percentage global, and edge cut.

With \eref{eq:perc_global_ext}, the graph actions defined by \tref{tab:gis_graph_actions},
edge cut data from \tref{tab:edgecut}, and mean percentage global values
(illustrated by dashed horizontal lines in \fref{fig:gis_g_l_traf}),
we test if our claim holds for the GIS dataset.

For long operations, two partitions, and random partitioning, 
\fref{fig:gis2_g_l_traf_long} shows that measured $\mathcal{T}^{\%}_G$ is 0.0517.
\eref{eq:perc_global_ext_gis2_long} confirms the calculated value matches closely to measurements.
\begin{eqnarray}\label{eq:perc_global_ext_gis2_long}
\mathcal{T}_{PG} & = & 1 \nonumber \\
\mathcal{T}_{L} & = & 8 \nonumber \\
ec(\Pi) & = & 0.50 \nonumber \\
\mathcal{T}^{\%}_G & = & \frac{1 \times 0.50}{8 + 1} \nonumber \\
& = & 0.0556 \nonumber \\
& \approx & 0.0517
\end{eqnarray}
For short operations, four partitions, and random partitioning, 
\fref{fig:gis4_g_l_traf_short} shows that measured $\mathcal{T}^{\%}_G$ is 0.0600.
While \eref{eq:perc_global_ext_gis4_short} results in a calculated value that approximately matches measurements.
\begin{eqnarray}\label{eq:perc_global_ext_gis4_short}
\mathcal{T}_{PG} & = & 1 \nonumber \\ 
\mathcal{T}_{L} & = & 8 \nonumber \\ 
ec(\Pi) & = & 0.75 \nonumber \\ 
\mathcal{T}^{\%}_G & = & \frac{1 \times 0.75}{8 + 1} \nonumber \\ 
& = & 0.0833 \nonumber \\ 
& \approx & 0.0600
\end{eqnarray}
The reason short operations displayed more fluctuations in their percentage global measurements is they were considerably shorter, 
meaning that the edge cut encountered by operations was more likely to diverge from the global average. 
As stated before the choice of the start point, and therefore the non-uniformity of the access pattern, 
has a higher impact on smaller traversals. 
The short operations of GIS tend to stay in highly connected city areas, 
and these areas are likely to be identified as clusters by DiDiC. 
A combination of these two observation is the reason for the difference in the mean values as well the differences between calculated $\mathcal{T}^{\%}_G$ and actual measurements. 


	\subsection{Twitter Social Network}		

Load balance results of the Twitter dataset are presented in \tref{tab:twitter_bal_std}.
Again, random partitioning created balanced vertex and edge distributions.
Unlike earlier results, the vertex and edge distributions created by DiDiC had visible imbalances,
especially with four partitions.
As explained in \sref{sec:partitioning_alg_didic}, DiDiC optimizes for modularity.
It attempts to locate natural clusters/communities in a graph, then assign those clusters to partitions.
If a dataset contains such clusters and they are of unequal sizes, 
the partitions created by DiDiC will likely be of unequal sizes too.
Edge distributions are significantly more imbalanced than vertex distributions because,
in our model, edges are stored on the same partition as their start vertex.
From the exponential out-degree distribution of the Twitter dataset (see \fref{fig:twitter_degree_dist_all}),
we know out-degree is low for most vertices and high for few.
With high probability, the partitions containing these high-out-degree vertices are those with most edges.

Looking at traffic distribution for the Twitter dataset, 
random and DiDiC partitioning performed worse than on other datasets.
This is correlated of the edge distribution imbalance.
Twitter access patterns selected start vertices with a probability proportionate to their out-degree,
if edges were not distributed evenly across partitions, out-degrees were not distributed evenly either.
With high probability, the partitions with greatest cumulative out-degree received most traffic.
\begin{table}[htbp]
\extrarowheight = 0.5mm
\begin{center}
\begin{tabular}{|c|c|c|c|c|}
\hline 
\textbf{Partitions} & \textbf{Partitioning} & \textbf{Traffic} & \textbf{Vertices} & \textbf{Edges} \\ 
\hline \hline

\multirow{2}{*}
{2} & 

Random &
5.85\,\% & 
0.29\,\% & 
0.73\,\% \\ \cline{2-5}

&
DiDiC & 
12.88\,\% & 
1.95\,\% & 
6.83\,\% \\ \hline

\multirow{2}{*}
{4} & 

Random &
7.71\,\% & 
0.40\,\% & 
0.89\,\% \\ \cline{2-5}

&
DiDiC & 
68.95\,\% & 
9.47\,\% & 
29.61\,\% \\ \hline

\end{tabular} 
\caption{Load balance dislayed as Coefficient of Variation --- Twitter}
\label{tab:twitter_bal_std}
\end{center}
\end{table}		

\begin{figure}[htbp]
 \centering
\subfloat[2 partitions]
{\includegraphics[scale=1.0]{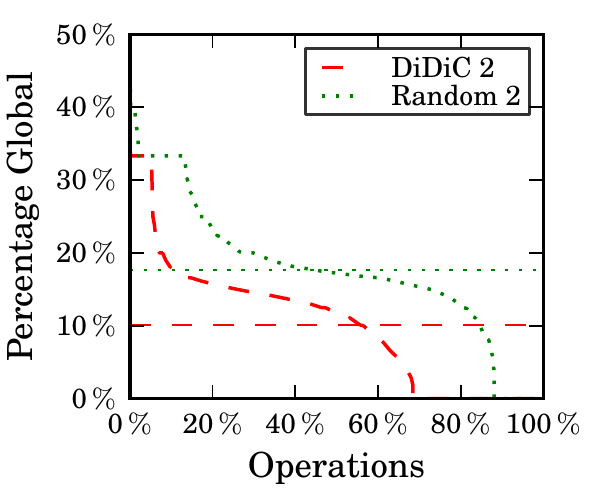}
\label{fig:twitter2_g_l_traf}}
\subfloat[4 partitions]
{\includegraphics[scale=1.0]{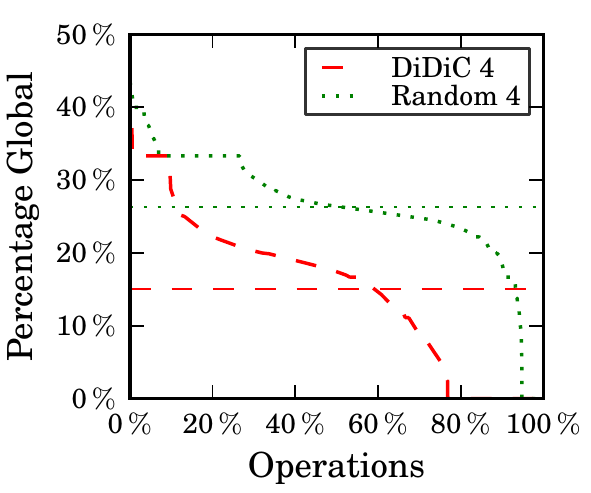}
\label{fig:twitter4_g_l_traf}}
\caption{Global traffic as percentage of total traffic --- Twitter}
\label{fig:twitter_g_l_traf} 
\end{figure}

\paragraph*{}
\fref{fig:twitter_g_l_traf} shows that, on the Twitter dataset, 
DiDiC reduced network traffic considerably when compared with random partitioning.
It did so by a lesser amount than on other datasets, but still provided an improvement of over 40\,\%.
Using \eref{eq:perc_global_ext}, the graph actions defined by \tref{tab:twitter_graph_actions},
edge cut data from \tref{tab:edgecut}, and mean percentage global values
(illustrated by dashed horizontal lines in \fref{fig:twitter_g_l_traf}),
we test if these results are consistent with previous reasoning.

For two partitions and random partitioning, 
\fref{fig:twitter2_g_l_traf} shows that measured $\mathcal{T}^{\%}_G$ is 0.1764.
\eref{eq:perc_global_ext_twitter2} confirms the calculated value matches closely to measurements.
\begin{eqnarray}\label{eq:perc_global_ext_twitter2}
\mathcal{T}_{PG} & = & 1 \nonumber \\
\mathcal{T}_{L} & = & 2 \nonumber \\
ec(\Pi) & = & 0.50 \nonumber \\
\mathcal{T}^{\%}_G & = & \frac{1 \times 0.50}{2 + 1} \nonumber \\
& = & 0.1667 \nonumber \\
& \approx & 0.1764
\end{eqnarray}
For four partitions and random partitioning, 
\fref{fig:twitter4_g_l_traf} shows that measured $\mathcal{T}^{\%}_G$ is 0.2627.
\eref{eq:perc_global_ext_twitter4} confirms the calculated value matches closely to measurements.
\begin{eqnarray}\label{eq:perc_global_ext_twitter4}
\mathcal{T}_{PG} & = & 1 \nonumber \\ 
\mathcal{T}_{L} & = & 2 \nonumber \\ 
ec(\Pi) & = & 0.75 \nonumber \\ 
\mathcal{T}^{\%}_G & = & \frac{1 \times 0.75}{2 + 1} \nonumber \\ 
& = & 0.2500 \nonumber \\ 
& \approx & 0.2627
\end{eqnarray}
In comparison with file system or GIS plots, the shape of these differs greatly.
The flat sections in upper-left and lower-right regions show operations that encountered edges of 
100\,\% edge cut and 0\,\% edge cut respectively.
It is probable that these operations started on vertices with relatively low out-degree,
meaning fewer edges were encountered, therefore increasing the probability of these outcomes occurring.
Similarly, the regions with high gradient --- inside from each flat section --- 
are the result of the same phenomenon, but to a lesser extent.

						
\section{Insert Experiment}

Results from the insert experiments are presented in this section
(refer to \sref{sec:dynamism} and \sref{sec:experiments} for more details).
Note that due to the excessive time required to execute the evaluation logs for the GIS long access pattern, they are excluded from this and subsequent experiments.

\newpage
	\subsection{File System}

\begin{figure}[htbp]
\begin{center}
\includegraphics[scale=1.0]{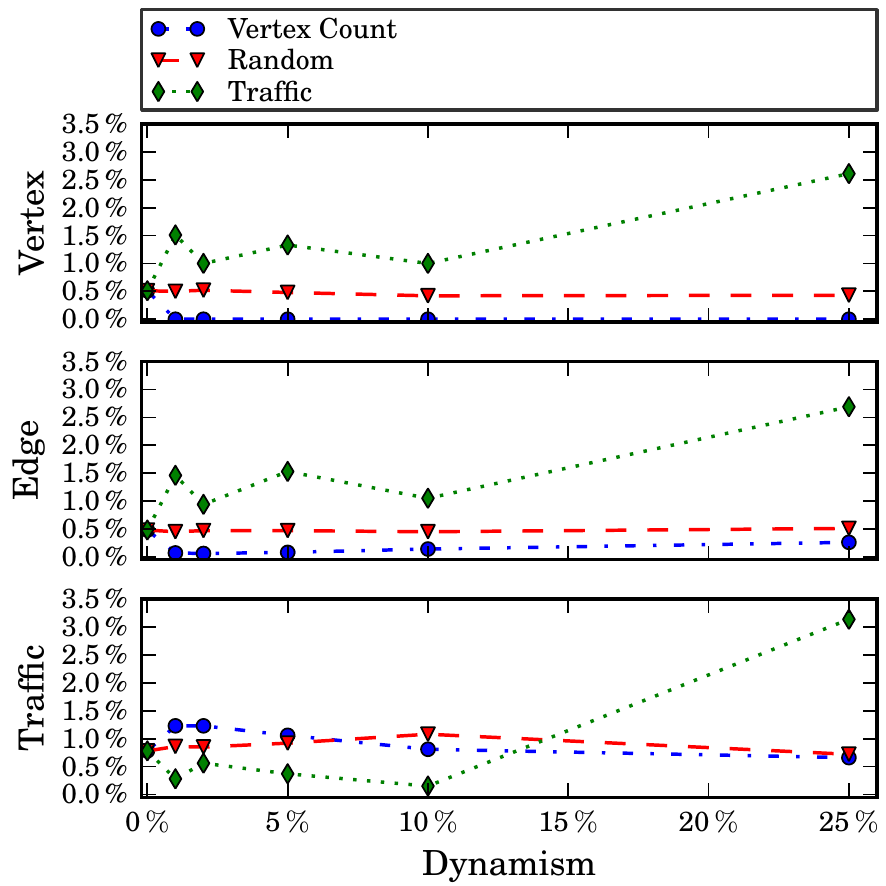} 
\caption{Load balance vs dynamism --- File System --- 4 partitions}
\label{fig:fstree4_insert_std_all}
\end{center}
\end{figure}

\fref{fig:fstree4_insert_std_all} shows how load balance of the file system dataset is affected by dynamism.
As expected, the vertex count insert method performed best with regards to balancing vertex and edge distributions.
However the random insert method also performed very well.
Given the experiment setup and the fact that random insert assigns vertices to any partition with equal probability,
it is expected to converge towards the balanced distribution.

Vertex and edge distributions created by the traffic insert method were more imbalanced.
As described in \sref{sec:dataset_and_access_fstree},
file system access patterns only made use of vertices modeling files or folders.
Event vertices were ignored, meaning these vertices were responsible for generating no traffic.
Moreover, over 50\,\% of vertices were Event vertices.
Therefore, when attempting to balance traffic distribution, 
many of the vertices moved by the traffic insert method had no effect on traffic distribution,
but did negatively impact vertex and edge distributions.
Edge distribution is shown to follow vertex distribution very closely.
This is because every vertex was equally likely to be moved, 
and all vertices of the same type had the same number of edges. 
Files had one out-going edge, folders had 35 out-going edges, and events have two out-going edges.

Regarding traffic distribution, 
because it started in a balanced state the vertex count and random insert methods did little to disturb it.
The reason for this was covered in \sref{sec:experiment_static_fstree}.
The traffic insert method performed worst here.
This is because it only considered past traffic when allocating vertices to partitions 
(the implementation is very simple).
If access patterns are not sufficiently predictable --- new patterns diverge from past patterns --- 
this insert method can negatively affect the traffic distribution.

\begin{figure}[htbp]
\begin{center}
\includegraphics[scale=1.0]{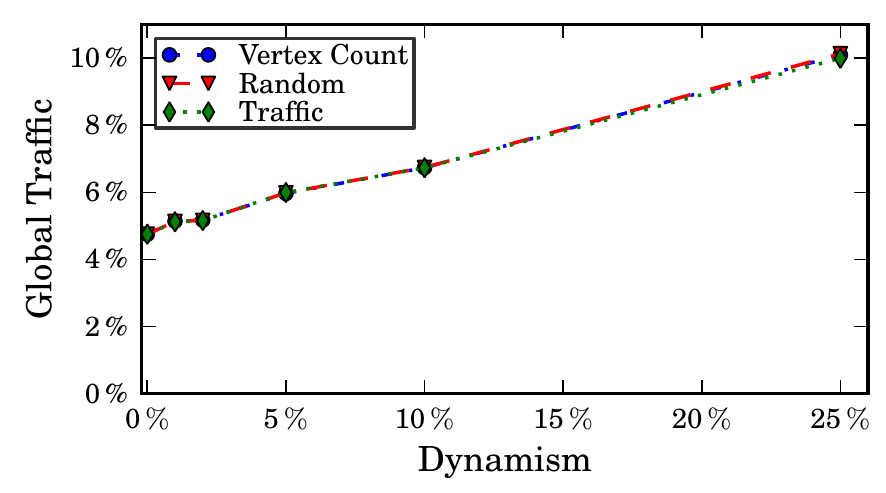} 
\caption{Network traffic vs dynamism --- File System --- 4 partitions}
\label{fig:fstree4_insert_comms_all}
\end{center}
\end{figure}

\newpage
\fref{fig:fstree4_insert_comms_all} shows how dynamism affects the amount of network traffic.
Measurements from initial experiments (\sref{sec:experiment_edge_cut})
have already shown that random partitioning resulted in a high edge cut.
The correlation between edge cut and network traffic explains why random insert resulted in increasing network traffic.
Interestingly, all insert methods behaved almost identically with regard to network traffic.
This can be explained by the fact that no insert method explicitly considered edge cut,
so none of them can expect to do better than the random insert method.

\newpage
	\subsection{Romania GIS}

\begin{figure}[htbp]
\begin{center}
\includegraphics[scale=1.0]{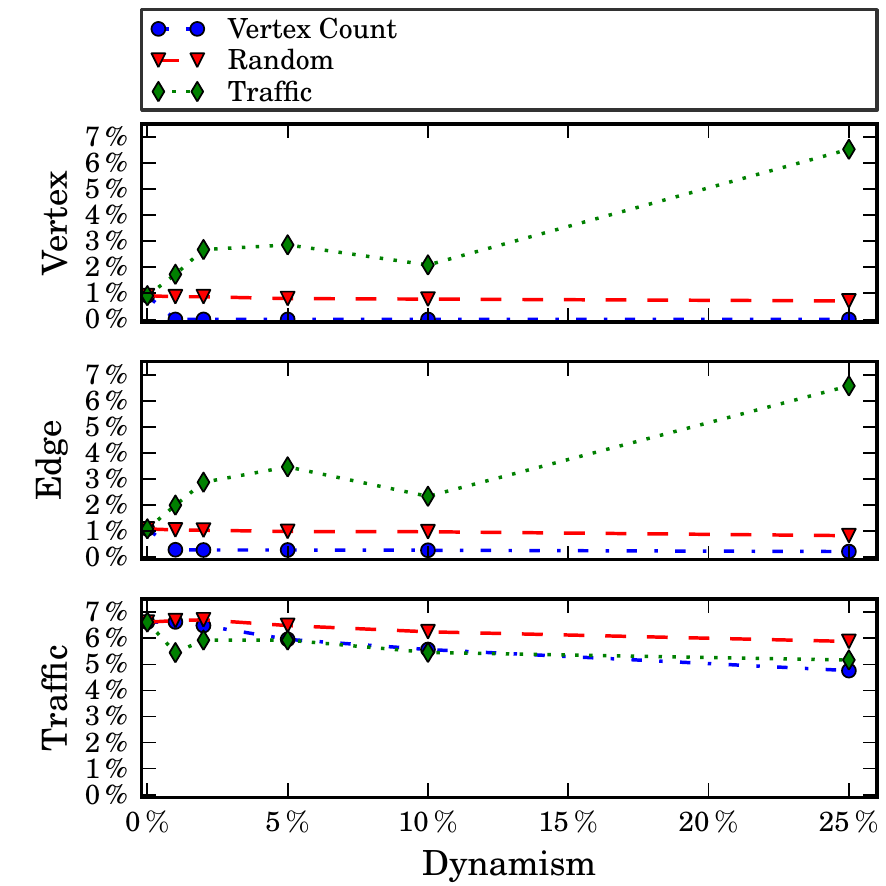} 
\caption{Load balance vs dynamism --- GIS --- 4 partitions}
\label{fig:gis4_insert_std_all}
\end{center}
\end{figure}

\begin{figure}[htbp]
\begin{center}
\includegraphics[scale=1.0]{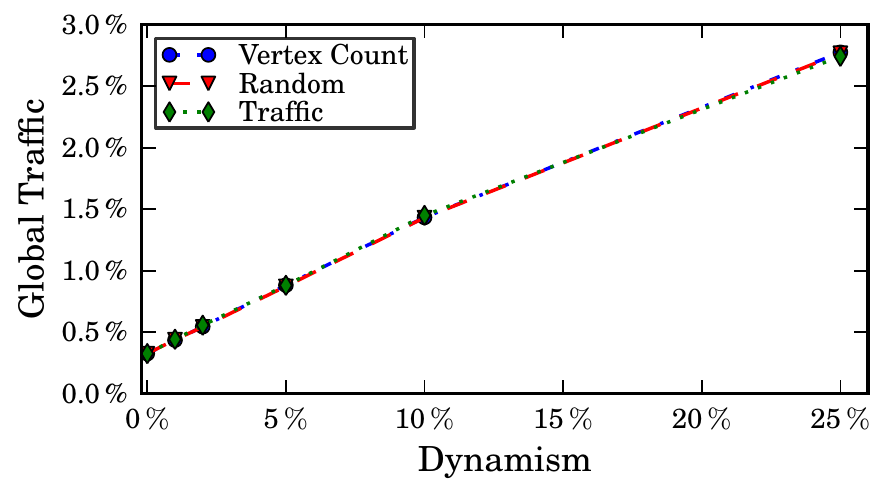} 
\caption{Network traffic vs dynamism --- GIS --- 4 partitions}
\label{fig:gis4_insert_comms_all}
\end{center}
\end{figure}

As shown in \fref{fig:gis4_insert_std_all}, 
load balance results of the GIS dataset are similar to the file system results.
Vertex count and random insert methods kept vertex and edge distributions balanced,
while traffic insert resulted in a greater imbalance.
Again, this is because GIS access patterns were non-uniform, 
meaning some vertices generated more traffic than others.
The traffic insert method was aware of traffic distribution across partitions,
but not of traffic distribution across vertices.
When attempting to balance traffic distribution, 
many of the vertices moved did not affect traffic distribution but did negatively impact vertex and edge distributions.
The traffic distribution results are also much like those of the file system dataset,
and caused by similar phenomenons.
The effect of dynamism on network traffic for the GIS dataset is shown in \fref{fig:gis4_insert_comms_all}.
These results closely mimic those of the file system dataset.

	\subsection{Twitter Social Network}		

\begin{figure}[htbp]
\begin{center}
\includegraphics[scale=1.0]{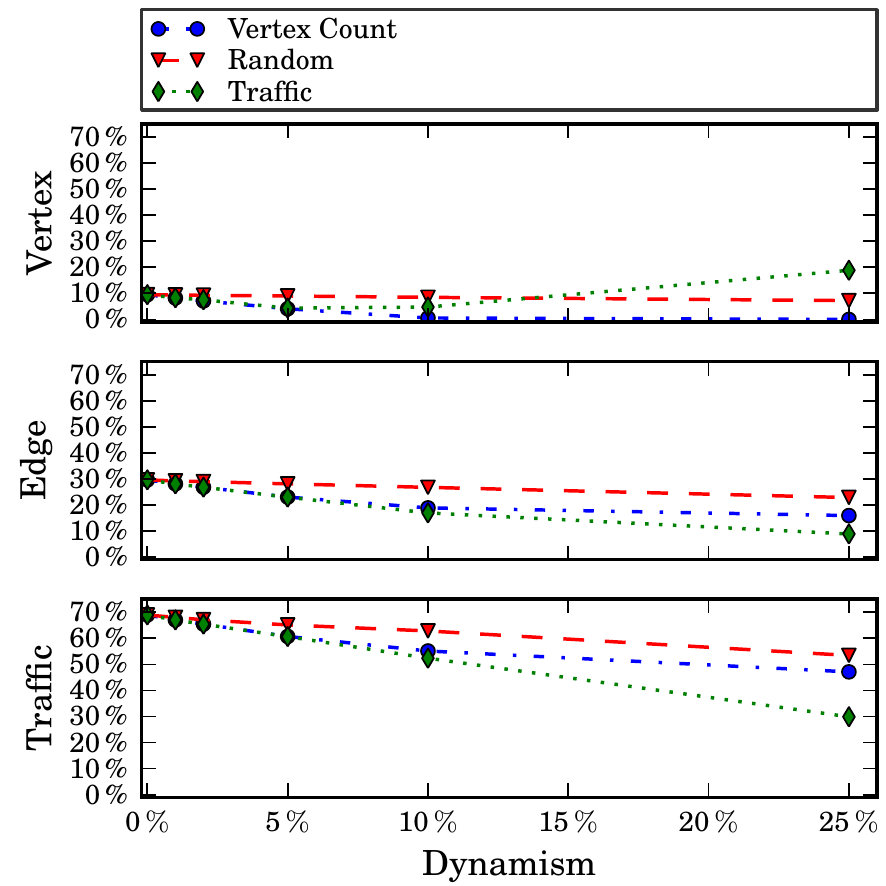} 
\caption{Load balance vs dynamism --- Twitter --- 4 partitions}
\label{fig:twitter4_insert_std_all}
\end{center}
\end{figure}

Load balance results for the Twitter dataset are presented in \fref{fig:twitter4_insert_std_all}.
Excluding the traffic insert method, everything is much the same as with our other datasets.
For the traffic insert method, as vertex imbalance increases, edge and traffic imbalance decreases.
The behavior of, both, edge and traffic distributions is related.
Twitter access patterns selected start vertices based on their out-degree.
Assuming a balanced vertex distribution, the out-degrees of vertices determines edge distribution across partitions.
Simply put, as the imbalance of edge distribution decreases so does the imbalance in traffic distribution.
		
\begin{figure}[htbp]
\begin{center}
\includegraphics[scale=1.0]{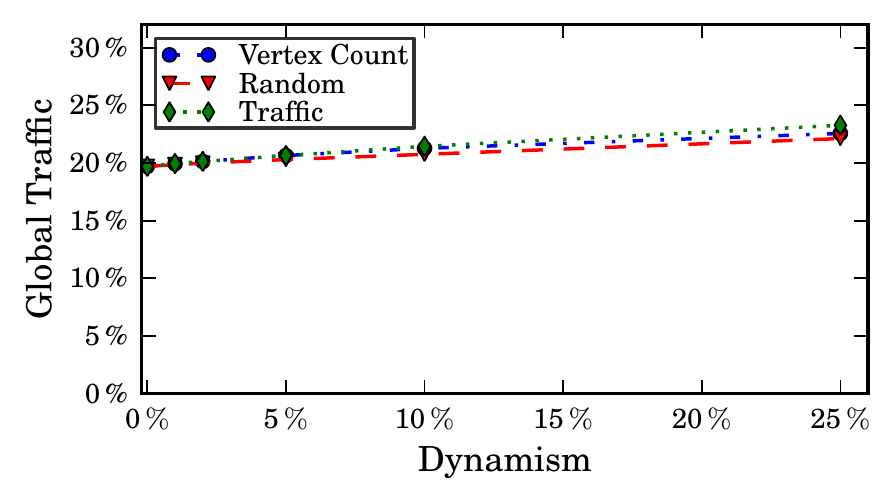} 
\caption{Network traffic vs dynamism --- Twitter --- 4 partitions}
\label{fig:twitter4_insert_comms_all}
\end{center}
\end{figure}

The effect of dynamism on network traffic for the Twitter dataset is shown in \fref{fig:twitter4_insert_comms_all}.
Once more closely mimicking the other datasets.

\section{Stress Experiment}

\begin{figure}[htbp]
 \centering
\subfloat[File System]
{\includegraphics[scale=1.0]{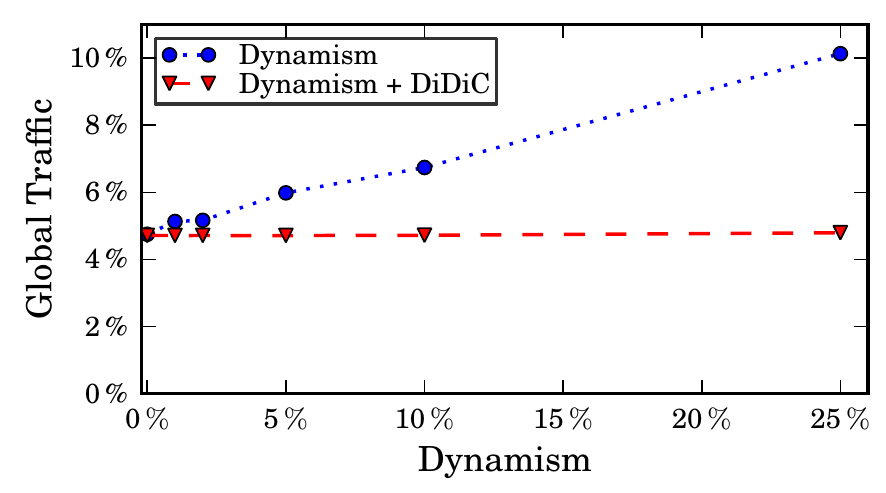}
\label{fig:fstree4_didic_stress_comms_all}}
\qquad
\subfloat[GIS]
{\includegraphics[scale=1.0]{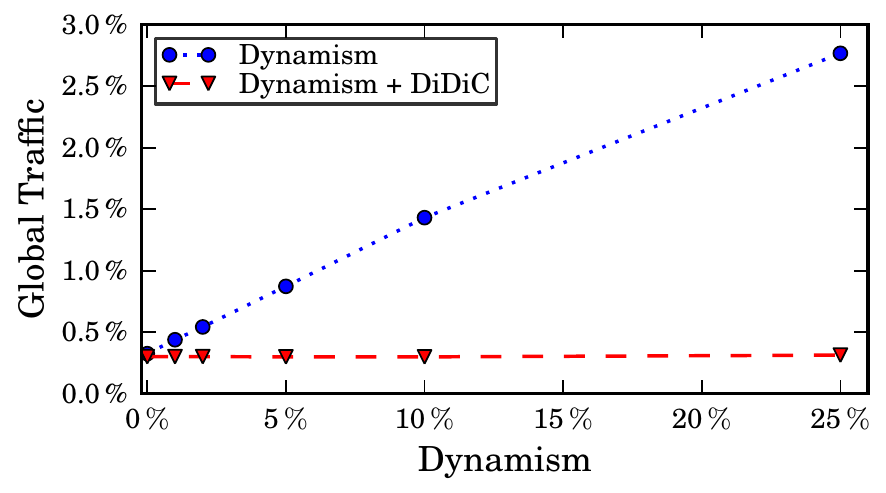}
\label{fig:gis4_didic_stress_comms_all}}
\qquad
\subfloat[Twitter]
{\includegraphics[scale=1.0]{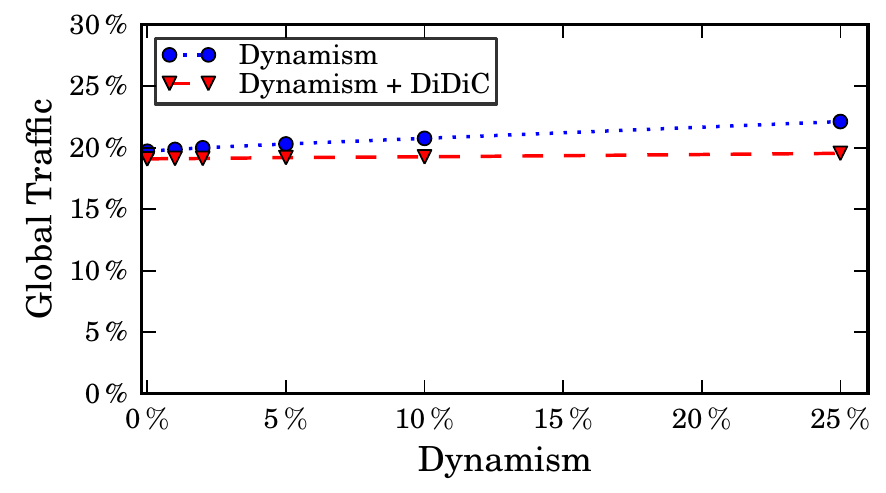}
\label{fig:twitter4_didic_stress_comms_all}}
\caption{Network traffic vs dynamism --- Stress experiment --- 4 partitions}
\label{fig:didic_stress_comms_all} 
\end{figure}

This experiment measured how well DiDiC could repair a partitioning after dynamism had degraded it.
Partitioning quality measurements were restricted to the quantity of generated network traffic
(refer to \fref{fig:didic_stress_comms_all}).

During the insert experiments, 
dynamism was applied to DiDiC partitionings with the goal of degrading their quality.
The degraded partitionings created as a result of that dynamism were used as input to this experiment.
Then one iteration of DiDiC was performed on each of them, and the improvement measured
(for more details refer to \sref{sec:dynamism} and \sref{sec:experiments}).

The results are very positive.
They indicate that, with respect to network traffic,
one iteration of DiDiC is sufficient to repair a partitioning that has been exposed to 
--- at least --- 25\,\% dynamism.

To create the partitionings used in static experiments, DiDiC was run for 100 iterations.
Furthermore, the authors of DiDiC performed up to 150 iterations while evaluating the algorithm.
For these reasons, 
it is probable that one iteration will not be sufficient when dynamism increases beyond a certain threshold.
Also, an important point to consider is that our dynamism only moves individual vertices,
but it never moves large connected subgraphs.
This allows DiDiC to more easily repair the partitioning,
because to diffuse load across a small number of misplaced vertices is less difficult than to do so across larger connected subgraphs
This --- at least in part --- explains why the results are so impressive.

\section{Dynamic Experiment}	

This experiment measured how effectively DiDiC could maintain the quality of a partitioning in the presence of ongoing dynamism. 

During execution of insert experiments the dynamism operations were logged to file, for future reuse.
Those dynamism logs were used in this experiment, executed intermittently between DiDiC iterations.
More specifically, 5\,\% dynamism was applied before each DiDiC iteration.
Experiments continued for six DiDiC iterations, terminating after the full 25\,\% dynamism had been applied
(for more details refer to \sref{sec:dynamism} and \sref{sec:experiments}).

\begin{figure}[htbp]
 \centering
\subfloat[File System]
{\includegraphics[scale=1.0]{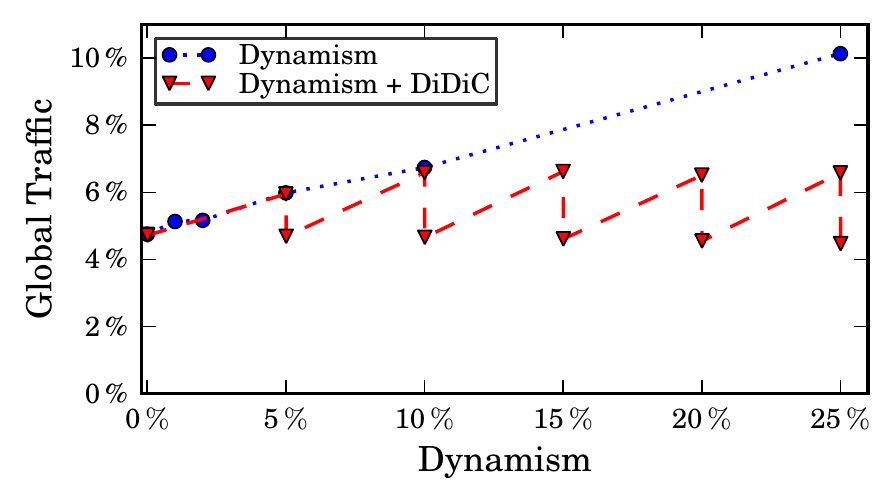}
\label{fig:fstree4_dynamic_comms_all}}
\qquad
\subfloat[GIS]
{\includegraphics[scale=1.0]{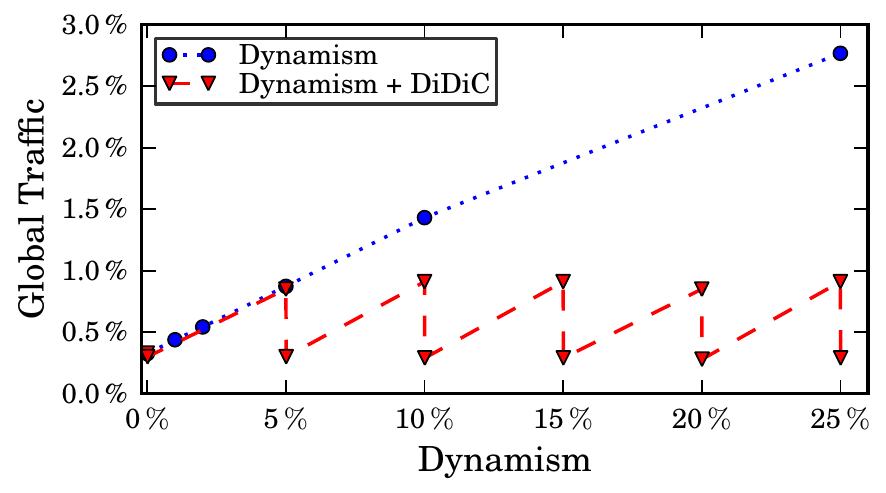}
\label{fig:gis4_dynamic_comms_all}}
\qquad
\subfloat[Twitter]
{\includegraphics[scale=1.0]{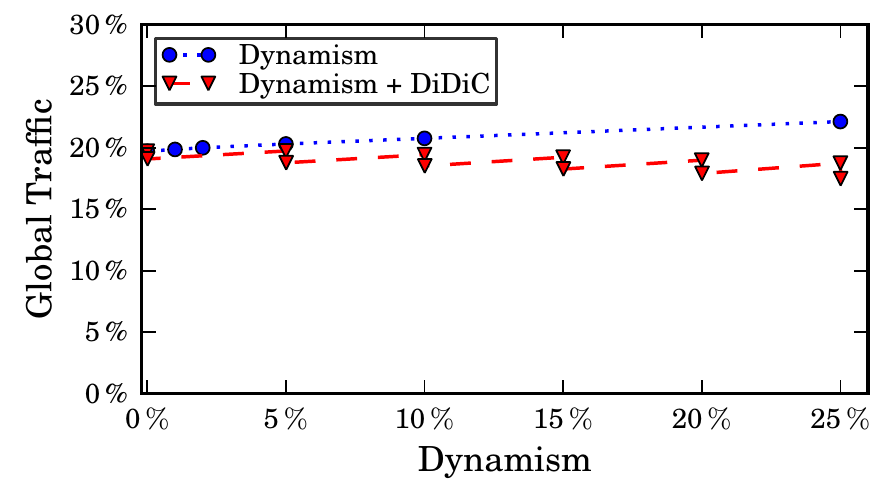}
\label{fig:twitter4_dynamic_comms_all}}
\caption{Network traffic vs dynamism --- Dynamic experiment --- 4 partitions}
\label{fig:dynamic_comms_all} 
\end{figure}

For the purpose of these experiments,
partitioning quality measurements were restricted to the quantity of generated network traffic.
Results are shown in \fref{fig:dynamic_comms_all}.
These results are impressive, showing that DiDiC not only maintained partitioning quality,
but slowly improved it as the experiment progressed.
Observe that, by selecting an appropriate interval for DiDiC execution,
an upper bound can be placed on the amount of degradation that a partitioning is permitted to experience.
As more dynamism is allowed to take place between DiDiC iterations, 
the level of partitioning degradation will increase.
Note that when setting this interval an important consideration needs to made, that of computation time.
With our sample datasets one DiDiC iteration required between 15--30 minutes to complete.

\section{Discussion}
We identified a correlation between edge cut, access patterns, and generated network traffic.
Note, this correlation existed in spite of the fact that all access patterns were non-uniform.
The correlation was formalized and supported by observations.

It was found that, when sufficient domain knowledge is available, 
application specific partitioning methods can yield very good results.
However, as shown by our Twitter dataset sufficient domain knowledge is not always attainable.
In these cases a graph partitioning algorithm like DiDiC may be suitable.

During evaluation DiDiC, using no domain knowledge, 
consistently produced partitionings of comparable quality to our hardcoded partitioning methods.
However, as shown in the results of our Twitter dataset,
the quality of partitionings returned by DiDiC depends on graph topology.
It is likely that this problem is not specific to DiDiC, but applies to graph partitioning in general.
Some graphs naturally contain clusters/communities of vertices that can be mapped to partitions.
Others are very dense (each vertex has connections to many other vertices), 
making it difficult to create a partitioning that results in a low edge cut.
Graphs with a random topology contain no community structures at all, any vertex may be connected to any other vertex with equal probability. Regardless of how these graphs are partitioned, a high edge cut is unavoidable.
Of our three datasets, Twitter had the lowest clustering coefficient.
With high probability, this contributed to the poor results produced by DiDiC.
Furthermore, Twitter represents the most complex graph evaluated in this work.
For example, in the GIS graph communities are clustered along two dimensions,
whereas in Twitter they are likely to comprise of many more dimensions,
making it more difficult to identify communities.


Results from our insert partitioning methods showed that, 
although they provided some benefits with regards to load balancing, 
more intelligent techniques are needed to impact edge cut. 
The random insert method showed a tendency to balance vertex distribution in all experiments. 
However, this is partly due to the way in which dynamism was generated.
In a production system the results may differ.

Regarding the maintenance of partitionings in the presence of dynamism,
it was shown that --- under the conditions of our evaluation environment ---
executing DiDiC for only one iteration was sufficient to repair a partitioning.
This was true even after that partitioning had been exposed to 25\,\% dynamism.
Although not the focus of our work, 
it is important to consider how long it takes for this amount of dynamism to occur.
The time available for an algorithm to update a partitioning is dependent on the level of dynamism in the system,
the lower the dynamism the greater the amount of time an algorithm may consume.
With our datasets one DiDiC iteration completed in 15--30 minutes.
However, the computation time of one iteration is linearly correlated to dataset size and partition count. 
Assuming that production systems store graphs several orders of magnitude larger than our sample datasets,
it may not be efficient to use DiDiC for database partitioning in production.
This does not invalidate the use of graph partitioning algorithms,
but highlights the importance of computational efficiency when selecting an algorithm.

\chapter{Conclusion}
\label{cha:conclusion}

%

In this work a set of abstractions were defined, which we consider necessary when partitioning graph databases.
These abstractions were then composed into a unifying framework and used for the design of our evaluation environment.

When performing traversals on partitioned graph datasets network traffic will inevitably be generated.
To illustrate this fact, 
we executed well-defined, 
non-uniform access patterns on various partitioned graph datasets, 
measuring the generated network traffic.
Results from our evaluations showed that datasets partitioned using a modularity-optimizing graph partitioning algorithm achieved a balanced load distribution, 
and generated significantly less network traffic than those that were partitioned randomly.
Additionally, we showed that application-specific partitioning methods can make use of domain knowledge to achieve very high quality partitionings. 
However these partitionings did not substantially outperform our graph partitioning algorithm results ---
which made no use of domain knowledge.

Due to its computational complexity, 
we conclude that DiDiC is well suited to partitioning graphs of a similar size to our sample datasets,
but impractical when working with larger graphs.
This does not invalidate the use of graph partitioning algorithms for partitioning graph databases.
Rather, it indicates that a faster modularity-optimizing algorithm must be used.
Results showed that modularity-optimizing algorithms effectively reduce network traffic while balancing load. 
As mentioned in \sref{sec:algorithm_overview}, numerous modularity-optimizing algorithms exist.



\section{Contributions}
%

We proposed the architecture of a novel graph partitioning framework,
capable of partitioning a dataset at the time of data insertion as well as during runtime to repair a partitioning that has degraded in quality.

A unique aspect of this work is the application of a graph partitioning algorithm to the problem of partitioning graph databases.
Behavior of the graph partitioning algorithm is examined in the presence of dynamism,
using datasets with different topologies, and of substantial size --- up to \numprint{785000} vertices and \numprint{1600000} edges.

To gauge partitioning quality we used various metric types,
including comparisons between theoretic quality metrics and measurements of generated network traffic and load distribution.
As an extension, 
we identified a correlation between these different metrics and presented it as a formal definition.
We are aware of no other evaluation on this topic, which is as comprehensive or of the same nature as ours.

%
%
%

\section{Future Work}
In this work the distinction between insert partitioning and runtime partitioning methods was introduced,
but few implementations of these methods were evaluated.
In the case of insert partitioning, the tested implementations were also very basic.
In a similar way that domain knowledge was used in our hardcoded partitioning methods,
it may be beneficial to consider graph topology during insert partitioning.
By knowing which region of the graph new entities are being written to,
insert operations can attempt to collocate new entities and their neighbors on the same partition.
Allowing for the insertion of subgraphs --- rather than individual entities --- 
could also be used to ensure all subgraph entities are written to the same partition.

Utilizing access patterns during runtime partitioning would be a natural extension of our work.
This may be possible by mapping access patterns to preexisting constraints of graph partitioning algorithms,
or via the development of new partitioning algorithms.
Likewise, comparisons and/or development of more graph partitioning algorithms would be interesting,
as would the implementation of these algorithms in a truly distributed environment --- rather than in a simulator.

To reduce complexity, data replication was never investigated in this work.
However, it may be possible to address many performance problems by designing intelligent replication schemes.
This may also lead to new problems, such as:
assessing if a replication scheme should operate at the granularity of individual graph entities, 
subgraphs, entire partitions, etc.;
finding optimal mappings of partitions to computers, such that network traffic is minimized;
and developing methods of enforcing consistency guarantees across replicas.

Lastly, for runtime partitioning to be feasible, an indexing structure with particular characteristics is required.
It is essential that entities can be efficiently stored and retrieved regardless of how often they move between partitions. 
For this purpose we suggest the development of a distributed indexing structure capable of performing data lookups via a primary dimension, and data placement via a secondary dimension.


\bibliography{references}
\bibliographystyle{plain}


\begin{appendix}
\appendixpage
\chapter{graph\_gen\_utils}
\label{apx:graph_gen_utils}

This library provides the functionality to load/store a Neo4j database instance from/to two different file formats,
Chaco and GML. 
This functionality was useful during implementation and testing of the graph partitioning algorithms.
Many sample graphs have been made publicly available in the Chaco format.
By using graph\_gen\_utils we were able to load them into Neo4j.
Being able to visualize these graphs, and the way an algorithm partitions them, 
is also beneficial during development.
The igraph \cite{ref:general50} library has rich graph visualization functionality and recognizes files in the GML format. 
Using graph\_gen\_utils we could export a Neo4j dataset to GML format and visualize it in igraph.

Graph partitioning algorithms are known to be computationally expensive,
but much performance can be gained by ensuring they execute in main memory --- rather than on disk.
Another feature of graph\_gen\_utils is the ability to load graphs into main memory,
it does this by providing an in-memory implementation, \texttt{MemGraph}, 
of the \texttt{GraphDatabaseService} interface.

The following list of descriptions covers a subset of the functions exposed by graph\_gen\_utils, 
along with brief explanations of them.

\begin{description}

\item[\texttt{applyPtnToNeo(GraphDatabaseService neo, Partitioner ptn)}:] 
\ \\ Assigns partition identifiers to the vertices in \texttt{neo} to simulate a partitioned database.
Partition allocation is defined by \texttt{ptn}.

\item[\texttt{writeNeoFromTopology(GraphDatabaseService neo, GraphTopology top)}:]
\ \\ Populates the Neo4j instance referenced by \texttt{neo}.
The number of vertices and edges, and the graph topology they form, is defined by \texttt{top}.
Topologies supported include fully connected, \texttt{GraphTopologyFullyConnected}, 
and random, \texttt{GraphTopologyRandom}.

\item[\texttt{writeNeoFromChaco(GraphDatabaseService neo, String inputChaco)}:] 
\ \\ Populates the Neo4j instance referenced by \texttt{neo}.
Contents of the Chaco file, \texttt{inputChaco}, 
defines the number of vertices and edges, and the graph topology they form.

\item[\texttt{writeNeoFromGML(GraphDatabaseService neo, String inputGml)}:] 
\ \\ Populates the Neo4j instance referenced by \texttt{neo}.
Contents of the GML file, \texttt{inputGML}, 
defines the number of vertices and edges, and the graph topology they form.

\item[\texttt{writeNeoFromTwitter(GraphDatabaseService neo, String inputTwitter)}:] 
\ \\ Populates the Neo4j instance referenced by \texttt{neo}.
Contents of the input file, \texttt{inputTwitter}, 
defines the number of vertices and edges, and the graph topology they form.
This function expects a proprietary file format,
it was developed only for importing our Twitter dataset during evaluation.

\item[\texttt{writeChaco(GraphDatabaseService neo, String outputChaco)}:] 
\ \\ Exports the Neo4j instance referenced by \texttt{neo}.
Contents of \texttt{neo} are persisted to file in the Chaco format.
The file destination is defined by \texttt{outputChaco}.

\item[\texttt{writeGML(GraphDatabaseService neo, String outputGML)}:] 
\ \\ Exports the Neo4j instance referenced by \texttt{neo}.
Contents of \texttt{neo} are persisted to file in the GML format.
The file destination is defined by \texttt{outputGML}.

\item[\texttt{readMemGraph(GraphDatabaseService neo)}:] 
\ \\ Loads the contents of a Neo4j instance, \texttt{neo}, into main memory, 
stored in a \texttt{MemGraph} object.

\end{description}	

\begin{figure}[htbp]
\begin{center}
\includegraphics[scale=1.0]{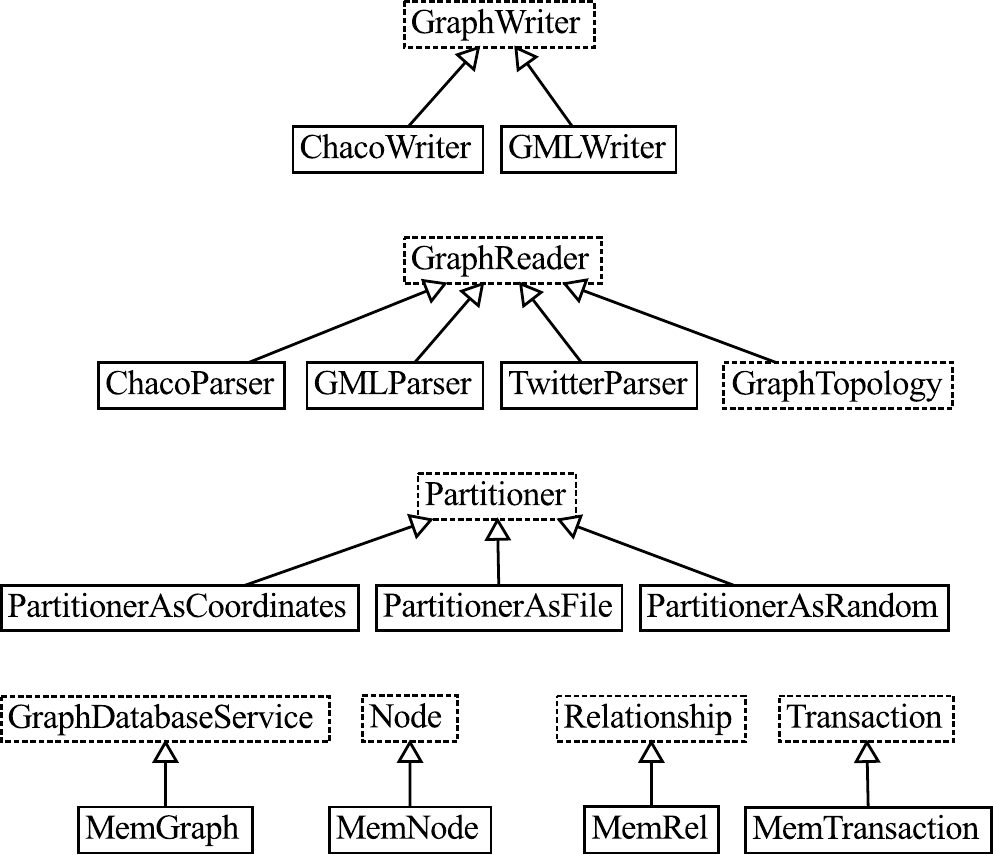} 
\caption{Simplified class diagram --- graph\_gen\_utils library}
\label{fig:graph_gen_utils}
\end{center}
\end{figure}

The class diagram in \fref{fig:graph_gen_utils} presents a subset of the classes in graph\_gen\_utils.
Interfaces and abstract classes are represented by boxes with dashed lines,
and all other classes by boxes with solid lines. Arrows with a white end denote inheritance.

\chapter{graph\_cluster\_utils}
\label{apx:graph_cluster_utils}

The graph\_cluster\_utils library contains prototypes of algorithms we explored.
These prototype algorithms implement the \texttt{PtnAlg} interface,
and use a \texttt{Conf} object to configure algorithm specific parameters.

As well as this, graph\_cluster\_utils implements prototypes for some other abstractions we defined in \fref{img:partitioned_graphdb}.
A Change-Log reader (\texttt{ChangeOpLogReader}) was implemented, 
which allows algorithms to adapt to dynamism.
Additionally, a Migration-Scheduler (\texttt{\texttt{Migrator}})
was used to decide when partitioning changes should be synchronized with the partitioned database
(\texttt{PGraphDatabaseService}).

\begin{figure}[htbp]
\begin{center}
\includegraphics[scale=1.0]{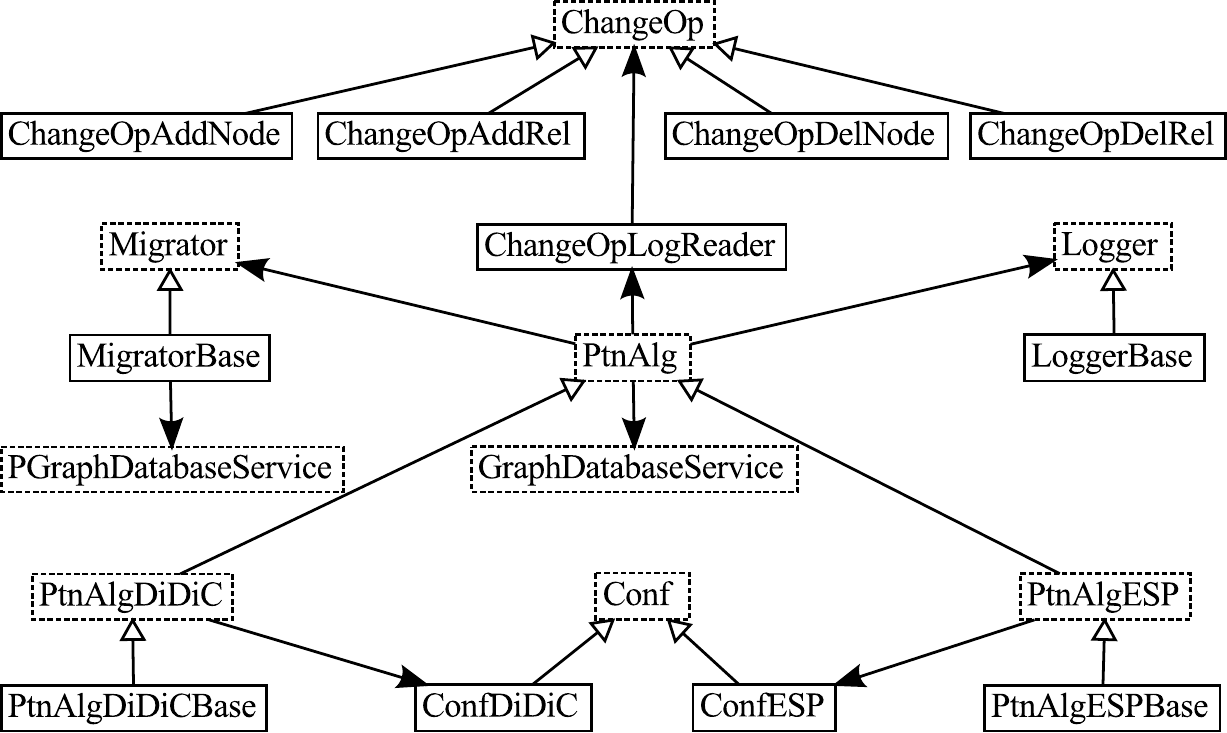} 
\caption{Simplified class diagram --- graph\_cluster\_utils library}
\label{fig:graph_cluster_utils}
\end{center}
\end{figure}

The class diagram in \fref{fig:graph_cluster_utils} presents a subset of the algorithms and other classes in graph\_cluster\_utils.
Interfaces and abstract classes are represented by boxes with dashed lines,
and all other classes by boxes with solid lines.
Arrows with a white end denote inheritance, and those with a black end denote composition.

\chapter{neo4j\_access\_simulator}
\label{apx:neo4j_access_simulator}

 
The neo4j\_access\_simulator library provides a simulation environment for creating, 
defining and executing evaluation logs on a Neo4j database. 
Individual accesses are performed by implementations of the \texttt{Operation} class.
Sequences of \texttt{Operation} instances as defined by an access pattern, and are created by \texttt{OperationFactory} implementations.
\texttt{Simulator} implementations take an \texttt{OperationFactory} as input, 
execute the \texttt{Operation}s it generates, 
then log those \texttt{Operation}s and their results to file. 
To rerun the same access patterns multiple times, 
\texttt{LogOperationFactory} implementations are capable of reading \texttt{Operation} definitions from a log file,
then generating those \texttt{Operation}s.
The \texttt{Rnd} class provides functionality for generating values according to random distributions. 
Each dataset has its own implementations of \texttt{OperationFactory},
\texttt{LogOperationFactory}, and \texttt{Operation}. 

\begin{figure}[htbp]
\begin{center}
\includegraphics[scale=1.0]{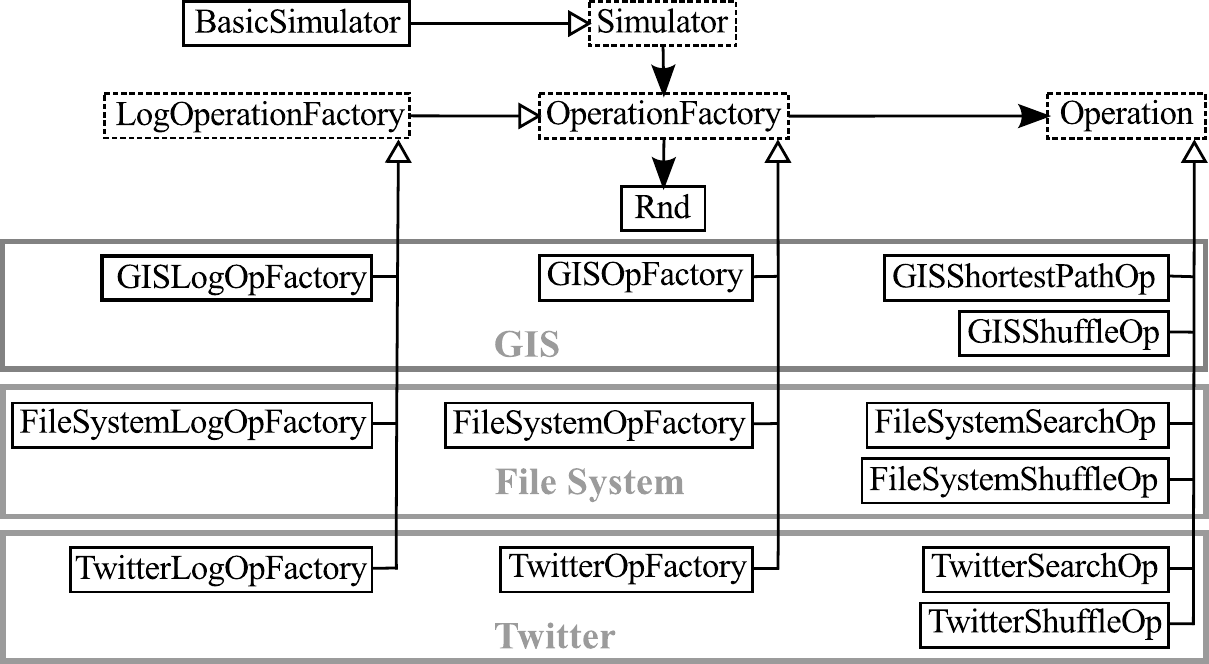} 
\caption{Simplified class diagram --- neo4j\_access\_simulator}
\label{fig:neo4j_access_simulator}
\end{center}
\end{figure}

The class diagram in \fref{fig:neo4j_access_simulator} illustrates the main classes of neo4j\_access\_simulator.
Interfaces and abstract classes are represented by boxes with dashed lines, 
and all other classes by boxes with solid lines. 
Arrows with a white end denote inheritance, and those with a black end denote composition.

\end{appendix}

\end{document}